\documentclass[aps,groupedaddress,floatfix,preprintnumbers,nofootinbib,eqsecnum]{revtex4}
\usepackage{amssymb,amsmath,graphicx,multirow}

\begin{document}

\preprint{UH511-1189-12}

\title{Phenomenological Constraints on Axion Models of Dynamical Dark Matter}
\author{Keith R. Dienes$^{1,2,3}$\footnote{E-mail address:  {\tt dienes@physics.arizona.edu}},
        Brooks Thomas$^{4}$\footnote{E-mail address:  {\tt thomasbd@phys.hawaii.edu}}}
\affiliation{
     $^1$ Physics Division, National Science Foundation, Arlington, VA  22230  USA\\
     $^2$ Department of Physics, University of Maryland, College Park, MD  20742  USA\\
     $^3$ Department of Physics, University of Arizona, Tucson, AZ  85721  USA\\
     $^4$ Department of Physics, University of Hawaii, Honolulu, HI 96822  USA}

\begin{abstract}
In two recent papers~\cite{DynamicalDM1,DynamicalDM2} we introduced 
``dynamical dark matter'' (DDM), a new framework for dark-matter physics 
in which the requirement of 
stability is replaced by a delicate balancing between lifetimes and cosmological 
abundances across a vast ensemble of individual dark-matter components whose collective 
behavior transcends that normally associated with traditional dark-matter candidates.  
We also presented an explicit model involving axions in large extra spacetime 
dimensions, and demonstrated that this model has all of the features necessary
to constitute a viable realization of the general DDM framework.
In this paper, we complete our study by performing a general analysis of all 
phenomenological constraints which are relevant to this bulk-axion DDM model.
Although the analysis in this paper is primarily aimed at our specific DDM model, 
many of our findings have important implications for bulk axion theories in general.  
Our analysis can also serve as a prototype for phenomenological studies of 
theories in which there exist large numbers of interacting and decaying particles.
\end{abstract}


\maketitle


\newcommand{\newc}{\newcommand}
\newc{\gsim}{\lower.7ex\hbox{$\;\stackrel{\textstyle>}{\sim}\;$}}
\newc{\lsim}{\lower.7ex\hbox{$\;\stackrel{\textstyle<}{\sim}\;$}}
\makeatletter
\newcommand{\biggg}{\bBigg@{3}}
\newcommand{\Biggg}{\bBigg@{4}}
\makeatother

\def\vac#1{{\bf \{{#1}\}}}

\def\beq{\begin{equation}}
\def\eeq{\end{equation}}
\def\beqn{\begin{eqnarray}}
\def\eeqn{\end{eqnarray}}
\def\calM{{\cal M}}
\def\calV{{\cal V}}
\def\calF{{\cal F}}
\def\half{{\textstyle{1\over 2}}}
\def\quarter{{\textstyle{1\over 4}}}
\def\ie{{\it i.e.}\/}
\def\eg{{\it e.g.}\/}
\def\etc{{\it etc}.\/}


\def\inbar{\,\vrule height1.5ex width.4pt depth0pt}
\def\IR{\relax{\rm I\kern-.18em R}}
 \font\cmss=cmss10 \font\cmsss=cmss10 at 7pt
\def\IQ{\relax{\rm I\kern-.18em Q}}
\def\IZ{\relax\ifmmode\mathchoice
 {\hbox{\cmss Z\kern-.4em Z}}{\hbox{\cmss Z\kern-.4em Z}}
 {\lower.9pt\hbox{\cmsss Z\kern-.4em Z}}
 {\lower1.2pt\hbox{\cmsss Z\kern-.4em Z}}\else{\cmss Z\kern-.4em Z}\fi}
\def\thbar{\bar{\theta}}
\def\fhatPQ{\hat{f}_{\mathrm{PQ}}}
\def\fPQ{f_{\mathrm{PQ}}}
\def\mPQ{m_{\mathrm{PQ}}}
\def\wtl{\widetilde{\lambda}}
\def\ta{\widetilde{a}}
\def\TBBN{T_{\mathrm{BBN}}}
\def\OmegaCDM{\Omega_{\mathrm{CDM}}}
\def\OmegaDM{\Omega_{\mathrm{CDM}}}
\def\Omegatot{\Omega_{\mathrm{tot}}}
\def\rhocrit{\rho_{\mathrm{crit}}}
\def\alMRE{a_{\lambda,\mathrm{M}}}
\def\aldotMRE{\dot{a}_{\lambda,\mathrm{M}}}
\def\tMRE{t_{\mathrm{MRE}}}
\def\TMRE{T_{\mathrm{MRE}}}
\def\tQCD{t_{\mathrm{QCD}}}
\def\tauMRE{\tau_{\mathrm{M}}}
\def\mPQdot{\dot{m}_{\mathrm{PQ}}}
\def\mPQddot{\ddot{m}_{\mathrm{PQ}}}
\def\mPQbar{\overline{m}_{\mathrm{PQ}}}
\def\mXdot{\dot{m}_X}
\def\mXddot{\ddot{m}_X}
\def\mXbar{\overline{m}_X}
\def\TRH{T_{\mathrm{RH}}}
\def\tRH{t_{\mathrm{RH}}}
\def\LambdaQCD{\Lambda_{\mathrm{QCD}}}
\def\fhatX{\hat{f}_X}
\def\tnow{t_{\mathrm{now}}}
\def\Omvac{\Omega_{\mathrm{vac}}^{(0)}}
\def\arcsinh{\mbox{arcsinh}}
\def\zRH{z_{\mathrm{RH}}}
\def\zMRE{z_{\mathrm{MRE}}}
\def\zinit{z_{\mathrm{init}}}
\def\tinit{t_{\mathrm{init}}}
\def\sinit{s_{\mathrm{init}}}
\def\sRH{s_{\mathrm{RH}}}
\def\sMRE{s_{\mathrm{MRE}}}
\def\sLS{s_{\mathrm{LS}}}
\def\snow{s_{\mathrm{now}}}
\def\BRgamma{\mathrm{BR}_{\lambda}^{(2\gamma)}}
\def\BRrad{\mathrm{BR}_{\lambda}^{(\mathrm{rad})}}
\def\te{t_{\mathrm{early}}}
\def\tl{t_{\mathrm{late}}}
\def\Ehi{E_{\mathrm{high}}}
\def\Elo{E_{\mathrm{low}}}
\def\tBBN{t_{\mathrm{BBN}}}
\def\tosc{t_{\mathrm{osc}}}
\def\Tosc{T_{\mathrm{osc}}}
\def\Tnow{T_{\mathrm{now}}}
\def\Tmax{T_{\mathrm{max}}}
\def\TLS{T_{\mathrm{LS}}}
\def\fhatX{\hat{f}_X}
\def\LambdaG{\Lambda_G}
\def\mX{m_X}
\def\tX{t_X}
\def\tG{t_G}
\def\OmegaDM{\Omega_{\mathrm{CDM}}}
\def\ntrans{n_{\mathrm{trans}}}
\def\nosc{n_{\mathrm{osc}}}
\def\ninf{n_{\mathrm{inf}}}
\def\nG{n_{G}}
\def\ndec{n_{\mathrm{dec}}}
\def\ncut{n_{\mathrm{cut}}}
\def\nexpl{n_{\mathrm{expl}}}
\def\tLS{t_{\mathrm{LS}}}
\def\Ech{E_{\mathrm{ch}}}
\def\rhorad{\rho_{\mathrm{rad}}}
\def\rhoradbar{\overline{\rho}_{\mathrm{rad}}}
\def\Omegavac{\Omega_{\mathrm{vac}}}
\def\etanow{\eta_\ast}
\def\lambdadec{\lambda_{\mathrm{dec}}}
\def\lambdatrans{\lambda_{\mathrm{trans}}}
\def\Omegatotnow{\Omega_{\mathrm{tot}}^\ast}
\def\Omegachinodec{\hat{\Omega}_\chi^\ast}
\def\Omegatotnodec{\hat{\Omega}_{\mathrm{tot}}^\ast}
\def\lambdafluc{\lambda_{\mathrm{fluc}}}
\def\Omegatotmis{\Omega_{\mathrm{tot}}^{(\mathrm{mis})}}
\def\Omegatotinf{\Omega_{\mathrm{tot}}^{(\mathrm{inf})}}
\def\Omegatotmisnow{\Omega_{\mathrm{tot}}^{\ast(\mathrm{mis})}}
\def\Omegatotinfnow{\Omega_{\mathrm{tot}}^{\st(\mathrm{inf})}}
\def\Omegatotjitternow{\widetilde{\Omega}_{\mathrm{tot}}^\ast}
\def\Omegaflucnow{\Omega_{\mathrm{fluc}}^\ast}
\newcommand{\Dsle}[1]{\hskip 0.09 cm \slash\hskip -0.28 cm #1}
\newcommand{\met}{{\Dsle E_T}}


\input epsf



\tableofcontents


\section{Introduction\label{sec:Introduction}}

 
Dynamical dark matter (DDM)~\cite{DynamicalDM1,DynamicalDM2} is a new framework for 
dark-matter physics in which the requirement of stability is replaced by a delicate 
balancing between lifetimes and cosmological abundances across a vast ensemble of 
individual dark-matter components.  Due to the range of lifetimes and abundances
of these components, their collective behavior transcends that normally associated with traditional dark-matter candidates.  In particular, quantities such as the total 
dark-matter relic abundance, the proportional composition of the ensemble in terms 
of its constituents, and the effective equation of state for the ensemble possess
a non-trivial time dependence beyond that associated with the expansion of the universe.
Indeed, from this perspective, DDM may be viewed as the most general possible framework for
dark-matter physics, and traditional dark-matter models are merely a limiting
case of the DDM framework in which the states which compose the dark sector are taken
to be relatively few in number and therefore stable.
  
In Ref.~\cite{DynamicalDM1}, we laid out the general theoretical features of the
DDM framework.  By contrast, in Ref.~\cite{DynamicalDM2}, we presented an explicit 
realization of the DDM framework: a model in which the particles which constitute the 
dark-matter ensemble are the KK excitations of an axion-like field propagating in the 
bulk of large extra spacetime dimensions.  We demonstrated that this model has 
all of the features necessary to constitute a viable realization of the general DDM 
framework.  In this paper, we complete our study 
by performing a general analysis of all phenomenological constraints which are 
relevant to this bulk-axion DDM model.
Although the analysis in this paper is primarily aimed at our specific DDM model, 
many of our findings have important implications for  
theories involving large extra dimensions in general.  Furthermore, our analysis 
can also serve as a prototype for phenomenological studies of theories in which 
there exist large numbers of interacting and decaying particles.

It is important to emphasize why a general analysis of this sort is necessary, 
given the existence of numerous prior studies
of the phenomenological and cosmological constraints on 
axions and axion-like fields, unstable relics, and the physical properties of 
miscellaneous dark-matter candidates. 
As discussed in Refs.~\cite{DynamicalDM1,DynamicalDM2},
such studies are typically applicable to dark sectors involving one or only a few 
fields, and are typically quoted in terms of limits on the mass, decay width, 
or couplings of any individual such field.  It is not at all obvious how such bounds
apply to a DDM {\it ensemble} --- a dark-matter candidate which is not 
characterized by a well-defined single mass, decay width, or set of couplings.       
For example, constraints on a cosmological population of unstable 
particles derived from big-bang-nucleosynthesis (BBN) considerations 
or bounds on distortions in the cosmic microwave background (CMB) are 
generally derived under the assumption that such a population 
comprises but a single particle species with a well-defined lifetime 
and branching fractions.  Such constraints are not directly 
applicable to a DDM ensemble (in which lifetimes are balanced against 
abundances), and must therefore be reexamined in this new context.     
    
In this paper, we shall develop methods for dealing with these issues
and for properly characterizing the constraints on models in which the dark-matter 
candidate is an ensemble of states rather than a single particle.
As we shall find, the presence of non-trivial mixings among the KK 
excitations in our DDM model gives rise to a number of surprising effects 
which are ultimately critical for its phenomenological viability.  One
of these is a so-called ``decoherence'' 
phenomenon~\cite{DDGAxions,DynamicalDM1,DynamicalDM2} which helps to explain 
how the dark matter in this model remains largely invisible to detection. 
Another is a suppression, induced by this mixing, of the couplings between the
lighter particles in the dark-matter ensemble and the fields of the 
Standard Model (SM).  As we shall see, these effects assert themselves
in a variety of phenomenological contexts and play a crucial role in 
loosening a battery of constraints which would otherwise prove extremely severe.    

This paper is organized as follows.
In Sect.~\ref{sec:AxionsInED}, we briefly summarize the physics of axions in
extra dimensions and review the notational conventions 
established in Ref.~\cite{DynamicalDM2}, which we once again adopt in this work.
In Sect.~\ref{sec:AbundanceConstraints}, we examine a number of processes, both 
thermal and non-thermal in nature, which contribute to the generation of a 
cosmological population of axions.  We calculate the rates associated 
with these processes and assess the relative importance of the associated production
mechanisms within different regions of model-parameter space.
In Sect.~\ref{sec:Bounds}, we then discuss the phenomenological, astrophysical, and
cosmological constraints relevant for bulk-axion DDM models and assess how
the parameter space of our model is bounded by each of these constraints.  In
Sect.~\ref{sec:Combined}, we summarize the collective consequences of these 
constraints on the parameter space of our bulk-axion DDM model.  Finally,
in Sect.~\ref{sec:Conclusions}, we discuss the implications of our results for 
future research.     


\section{Generalized Axions in Extra Dimensions: A Review\label{sec:AxionsInED}}      


In this section, we provide a brief review of the physics of generalized 
axions in extra dimensions.  (More detailed reviews can be found in 
Refs.~\cite{DDGAxions,DynamicalDM2}.)  By ``generalized axion,'' we mean any 
pseudo-Nambu-Goldstone boson which receives its mass from instanton effects
related to a non-Abelian gauge group $G$ which confines at some scale $\Lambda_G$.     
Note that the ordinary QCD axion~\cite{PecceiQuinn,WeinbergWilczekAxion}
is a special case of this, in which $G$ is 
identified with $SU(3)$ color and $\Lambda_G$ is identified with 
$\LambdaQCD\approx 250$~MeV.  However, in this paper, we shall leave these 
scales arbitrary in order to give our analysis a wider range of applicability.  
We will also assume the existence of a global Abelian symmetry $U(1)_X$ which
plays the role played by the Peccei-Quinn symmetry $U(1)_{\mathrm{PQ}}$ in the
specific case of a QCD axion.  

Our goal in this paper is to study the phenomenological constraints 
that arise when a generalized axion is allowed to propagate in the 
bulk~\cite{DDGAxions} of a theory with extra spacetime 
dimensions~\cite{ADD,ADDPhenoBounds}. 
In particular, we consider the case in which the axion propagates in a single, 
large, flat extra dimension compactified on a $S_1/\IZ_2$ orbifold of 
radius $R$.  The fields of the SM are assumed to be restricted to a 
brane located at $x_5=0$.  We also assume that the additional 
non-Abelian gauge group $G$ is restricted to the brane at $x_5=0$.  
At temperatures $T \gg \Lambda_G$, the effective action for a bulk axion
in five dimensions can be written in
the form
\begin{equation}
   S_{\mathrm{eff}} ~=~ \int d^4x\int_0^{2\pi R} dx_5
     \left[\frac{1}{2}\partial_M a \partial^M a + 
     \delta(x_5)\,\big(\mathcal{L}_{\mathrm{brane}}+\mathcal{L}_{\mathrm{int}}\big)\right]~,
\end{equation}
where `$a$' denotes our five-dimensional axion field, 
$\mathcal{L}_{\mathrm{brane}}$ contains the terms involving the brane fields alone, 
and $\mathcal{L}_{\mathrm{int}}$ contains the interaction terms coupling the 
brane-localized fields to the five-dimensional axion.  The second of these terms is
given by
\begin{equation}
  \mathcal{L}_{\mathrm{int}} ~=~ 
     \frac{g_G^2\xi}{32\pi^2f_X^{3/2}} a \mathcal{G}_{\mu\nu}^a\tilde{\mathcal{G}}^{a\mu\nu}+
     \sum_i\frac{c_i}{f_X^{3/2}}
     (\partial_\mu a)\overline{\psi}_i\gamma^\mu\gamma^5\psi_i 
     +\frac{g_s^2 c_g^2}{32\pi^2f_X^{3/2}} a G_{\mu\nu}^a\tilde{G}^{a\mu\nu}
     +\frac{e^2 c_{\gamma}}{32\pi^2f_X^{3/2}} 
     a F_{\mu\nu}\tilde{F}^{\mu\nu} +\ldots~,
  \label{eq:HighT5DAxionAction}
\end{equation} 
where $F_{\mu\nu}$, $G_{\mu\nu}^a$, and $\mathcal{G}_{\mu\nu}^a$ are the field strengths 
respectively associated with the $U(1)_{\mathrm{EM}}$, $SU(3)$ color, and $G$ gauge
groups; $\tilde{F}_{\mu\nu}$, $\tilde{G}_{\mu\nu}^a$, and $\tilde{\mathcal{G}}^a_{\mu\nu}$ are
their respective duals; $e$, $g_s$, and $g_G$ are the respective coupling constants for 
these groups; $f_X$ is the fundamental five-dimensional scale associated with the 
breaking of the $U(1)_X$ symmetry; $c_\gamma$, $c_g$, and $c_i$ are coefficients 
which respectively parametrize the coupling strength of the five-dimensional axion 
field to the photon, gluon, and fermion fields of the SM; and $\xi$ is an $\mathcal{O}(1)$
coefficient which depends on the specifics of the axion model in question.  Note that in 
Eq.~(\ref{eq:HighT5DAxionAction}), we have displayed terms 
involving only the light fields of the SM (\ie, the photon, 
gluon, and light fermion fields), as couplings to the heavier SM fields will not 
play a significant role in our phenomenological analysis.   
   
The five-dimensional axion field can be
represented as a tower of four-dimensional KK excitations via the decomposition
\begin{equation}
  a(x^\mu,x_5) ~=~ \frac{1}{\sqrt{2\pi R}}\sum_{n=0}^{\infty} 
     r_n a_n(x^\mu)\cos\left(\frac{nx_5}{R}\right)~,
  \label{eq:AxionModeDecomp}
\end{equation}    
where the factor
\begin{equation}
  r_n ~\equiv~ \begin{cases}
  1 &\mathrm{for~} n=0\\
  \sqrt{2}  &\mathrm{for~} n>0
  \end{cases}
  \label{eq:rmDef}
\end{equation}
ensures that the kinetic term for each mode is canonically normalized.
Substituting this expression into Eq.~(\ref{eq:HighT5DAxionAction}) and 
integrating over $x_5$, we obtain
\begin{eqnarray}
  S_{\mathrm{eff}} &=& \int d^4x\Bigg[\sum_{n=0}^\infty
     \bigg(\frac{1}{2}\partial_\mu a_n \partial^\mu a_n 
     +\frac{g_G^2\xi}{32\pi^2\fhatX} 
     r_n a_n \mathcal{G}^a_{\mu\nu}\tilde{\mathcal{G}}^{a\mu\nu}
     +\sum_i\frac{c_i}{\fhatX}
     r_n(\partial_\mu a_n)\overline{\psi}_i\gamma^\mu\gamma^5\psi_i 
     \nonumber\\ & & ~~~~~~~~~~~     
     +~\frac{g_s^2c_g}{32\pi^2\fhatX} 
     r_n a_n G^a_{\mu\nu}\tilde{G}^{a\mu\nu}
     +\frac{e^2 c_{\gamma}}{32\pi^2\fhatX} 
     r_n a_n F_{\mu\nu}\tilde{F}^{\mu\nu}\bigg)
     -V(a)\Bigg]~,
  \label{eq:HighT4DAxionAction}
\end{eqnarray}  
where the axion potential is given by 
\begin{equation}
  V(a) ~=~ \sum_{n=0}^\infty\frac{1}{2}\frac{n^2}{R^2}a_n^2~,
\end{equation}
and where the quantity $\fhatX$, defined by the relation
\begin{equation}
  \fhatX^2 ~\equiv~ 2 \pi Rf_X^3~,
  \label{eq:fhatInTermsOff}
\end{equation}
represents the effective 
four-dimensional $U(1)_X$-breaking scale.  Note that each mode in the KK tower couples
to the SM fields with a strength inversely proportional to $\fhatX$.
Also note that at these scales, the axion mass-squared matrix 
\begin{equation}
  \mathcal{M}^2_{mn} ~\equiv~ \left.\frac{\partial^2 V(a)}{\partial a_m\partial a_n}
    \right|_{\langle a\rangle}
\end{equation}  
is purely diagonal.

At scales $T \lesssim \Lambda_G$, an additional contribution 
to the effective axion potential arises due to
instanton effects.  In this regime, the potential is modified to
\begin{equation}
   V(a) ~=~ \sum_{n=0}^\infty\frac{1}{2}\frac{n^2}{R^2}a_n^2+
          \frac{g_G^2}{32\pi^2}\Lambda_G^4
          \left[1-\cos\left(\frac{\xi}{\fhatX}
          \sum_{n=0}^\infty r_n a_n + \overline{\Theta}_G\right)\right]~, 
  \label{eq:InstantonPotential}
\end{equation}     
where $\overline{\Theta}_G$ is the analogue of the QCD theta-parameter  
$\overline{\Theta}$.  This results in a modification of the axion 
mass-squared matrix to
\begin{equation}
  \mathcal{M}^2_{mn} ~=~ n^2M_c^2\delta_{mn}
  +\frac{g_G^2\xi^2}{32\pi^2}\frac{\Lambda_G^4}{\fhatX^2}
  r_mr_n~,
  \label{eq:MassMixMatmn}
\end{equation}
where $M_c \equiv 1/R$ is the compactification scale. 
This matrix above takes the form~\cite{DDGAxions}
\begin{equation}
  \mathcal{M}^2 ~=~ \mX^2\left(\begin{array}{ccccc}
  1 & \sqrt{2} & \sqrt{2} & \sqrt{2} & \ldots \\
  \sqrt{2} & 2 + y^2 & 2 & 2 & \ldots \\
  \sqrt{2} & 2 & 2+4y^2 & 2 & \ldots \\
  \sqrt{2} & 2 & 2 & 2+9y^2 & \ldots \\
  \vdots & \vdots & \vdots & \vdots & \ddots 
  \end{array}\right)~,
  \label{eq:AxionMassMatrixExplicit}
\end{equation}   
where 
\begin{equation}
  y~\equiv~ \frac{M_c}{\mX}~~~~~~~ \mathrm{and}~~~~~~~
  \mX^2 ~\equiv~ \frac{g_G^2\xi^2}{32\pi^2}\frac{\Lambda_G^4}{\fhatX^2}~.
  \label{eq:DefsOfyandmPQ}
\end{equation}
The eigenvalues $\lambda^2$ of this mass-squared matrix are the
solutions to the transcendental equation
\begin{equation}
  \frac{\pi\lambda\mX}{y}\cot\left(\frac{\pi\lambda}{\mX y}\right) ~=~ \lambda^2~.
  \label{eq:TranscendentalEqForLambdas}
\end{equation}
The corresponding normalized mass eigenstates $a_\lambda$ are related to the 
KK-number eigenstates $a_n$ via 
\begin{equation}
  a_\lambda ~=~ \sum_{n=0}^\infty U_{\lambda n} a_n
    ~\equiv~\sum_{n=0}^\infty\left(
    \frac{r_n\widetilde{\lambda}^2}{\widetilde{\lambda}^2-n^2y^2}\right)A_\lambda a_n~,
  \label{eq:DefOfalambda}
\end{equation}
where $\widetilde{\lambda}\equiv\lambda/\mX$.  The dimensionless quantity $A_\lambda$ is
given by
\begin{equation}
  A_\lambda ~\equiv~ \frac{\sqrt{2}}{\wtl}\left[1+ \wtl^2 + \pi^2/y^2\right]^{-1/2}~.
  \label{eq:DefOfCapitalAlambda}
\end{equation}
and obeys the sum rules~\cite{DDGAxions}
\begin{equation}
  \sum_\lambda A_\lambda^2 ~=~ 1~, ~~~~~~
  \sum_\lambda \wtl^2A_\lambda^2 ~=~ 1~.
  \label{eq:AlambdaSqdID}
\end{equation}  

For $T\ll \hat{f}_X$, rewriting Eq.~(\ref{eq:HighT4DAxionAction}) in terms of the 
$a_\lambda$ and expanding the axion potential given in 
Eq.~(\ref{eq:InstantonPotential}) out to $\mathcal{O}(a_\lambda^6/\fhatX^6)$ 
yields the effective action 
\begin{eqnarray}
  S_{\mathrm{eff}} &=& \int d^4x
     \Bigg[\sum_{\lambda}\bigg(\frac{1}{2}\partial_\mu a_\lambda 
     \partial^\mu a_\lambda -\frac{1}{2}\wtl^2\mX^2 a_\lambda^2
     +\frac{e^2 c_{\gamma}\wtl^2A_\lambda}{32\pi^2\fhatX} 
     a_\lambda F_{\mu\nu}\tilde{F}^{\mu\nu}
     +\frac{g_s^2 c_g\wtl^2A_\lambda}{32\pi^2\fhatX} 
     a_\lambda G_{\mu\nu}^a\tilde{G}^{\mu\nu a}  
     \nonumber\\ & & ~~~~~ 
     + \sum_i\frac{c_i\wtl^2A_\lambda}{\fhatX}
     (\partial_\mu a_\lambda)\overline{\psi}_i\gamma^\mu\gamma^5\psi_i\bigg) 
     +\frac{g_G^2\xi^4\Lambda_G^4}{768\pi^2\fhatX^4}
     \sum_{\lambda_i,\lambda_j,\lambda_k,\lambda_\ell}
     \hspace{-0.4cm}\wtl_i^2\wtl_j^2\wtl_k^2\wtl_\ell^2
     A_{\lambda_i}A_{\lambda_j}A_{\lambda_k}A_{\lambda_\ell}
     a_{\lambda_i}a_{\lambda_j}a_{\lambda_k}a_{\lambda_\ell}
     \Bigg]~.
  \label{eq:ActionInMassEigenbasis}
\end{eqnarray}
Of course, the interaction term between the $a_\lambda$ and the gluon field
is only a useful description of the physics at temperatures above 
the quark-hadron phase transition at $T \sim \LambdaQCD$.  At 
temperatures below this threshold, this interaction term 
gives rise to an effective Lagrangian containing interactions 
between the $a_\lambda$ and various hadrons, including the proton $p$,
the neutron $n$, and the charged and neutral pions $\pi^\pm$ and $\pi^0$.
This Lagrangian takes the form 
\begin{eqnarray}
   \mathcal{L}_{\mathrm{had}} &=& \wtl^2A_\lambda
       \frac{C_{a\pi}}{f_\pi\fhatX}(\partial_\mu a_\lambda)
       \Big[(\partial^\mu \pi^+)\pi^-\pi^0 + (\partial^\mu \pi^-)\pi^+\pi^0
       -2(\partial^\mu \pi^0)\pi^+\pi^-\Big] ~+~
       \wtl^2A_\lambda\frac{C_{an}}{\fhatX}(\partial_\mu 
         a_\lambda)\overline{n}\gamma^\mu\gamma^5n \nonumber\\ & & ~+~
         \wtl^2A_\lambda\frac{C_{ap}}{\fhatX}(\partial_\mu 
         a_\lambda)\overline{p}\gamma^\mu\gamma^5p ~+~ 
         i\wtl^2A_\lambda\frac{C_{a\pi N}}{f_\pi\fhatX}(\partial_\mu a_\lambda)
         \Big[\pi^+\overline{p}\gamma^\mu n - \pi^-\overline{n}\gamma^\mu p\Big]~,
   \label{eq:HadronAxionCouplings5D}
\end{eqnarray} 
where the coefficients $C_{a\pi}$, $C_{an}$, \etc, depend on the details of the
theory.  For example, for a ``hadronic'' QCD axion~\cite{KSVZ} (\ie, a QCD axion which 
does not couple directly to the SM quarks), the coefficients $C_{ap}$ and $C_{an}$, 
which determine the strength of the axion-nucleon-nucleon interactions, are   
\begin{equation}
  C_{ap} ~=~ 0.24 \left(\frac{z}{1+z}\right) + 
       0.15 \left(\frac{z-2}{1+z}\right) + 0.02~,
  ~~~~~~~~
  C_{an} ~=~ 0.24 \left(\frac{z}{1+z}\right) + 
      0.15 \left(\frac{1-2z}{1+z}\right) + 0.02~,
\label{eq:CapAndCan}\\
\end{equation}
where $z=m_u/m_d\approx 0.56$ is the ratio of the up-quark and down-quark masses.
Likewise, the coefficients $C_{a\pi N}$ and $C_{a\pi}$ for such an axion are 
\begin{equation}
  C_{a\pi N} ~=~ \frac{1-z}{2\sqrt{2}(1+z)}~,~~~~~~~
  ~~~~~~C_{a\pi} ~=~ \frac{1-z}{3(1+z)}~,
\label{eq:CaNAndCapi}
\end{equation}
where $f_{\pi}\approx 93$~MeV is the pion decay constant and 
$m_\pi\approx 135.0$~MeV is the neutral pion mass.
      
Before concluding this review,
we note that the effective coupling coefficients $c_\gamma$, $c_g$, and $c_i$ 
appearing in Eq.~(\ref{eq:HighT5DAxionAction}) are highly 
model-dependent.  They need not be $\mathcal{O}(1)$, and in many theories 
any of them may vanish outright.  Indeed it has been 
argued~\cite{StringAxiverse} that the existence of axions and 
axion-like fields which couple to electromagnetism but not to $SU(3)$ 
color is a generic feature of certain extensions of the SM, including 
string theory.  In assessing the constraints on our bulk-axion DDM model, 
we shall therefore focus primarily on a ``photonic'' axion of this sort --- \ie, 
a general axion with $c_g = 0$ and $c_\gamma \neq 0$.  However, we shall also  
discuss how such phenomenological constraints are modified in the
case of a so-called ``hadronic'' axion with non-vanishing values for both 
$c_g$ and $c_\gamma$.  We note that additional subtleties arise in this latter case, 
due to non-trivial mixings between the $a_n$ and other pseudoscalars present in 
the theory which also necessarily couple to $G_{\mu\nu}^a\widetilde{G}^{a\mu\nu}$.  
These include hadrons such as $\pi^0$ and $\eta$, as well as any other axions 
in the theory which play a role in 
addressing the strong-CP problem~\cite{PecceiQuinn,WeinbergWilczekAxion}.
In discussing constraints on hadronic axions, we shall implicitly assume that the 
full mass-squared matrix for the theory is such that the relationship between
the $a_n$ and the mass eigenstates $a_\lambda$ defined in 
Eq.~(\ref{eq:DefOfalambda}) is not significantly disturbed.  
Indeed, given the inherently large number of independent scales and couplings 
that emerge in scenarios involving multiple axions and other pseudoscalars, this 
is not an unreasonable assumption; moreover, it is straightforward to show in a 
general way that these favorable conditions can always be arranged for certain 
sets of axion and pseudo-scalar mixings.  Such an assumption thereby enables us
to perform our phenomenological analysis in a model-independent way.


\section{Axion Production in the Early Universe\label{sec:AbundanceConstraints}}


Axions and axion-like fields can be produced via a number of different mechanisms 
in the early universe.  For example, these particles can be produced thermally, 
via their interactions with the SM fields in the radiation bath.  In
addition, a number of non-thermal mechanisms exist through which a sizable 
population of axions also may be generated.  These include production via 
vacuum misalignment, production from the decays of cosmic strings and other 
topological defects, and production from the out-of-equilibrium decays of 
other, heavier fields in the theory.  This last mechanism is particularly 
relevant in the context of the DDM models, since, by assumption, the 
dark sector in such models involves large numbers of unstable fields with 
long lifetimes.  Indeed, in the axion DDM model under consideration in this paper,   
a non-thermal population of any $a_\lambda$ may be produced via the decays 
of both heavier KK gravitons and other heavier $a_\lambda$.

In Ref.~\cite{DynamicalDM2}, we focused on misalignment production as the 
primary mechanism responsible for establishing a cosmological population of 
dark axions.  In order for the results for the relic abundances $\Omega_\lambda$ 
of the $a_\lambda$ obtained there to be valid, the contributions from all 
of the alternative production mechanisms mentioned above must be 
subdominant for each $a_\lambda$.  Therefore, in this section, we examine 
each of the relevant axion-production mechanisms in turn, beginning with a 
brief review of the results for misalignment production itself.  Since 
phenomenological constraints on scenarios involving large, flat extra dimensions 
prefer that the reheating temperature $T_R$ associated with cosmic inflation be 
quite low~\cite{ADDPhenoBounds}, we will hereafter operate within the context
of a low-temperature-reheating (LTR) cosmology with 
$T_R \sim \mathcal{O}(\mathrm{MeV})$.  Within such a cosmological context 
and within the region of model-parameter space in which misalignment 
production yields a total relic abundance $\Omegatot$ comparable to the observed 
dark-matter relic abundance $\OmegaCDM$, we demonstrate that the contributions to each 
$\Omega_\lambda$ from all other production mechanisms are indeed subdominant.


\subsection{Axion Production from Vacuum Misalignment\label{sec:MisalignmentProd}}


We begin our discussion of axion production in the early universe with
a brief review of the misalignment mechanism and its implications for 
axion DDM models.  (A more detailed discussion can be found in 
Ref.~\cite{DynamicalDM2}.)  As we shall see, this mechanism turns out 
to be the dominant production mechanism for dark-matter axions in such models.

At temperatures $T\gg \Lambda_G$, the 
only contributions to the axion mass-squared matrix are the
contributions from the KK masses.  Since these contributions to 
$\mathcal{M}^2$ are diagonal in the KK eigenbasis, no mixing occurs, 
and the KK eigenstates are the mass eigenstates of the theory.  
The potential for each $a_n$ with $n \neq 0$ is therefore 
non-vanishing, due to the presence of the KK masses, and is 
minimized at $a_n = 0$.  However, the potential for the zero mode $a_0$ 
vanishes.  In the absence of a potential for $a_0$, there is no preferred 
vacuum expectation value (VEV) $\langle a_0\rangle$ which minimizes $V(a_0)$.
It therefore follows that immediately following the phase transition at 
$T\sim \hat{f}_X$, the universe comprises a set of domains, each with
a different homogeneous background value for the axion field which may
be expressed in terms of a ``misalignment angle'' 
$\theta \equiv \langle a_0\rangle/\fhatX$.
This angle is generically expected to be $\mathcal{O}(1)$ in any particular
domain, but could also be smaller.  We assume here that 
$H_I \lesssim 2\pi \hat{f}_X$, where $H_I$ is the value of the Hubble parameter
during inflation, and therefore that the value of $\theta$ is 
uniform over our present Hubble volume.  In this case, we find that
\begin{equation}
  \langle a_0\rangle = \theta\fhatX~, ~~~~~~~~~~~~~ 
  \langle a_n\rangle = 0 \mathrm{~~~~for~~~} n\neq 0~. 
  \label{eq:a0initcondits}
\end{equation}

Note that the above discussion is strictly valid only in the limit in which the 
Hubble volume is taken to infinity.  In reality, the presence of a finite Hubble volume 
limits our ability to distinguish fields with wavelengths larger than the Hubble 
radius from true background values.  Because of this ambiguity,  
all $a_n$ for which $n/R \lesssim H_I$ can also acquire $\mathcal{O}(1)$ 
background values after $U(1)_X$ breaking.  
In Sect.~\ref{sec:InflationScale}, we will analyze the phenomenological 
consequences of this effect in detail and derive conditions under which
it can be safely neglected.  As we shall demonstrate, it turns out that within our 
preferred region of parameter space, these conditions involve only
mild restrictions on the cosmological context into which our model is 
embedded.  We will therefore assume from this point forward that the 
$\langle a_n\rangle$ in our model are given by Eq.~(\ref{eq:a0initcondits}).

At temperatures down to $T \sim \Lambda_G$, the $\langle a_n\rangle$ 
remain fixed at these initial values.  At lower temperatures, however, 
the situation changes as instanton effects generate a potential for 
the axion KK modes.  Indeed, in the regime in which $T \ll \Lambda_G$ and 
the brane mass engendered by this potential has attained the constant, 
low-temperature value $\mX$ given in Eq.~(\ref{eq:DefsOfyandmPQ}), the 
time-evolution of each field $a_\lambda$ is governed by an equation of the form 
\begin{equation}
   \ddot{a}_\lambda + \frac{\kappa}{t}\dot{a}_\lambda +
     \Gamma_\lambda\dot{a}_\lambda + \lambda^2 a_\lambda  ~=~ 0~,
   \label{eq:TheDoubleDotEqnWithGammaLambda}
\end{equation}  
where each dot denotes a time derivative, and where 
\begin{equation}
  \kappa ~\equiv~ 
     \begin{cases}
     3/2 &\mbox{in radiation-dominated (RD) eras}\\
     2 & \mbox{in matter-dominated (MD) eras}~. 
     \end{cases}
  \label{eq:DefOfkappaForH}
\end{equation} 
When $\lambda\lesssim 3H/2$, the solution to this equation remains 
approximately constant.   This implies that the energy density stored in 
$a_\lambda$ scales approximately like vacuum energy during this epoch.  However, 
at later times, when $\lambda\gtrsim 3H/2$, we see that $a_\lambda$ 
oscillates coherently around the minimum of its potential, with oscillations 
damped by a ``friction'' term with coefficient $3H + \Gamma_\lambda$.  
During this latter epoch, the energy density stored in $a_\lambda$ scales 
like massive matter.

At temperatures $T \sim \Lambda_G$, the evolution of $a_\lambda$ depends more
sensitively on the explicit time-dependence of the brane mass $\mX(t)$.
In what follows, we adopt a ``rapid-turn-on'' approximation, in which 
the instanton potential is assumed to turn on instantaneously at $t = t_G$,
where $t_G$ is the time at which the confining transition for the gauge 
group $G$ occurs.  In this approximation, $\mX(t)$ takes the form of a 
Heaviside step function:  
\begin{equation}
  \mX(t) ~=~ \mX \Theta(t - \tG)~.
  \label{eq:Heaviside}
\end{equation}
In this approximation, the $a_n$ remain 
fixed at the initial values given in Eq.~(\ref{eq:a0initcondits}) so 
long as $t < t_G$.  At $t=t_G$, the brane mass immediately assumes its  
constant, late-time value $\mX$.  Since only $a_0$ is populated
immediately prior to the phase transition at $t_G$, each of the $a_\lambda$
initially acquires a background value proportional to 
its overlap with $a_0$: 
\begin{equation}
  \langle a_\lambda(t_G)\rangle ~=~ \theta\fhatX A_\lambda~,~~~~~~
  \langle \dot{a}_\lambda(t_G)\rangle  ~=~ 0~.
  \label{eq:alambdaInitCondits}
\end{equation}
Subsequently, after the $a_\lambda$ have been populated, each begins 
oscillating at a characteristic time scale 
\begin{equation}
  t_\lambda ~\equiv~ \max\left\{\frac{\kappa_\lambda}{2\lambda},\tG\right\}~,
  \label{eq:tlambdaInBothRegimes}
\end{equation}
where $\kappa_\lambda$ is the value of $\kappa$ corresponding to the epoch during 
which this oscillation begins.
At late times $t \gg t_\lambda$, when these oscillations become rapid compared to 
the rate of change of $\langle a_\lambda \rangle$ and the virial approximation is 
therefore valid, one finds that the energy density $\rho_\lambda$ stored in each mode 
is given by 
\begin{equation}
  \rho_\lambda(t) ~=~ \frac{1}{2}\theta^2\fhatX^2 \lambda^2 
      A_\lambda^2\left(\frac{t_\lambda}{t}\right)^{\kappa_\lambda}
      e^{-\Gamma_\lambda(t-t_G)}
  \label{eq:RhoOftEqnWithR}
\end{equation}  
during the epoch in which the oscillation began.  Computing $\rho_\lambda$ during 
subsequent epochs is simply a matter of applying Eq.~(\ref{eq:RhoOftEqnWithR}) 
iteratively with the appropriate boundary conditions at each transition point.  
Consequently, in the LTR cosmology, we have~\cite{DynamicalDM2}
\begin{equation}
   \rho_\lambda^{\mathrm{LTR}}(t)~\approx~ 
      \frac{1}{2}\theta^2\fhatX^2 \lambda^2 A_\lambda^2
      e^{-\Gamma_\lambda(t-\tG)}\times\begin{cases}
      \displaystyle \vspace{0.25cm}
      \left(\frac{t_\lambda}{t}\right)^2~~ & t_\lambda~\lesssim~ t~\lesssim~\tRH \\
      \displaystyle \vspace{0.25cm} 
      \left(\frac{t_\lambda^2}{\tRH^{1/2}\,t^{3/2}}\right)~~ & \tRH~\lesssim~t~\lesssim~\tMRE\\
      \displaystyle
      \left(\frac{t_\lambda^2\,\tMRE^{1/2}}{t^2\,\tRH^{1/2}}\right)~~ &  t~\gtrsim~\tMRE~,
      \end{cases}
   \label{eq:RhoLambdaInLTRCosmo}
\end{equation}
where $\tRH$ denotes the reheating time --- \ie, the time at which $T = \TRH$, and the universe 
transitions from an initial epoch of matter domination by the coherent oscillations of
the inflaton field to the usual radiation-dominated era.

Given the energy-density expression in
(\ref{eq:RhoLambdaInLTRCosmo}), it is straightforward to obtain the
relic abundance $\Omega_\lambda \equiv \rho_\lambda/\rhocrit$ for each
$a_\lambda$, where $\rhocrit \equiv 3M_P^2 H^2$.  For the heavier modes 
in the tower, for which $t_\lambda = \tG$, one finds
\begin{equation}
  \Omega_\lambda^{\mathrm{LTR}} ~\approx~
    3\left(\frac{\theta \fhatX \mX}{M_P}\right)^2  
      \tG^2  
      \left[1+\frac{\lambda^2}{\mX^2}+
      \frac{\pi^2\mX^2}{M_c^2}\right]^{-1}
      e^{-\Gamma_\lambda(t-\tG)}\times
      \begin{cases}
      \displaystyle\frac{1}{4} \vspace{0.25cm}~~
      & 1/\lambda ~\lesssim~ t~ \lesssim~\tRH \\
      \displaystyle\frac{4}{9}\left(\frac{t}{\tRH}\right)^{1/2} \vspace{0.25cm}~~
      & \tRH ~\lesssim~ t ~\lesssim~ \tMRE \\ 
      \displaystyle\frac{1}{4}\left(\frac{\tMRE}{\tRH}\right)^{1/2}~~
      & t ~\gtrsim~ \tMRE~.    
    \end{cases}
  \label{eq:OmegaLambdaOftEqnLTRtG}
\end{equation}
For the modes in the tower for which $t_\lambda > \tG$, the corresponding result is
\begin{equation}
  \Omega_\lambda^{\mathrm{LTR}} ~\approx~
    3\left(\frac{\theta \fhatX \mX}{M_P}\right)^2 
      \lambda^{-2}
      \left[1+\frac{\lambda^2}{\mX^2}+
      \frac{\pi^2\mX^2}{M_c^2}\right]^{-1}
      e^{-\Gamma_\lambda(t-\tG)}\times
      \begin{cases}
     \displaystyle\frac{1}{4} \vspace{0.25cm}~~
      & 1/\lambda ~\lesssim ~t ~\lesssim~ \tRH \\
      \displaystyle\frac{4}{9}\left(\frac{t}{\tRH}\right)^{1/2} \vspace{0.25cm}~~
      & \tRH ~\lesssim~ t ~\lesssim~ \tMRE \\ 
      \displaystyle\frac{1}{4}\left(\frac{\tMRE}{\tRH}\right)^{1/2}~~
      & t~\gtrsim ~\tMRE~.    
    \end{cases}
  \label{eq:OmegaLambdaOftEqnLTRtlambda}
\end{equation}
 
The total contribution $\Omegatot$ to the dark-matter relic abundance from the
axion tower is simply the sum over these individual contributions.
While the generic behavior of $\Omegatot$ as a function of $\fhatX$, $M_c$, and 
$\Lambda_G$ is somewhat complicated, simple analytical results can be obtained
in certain limiting cases of physical importance.  For example, let us 
consider the limit in which $t_\lambda = t_G$ for all modes in the tower and 
$H_I$ is sufficiently large that none of the $a_\lambda$ which would otherwise 
contribute significantly to $\Omegatot$ begin oscillating before the end of 
inflation.  In this limit, all of the $\Omega_\lambda$ take the form given in
Eq.~(\ref{eq:OmegaLambdaOftEqnLTRtG}), and one finds that the present-day value 
of $\Omegatot$, here denoted $\Omegatotnow$, is given by the simple closed-form 
expression~\cite{DynamicalDM2} 
\begin{equation}
  \Omegatotnow ~\approx~
    \frac{3}{256\pi^2}(g_G\xi)^2\left(\frac{\theta \Lambda_G^2}{M_P}\right)^2
      \tG^{3/2}\tMRE^{1/2}
      \left(\frac{\tG}{\tRH}\right)^{1/2}~.
  \label{eq:OmegaLambdaShelfLimit}
\end{equation}  
In the opposite limit, when all of modes which contribute significantly toward $\Omegatotnow$
begin oscillating at $t_\lambda > t_G$ and have oscillation-onset times 
which depend on $\lambda$ and are therefore staggered in time, $\Omega_\lambda$ 
is given by Eq.~(\ref{eq:OmegaLambdaOftEqnLTRtlambda}) for all $a_\lambda$.  In this limit,  
the expression for $\Omegatotnow$ reduces to~\cite{DynamicalDM2}  
\begin{equation}
    \Omegatotnow ~\approx~ \frac{3}{8} \left(\frac{\theta \fhatX}{M_P}\right)^2
     \left(\frac{\tMRE}{\tRH}\right)^{1/2}~.
  \label{eq:OmegaLambdaStaggeredLimit}
\end{equation}   

The preferred region of parameter space from the perspective of dark-matter
phenomenology is that within which $\Omegatotnow$ represents an $\mathcal{O}(1)$ 
fraction of the dark-matter relic abundance inferred from WMAP data~\cite{WMAP}: 
\begin{equation}
  \Omega_{\mathrm{CDM}} h^2 ~=~ 0.1131 \pm 0.0034~,
  \label{eq:OmegaWMAP}
\end{equation}
where $h \approx 0.72$ is the Hubble constant.  From a dynamical dark-matter
perspective, it is also preferable that the full axion tower contribute 
meaningfully to $\Omegatotnow$.  For an $\mathcal{O}(1)$ value
of the misalignment angle $\theta$ and a reheating temperature 
within the preferred range $T_R \sim 4 - 30$~MeV for theories with large 
extra dimensions, one finds~\cite{DynamicalDM2} that these two conditions are realized 
for $\fhatX \sim 10^{14} - 10^{15}$~GeV and $\Lambda_G \sim 10^2 - 10^5$~GeV, 
provided that $M_c$ is small enough that $y \lesssim 1$.
Within this region of parameter space, the $t_\lambda$ of all $a_\lambda$ 
which contribute meaningfully toward $\Omegatotnow$ are staggered in time,
and therefore the lighter modes yield a proportionally greater contribution
to that total abundance.  We will often focus our attention on 
this particular region of parameter space when discussing constraints on
axion DDM models.


\subsection{Axion Production from Particle Decays\label{sec:IntraensembleProd}}


Another mechanism by which a non-thermal population of relic particles may 
be generated in the early universe is through the decays of heavier, 
unstable relics.  In scenarios involving extra dimensions, these relics 
include the higher KK modes of any fields which propagate within at
least some subspace of the extra-dimensional bulk.  For example,    
since the graviton field necessarily propagates throughout the entirety of the
bulk, a population of unstable KK gravitons 
is a generic feature of all such scenarios.  
In the minimal bulk-axion DDM model under consideration here, the 
unstable relics whose decays can serve as a source for any given $a_\lambda$ 
include these KK gravitons as well as other, heavier $a_\lambda$.  Moreover, since
these fields span a broad range of masses from the sub-eV to multi-TeV scale and 
beyond, one would expect the population of axions produced by their collective 
decays to possess by a highly non-trivial phase-space distribution.
However, as we shall demonstrate below, the total contribution
$\Gamma^{(\mathrm{IE})}$ to the decay rate of any $a_\lambda$ from 
{\it intra-ensemble}\/ decays (\ie, decays to final states which include one or 
more dark-sector fields in addition to any visible-sector fields that might also 
be present) is far smaller than that from decays to final states involving 
visible-sector fields alone. 
That the total branching fraction for intra-ensemble decays is negligible 
suggests that the population of axions produced by such decays will, in general,
be quite small.  Thus, provided the initial abundances of the $a_\lambda$ are 
set by some mechanism such as vacuum misalignment for which the $\Omega_\lambda$ 
of the heavier $a_\lambda$ are initially similar to or smaller than those of the
light fields, it is reasonable to assume that the contributions from intra-ensemble
decays are subleading and may therefore be safely neglected.    

One class of processes which contribute to $\Gamma^{(\mathrm{IE})}$ are those which
arise due to the axion self-interactions implied by the final term in  
Eq.~(\ref{eq:ActionInMassEigenbasis}).  The leading such contribution comes 
from three-body decay processes of the form 
$a_\lambda\rightarrow a_{\lambda_1}a_{\lambda_2}a_{\lambda_3}$. 
An upper bound on the total contribution $\Gamma(a_\lambda\rightarrow 3a)$ 
to the decay width of a given $a_\lambda$ from all kinematically allowed decays
of this form was derived in Ref.~\cite{DynamicalDM2}:
\begin{equation}
   \Gamma(a_\lambda\rightarrow 3a) ~\leq~ \frac{g_G^4\xi^8}{45(4\pi)^7}
     \frac{\lambda^4}{M_c^3}\left(\frac{\LambdaG}{\fhatX}\right)^8~.
   \label{eq:Gammaato3a} 
\end{equation}
It was also shown in Ref.~\cite{DynamicalDM2} that the partial width of the 
$a_\lambda$ to a pair of photons is given by
\begin{equation}
  \Gamma(a_\lambda\rightarrow \gamma\gamma) ~=~ G_\gamma (\wtl^2A_\lambda)^2
  \frac{\lambda^3}{\fhatX^2}~,
  \label{eq:PartialWidthToPhotons}
\end{equation}
with $G_\gamma \equiv c_\gamma^2\alpha^2/256\pi^3$, where 
$\alpha \equiv e^2/4\pi$ is the fine-structure constant.
Within the preferred region of parameter space discussed above, in which 
$\fhatX \sim 10^{14} - 10^{15}$~GeV and $\Lambda_G \gtrsim 10^2 - 10^5$~GeV,
we see that $\Gamma_\lambda(a\rightarrow 3a)$ is negligible compared to 
$\Gamma(a_\lambda\rightarrow \gamma\gamma)$.
It then follows that $\Gamma_\lambda(a\rightarrow 3a)$
represents a vanishingly small contribution to the total decay width 
$\Gamma_\lambda$ of any $a_\lambda$ in any theory with an $\mathcal{O}(1)$
value of $c_\gamma$.  We therefore conclude that decays of the form
$a_\lambda\rightarrow a_{\lambda_1}a_{\lambda_2}a_{\lambda_3}$ 
do not play a significant role in the phenomenology of 
realistic bulk-axion models of dynamical dark matter. 
   
In addition to these decays, however, an additional set of decay channels --- those
involving lighter graviton or radion fields in the final state --- are also 
open to the $a_\lambda$.  In order to assess whether such decay 
channels are capable of yielding a significant contribution to the relic abundance 
of any of the $a_\lambda$, we begin by identifying the relevant interactions 
among the modes in the KK graviton and axion towers.  
Since we are considering the case of a flat extra dimension and assuming fluctuations
of the metric to be small, it is justified to work in the regime of linearized gravity.
The relevant term in the five-dimensional action is therefore
\begin{equation}
  S ~=~ -\int d^4x\int^{2\pi R}_0 dy 
    \frac{1}{M_5^{3/2}}T_{MN}h^{MN}~,
    \label{eq:Relevant5DActionTerm}
\end{equation}
where $T_{MN}$ is the stress-energy tensor, and $h_{MN}$ is the metric perturbation
defined according to the relation
\begin{equation}
  g_{MN} ~=~ \eta_{MN} + \frac{2}{M_5^{3/2}}h_{MN}~.
  \label{eq:LinearExpandgmunu}
\end{equation}   

The piece of the stress-energy tensor which involves the five-dimensional axion field 
$a$ includes both a bulk contribution and a contribution arising from terms in the 
interaction Lagrangian which involve interactions of the axion with the brane-localized
fields of the SM.  The bulk contribution is given by    
\begin{eqnarray}
  T_{MN}^{\mathrm{bulk}} ~=~
      \partial_Ma\partial_Na -\frac{1}{2}\eta_{MN}(\partial_{P}a\partial^Pa)~.
      \label{eq:TMNBulk}
\end{eqnarray}
Upon KK decomposition, this contribution, when coupled to
$h_{MN}$ as in Eq.~(\ref{eq:Relevant5DActionTerm}), 
gives rise to three-point interactions between a KK graviton or radion 
field and a pair of $a_\lambda$.  These interactions lead to decays of the form 
$a_\lambda\rightarrow G_{\mu\nu}^{(m)} a_{\lambda'}$, where $G_{\mu\nu}^{(m)}$
denotes a KK graviton with KK mode number $m$.  In the absence of an instanton-induced 
brane mass term $\mX$ for the axion field
(\ie, in the $\mX \rightarrow 0$ limit, in which
all mixing between the axion KK modes vanishes and $a_\lambda\rightarrow a_n$),
KK-momentum conservation would imply that only a single, marginal decay
channel would exist for each $a_\lambda$.  Hence the contribution to $\Gamma_\lambda$
from such decays can be neglected.  However, the instanton contribution to the axion 
mass-squared matrix violates KK-momentum conservation, and therefore, despite the fact
that these axion-axion-graviton interactions are Planck-suppressed, they can still
potentially contribute significantly to $\Gamma_\lambda$, due to the large number
of modes into which each $a_\lambda$ can decay.   
    
By contrast, the brane-localized contribution, which is given by
\begin{eqnarray}
  T_{MN}^{\mathrm{brane}} &=& \delta(y)\delta_M^{\mu}\delta_N^\nu\bigg[
      \frac{1}{2}\sum_i \frac{c_i}{f_X^{3/2}}
           \Big[(\partial_\mu a) \overline{\psi}_i\gamma_\nu\gamma^5\psi_i +
           (\partial_\nu a) \overline{\psi}_i\gamma_\mu\gamma^5\psi_i
           -2\eta_{\mu\nu}(\partial_\rho a)\overline{\psi}_i\gamma^\rho\gamma^5\psi_i\Big]
      \nonumber \\ & &
      +~\frac{c_\gamma e^2}{32\pi^2 f_X^{3/2}}a \left(4\tilde{F}_{\mu\rho}F_{\nu}^{~\rho}
          - \eta_{\mu\nu}\tilde{F}^{\rho\sigma}F_{\rho\sigma}\right)
      + \frac{\xi g_s^2}{32\pi^2 f_X^{3/2}}a \left(4\tilde{G}_{\mu\rho}^aG_{\nu}^{~\rho a}
          - \eta_{\mu\nu}\tilde{G}^{\rho\sigma a}G_{\rho\sigma}^a\right)
          \bigg]~,
      \label{eq:TMNBrane}
\end{eqnarray}
leads to four-, five-, and six-point interactions between the graviton field, 
the $a_\lambda$, and the various SM fields.  These interactions take the same form
as those discussed in Section~\ref{sec:IntraensembleProd}, save that each vertex involves 
the coupling of an additional KK graviton and is suppressed, relative to the 
corresponding interaction involving the axion and SM fields alone, by an additional 
factor of $M_P$.  The rates for such interactions will therefore always be much 
smaller than those calculated in Section~\ref{sec:IntraensembleProd}.  Indeed, even the 
total contribution to the decay rate of a given $a_\lambda$ from such processes, summed
over graviton KK modes, will still be suppressed by a factor of roughly $M_{5}$, where
$M_5$ denotes the five-dimensional Planck scale, relative to the contribution from decays
to SM fields alone.  It is therefore sufficient, at least for our present purposes, to 
neglect $T_{MN}^{\mathrm{brane}}$ and to focus solely on the interactions arising from 
the bulk contribution $T_{MN}^{\mathrm{bulk}}$.

We begin our analysis of axion-axion-graviton interactions
by expanding the five-dimensional axion field, as well as the
various components $h_{\mu\nu}$, $h_{\mu 5}$, and $h_{55}$ of the metric
perturbation $h_{MN}$, in terms of KK modes.  The mode expansion of the
axion field for the orbifold compactification considered here was given in 
Eq.~(\ref{eq:AxionModeDecomp}); the mode expansions of 
$h_{\mu\nu}$, $h_{\mu 5}$, and $h_{55}$ are analogously given by     
\begin{eqnarray}
  h_{\mu\nu} &=& \frac{1}{\sqrt{2\pi R}}\sum_{m=0}^\infty r_m 
         h_{\mu\nu}^{(m)}\cos\left(\frac{my}{R}\right)\nonumber\\
  h_{\mu 5} &=& \frac{1}{\sqrt{2\pi R}}\sum_{m=1}^\infty r_m 
         h_{\mu 5}^{(m)}\sin\left(\frac{my}{R}\right)\nonumber\\
  h_{55} &=& \frac{1}{\sqrt{2\pi R}}\sum_{m=0}^\infty r_m 
         h_{55}^{(m)}\cos\left(\frac{my}{R}\right)~.
  \label{eq:KKModeDecompsDefs}
\end{eqnarray} 
Note in particular that $h_{\mu 5}$ must be odd
with respect to the parity transformation $x_5 \rightarrow -x_5$.
Upon substituting these KK-mode decompositions into the linearized-gravity action
given in Eq.~(\ref{eq:Relevant5DActionTerm}) and integrating over
$y$, we find that the terms in the effective, four-dimensional interaction 
Lagrangian which govern the interactions between the graviton and axion KK modes
consist of the following three contributions:
\begin{eqnarray}
  \int_0^{2\pi R}\frac{h_{\mu\nu}T_{\mathrm{bulk}}^{\mu\nu}}{M_5^{3/2}}dy &=&
     \sum_{m,n,p = 0}^\infty \frac{r_m r_n r_p}{4M_P} h^{(m)}_{\mu\nu}
     \bigg[
     \Big(2\partial^\mu a^{(n)}\partial^\nu a^{(p)} 
     -\eta^{\mu\nu} \partial^\rho a^{(n)}\partial_\rho a^{(p)}\Big)
     \Delta_{mnp}^+\bigg.\nonumber\\ & & \bigg.
     +\eta^{\mu\nu}\left(\frac{np}{R^2}\right)a^{(n)}a^{(p)}
     \Delta_{mnp}^-\bigg]\nonumber\\
  \int_0^{2\pi R}\frac{h_{55}T_{\mathrm{bulk}}^{55}}{{M_5^{3/2}}}dy &=& 
     \sum_{m,n,p = 0}^\infty \frac{r_mr_nr_p}{4M_P} h^{(m)}_{55}
     \bigg[\partial^\rho a^{(n)}\partial_\rho a^{(p)}\Delta_{mnp}^+
     + \left(\frac{np}{R^2}\right)a^{(n)}a^{(p)}\Delta_{mnp}^-
     \bigg]\nonumber\\
  \int_0^{2\pi R}\frac{h_{\mu 5}T_{\mathrm{bulk}}^{\mu 5}}{M_5^{3/2}}dy &=& 
     \sum_{m = 1}^\infty \sum_{n,p = 0}^\infty \frac{r_mr_nr_p}{2M_P}\left(\frac{p}{R}\right)
     h_{\mu 5}^{(m)}(\partial^\mu a^{(n)})a^{(p)}\Delta^-_{nmp}~,
     \label{eq:IntTermsHmunu}
\end{eqnarray} 
where
\begin{equation}
  \Delta_{mnp}^\pm ~\equiv~ \Big[\delta_{m,n-p}+\delta_{m,p-n}\Big]\pm
  \Big[\delta_{m,n+p}+\delta_{m,-n-p}\Big]~.
\end{equation}

For the purposes of computing Feynman diagrams, it is convenient to work in
the unitary gauge, in which the $h_{\mu 5}^{(m)}$ and $h_{55}^{(m)}$ fields 
with $m>0$ are set to zero by the five-dimensional gauge transformations 
$g_{MN}\rightarrow g_{MN} + \partial_M \epsilon_N + \partial_N \epsilon_M$,
where $\epsilon_M$ is the gauge parameter. 
In this gauge, the contributions in the second and third line of 
Eq.~(\ref{eq:IntTermsHmunu}) vanish (save for the interactions between
the axion KK modes and the radion field $h_{55}^{(0)}$, which will be
discussed in due time), and the physical, gauge-invariant graviton fields 
\begin{equation}
  G_{\mu\nu}^{(m)} ~\equiv~ h^{(m)}_{\mu\nu} + \left(\frac{R}{m}\right)
    \Big[\partial_\mu h_{\nu 5}^{(m)} + \partial_\nu h^{(m)}_{\mu 5}\Big]
    - \left(\frac{R^2}{m^2}\right)\partial_\mu\partial_\nu h_{55}^{(m)} 
  \label{eq:DefOfGmunu}
\end{equation} 
reduce to $h_{\mu\nu}^{(m)}$ for all $m > 0$.
The relevant part of the effective Lagrangian consequently reduces to   
\begin{equation}
  \mathcal{L}_{\mathrm{int}}^{(m>0)} ~=~
     -\sum_{m=1}^\infty\sum_{n,p=0}^\infty\frac{r_nr_p}{2\sqrt{2}M_P} 
     h_{\mu\nu}^{(m)}\bigg[
     \Big(2\partial^\mu a^{(n)}\partial^\nu a^{(p)} 
     -\eta^{\mu\nu} \partial^\rho a^{(n)}\partial_\rho a^{(p)}\Big)
     \Delta_{mnp}^+
     +\eta^{\mu\nu}\left(\frac{np}{R^2}\right)a^{(n)}a^{(p)}
     \Delta_{mnp}^-\bigg]~.
     \label{eq:IntTermsGmunu}
\end{equation}  
The expression in Eq.~(\ref{eq:IntTermsGmunu}) can be 
rewritten in terms of the mass eigenstates $a_\lambda$ via the mixing 
matrix $U_{\lambda n}$ in Eq.~(\ref{eq:DefOfalambda}).  The result is   
\begin{eqnarray}
  \mathcal{L}_{\mathrm{int}}^{(m>0)} &=&
     -\sum_{m=1}^\infty\sum_{n,p=0}^\infty\sum_{\lambda,\lambda'}
       \frac{r_nr_p}{2\sqrt{2}M_P} h_{\mu\nu}^{(m)}U_{n\lambda}^\dagger
       U_{p\lambda'}^\dagger\bigg[
       \Big(2\partial^\mu a_\lambda\partial^\nu a_{\lambda'}
       -\eta^{\mu\nu} \partial^\rho a_\lambda\partial_\rho a_{\lambda'}\Big)
       \Delta_{mnp}^+
       +\eta^{\mu\nu}\left(\frac{np}{R^2}\right)a_\lambda a_{\lambda'}
       \Delta_{mnp}^-\bigg]~~\nonumber\\ &=&
     -\sum_{m=1}^\infty\sum_{n=0}^\infty\sum_{\lambda,\lambda'}
       \frac{r_n}{2\sqrt{2}M_P} h_{\mu\nu}^{(m)}
       U_{n\lambda}^\dagger\Bigg\{\Bigg.
       \Big(2\partial^\mu a_\lambda\partial^\nu a_{\lambda'}
       -\eta^{\mu\nu} \partial^\rho a_\lambda\partial_\rho a_{\lambda'}\Big)
       \nonumber\\ & & ~~\times~
       \Big(r_{n-m}U^\dagger_{n-m,\lambda'} + r_{n+m}U^\dagger_{n+m,\lambda'}
       +r_{m-n}U^\dagger_{m-n,\lambda'} + r_{-n-m}U^\dagger_{-n-m,\lambda'}\Big)      
       +\eta^{\mu\nu} \frac{n}{R^2} a_\lambda a_{\lambda'}\nonumber\\ & &  \Bigg. 
       ~~~~\times~
       \Big[(n-m)\big(r_{n-m}U^\dagger_{n-m,\lambda'} + r_{m-n}U^\dagger_{m-n,\lambda'}\big) 
       +(n+m)\big(r_{n+m}U^\dagger_{n+m,\lambda'}+r_{-n-m}U^\dagger_{-n-m,\lambda'}\big)\Big]
       \Bigg\}~,
     \label{eq:IntLagGmunuLambdaBasis}
\end{eqnarray}
where in going from the first equality to the second we have exploited the Kronecker
deltas in $\Delta^{\pm}_{mnp}$ to evaluate the sum over $p$.  It should be noted that 
in the notation employed in the above expression, $U_{n\lambda}^\dagger = 0$ by 
definition for $n<0$.  The sum over $n$ in Eq.~(\ref{eq:IntLagGmunuLambdaBasis})
can also be performed analytically, and the resulting, final expression for the 
Lagrangian in terms of the $a_\lambda$ is found to be
\begin{equation}
\mathcal{L}_{\mathrm{int}}^{(m>0)} ~=~
     -\sum_{m=1}^\infty\sum_{\lambda,\lambda'}
       \frac{1}{2\sqrt{2}M_P} h_{\mu\nu}^{(m)}\bigg[
       \Big(2\partial^\mu a_\lambda\partial^\nu a_{\lambda'}
       -\eta^{\mu\nu} \partial^\rho a_\lambda\partial_\rho a_{\lambda'}\Big)
       C^{(1)}_{m\lambda\lambda'}
       +\eta^{\mu\nu}M_c^2a_\lambda a_{\lambda'}
       C^{(2)}_{m\lambda\lambda'}\bigg]~,
     \label{eq:IntLagGmunuLambdaBasisFinal}
\end{equation}
where the coefficients $C^{(1)}_{m\lambda\lambda'}$ and $C^{(2)}_{m\lambda\lambda'}$
are given by
\begin{eqnarray}
   C^{(1)}_{m\lambda\lambda'} &=&
     \frac{-8m^2y^2\wtl^2\wtl'^2A_\lambda A_{\lambda'}}
       {m^4y^4-2m^2y^2(\wtl^2+\wtl'^2)+(\wtl^2-\wtl'^2)^2}\nonumber\\
   C^{(2)}_{m\lambda\lambda'} &=& 
     \frac{4\wtl^2\wtl'^2[m^2y^2(\wtl^2+\wtl'^2)-(\wtl^2-\wtl'^2)^2]
       A_\lambda A_{\lambda'}}
       {y^2[m^4y^4-2m^2y^2(\wtl^2+\wtl'^2)+(\wtl^2-\wtl'^2)^2]}~.
\end{eqnarray}

From the interaction Lagrangian in Eq.~(\ref{eq:IntLagGmunuLambdaBasisFinal}), it
is straightforward to obtain the Feynman rule for the graviton-axion-axion
interaction vertex in the unitary gauge:  
\begin{eqnarray*}
\resizebox{1.65in}{!}{\includegraphics{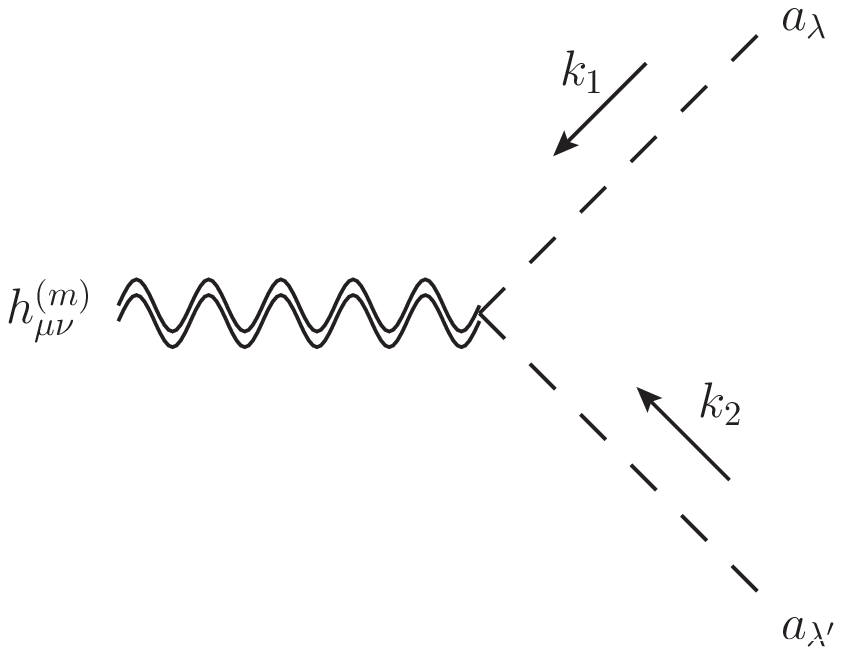}}
   &\raisebox{1.75cm}{$\displaystyle ~=~
   -\frac{i}{\sqrt{2}M_P}\bigg[
    \left(k_{1\mu}k_{2\nu}+k_{1\nu}k_{2\mu}-\eta_{\mu\nu}k_1\cdot k_2\right)
    C^{(1)}_{m\lambda\lambda'} -\eta_{\mu\nu}M_c^2C^{(2)}_{m\lambda\lambda'}
    \bigg]$~.}
\end{eqnarray*} 
Using this vertex rule along with the graviton-polarization sum 
rule given in Refs.~\cite{GravFeynmanRulesGiudice,GravFeynmanRulesHan},
we find
\begin{eqnarray}
  |\mathcal{M}(a_\lambda\rightarrow h_{\mu\nu}^{(m)}a_{\lambda'})|^2&=&
  \frac{\big(C^{(1)}_{m\lambda\lambda'}\big)^2}{12 M_P^2 (mM_c)^4}
  \Big[\lambda^4+(mM_c)^4+\lambda'^4-2(mM_c)^2\lambda^2
    -2(mM_c)^2\lambda'^2-2\lambda^2\lambda'^2\Big]^2\nonumber\\ & = &
  \frac{16}{3}\left(\frac{\mX^4}{M_P^2}\right)(\wtl^2A_\lambda)^2
    (\wtl'^2A_{\lambda'})^2
\end{eqnarray}
for $\lambda' < \lambda$.
Consequently, the partial width of $a_\lambda$ from such a decay is
\begin{equation}
  \Gamma(a_\lambda\rightarrow h_{\mu\nu}^{(m)}a_{\lambda'}) ~=~
    \frac{\mX^4}{3\pi\lambda^3M_P^2}
     (\wtl^2A_\lambda)^2(\wtl'^2A_{\lambda'})^2
      \Big[\lambda^4+(mM_c)^4+\lambda'^4-2(mM_c)^2(\lambda^2
      +\lambda'^2)-2\lambda^2\lambda'^2\Big]^{1/2}~.
  \label{eq:GammaToGravAx}
\end{equation}

Once again, in order to obtain the full contribution
$\Gamma(a_\lambda\rightarrow h_{\mu\nu}a)$ to $\Gamma_\lambda$
from decays of the form $a_\lambda\rightarrow h_{\mu\nu}^{(m)}a_{\lambda'}$,
it is necessary to sum over all combinations of 
final-state graviton and axion modes which are kinematically accessible.
As before, we will approximate the mode
sums over both $m$ and $\lambda'$ as integrals.  This yields the result 
\begin{eqnarray}
  \Gamma(a_\lambda\rightarrow h_{\mu\nu}a) &\lesssim& 
    \frac{4\mX^4(\wtl^2A_\lambda)^2}{3\pi\lambda^3 M_c M_P^2}
    \int_{\lambda_0}^\lambda d\lambda'\int_0^{(\lambda - \lambda')/M_c}   
    dm (\wtl'^2A_{\lambda'})^2\nonumber \\     
    & & ~~~~~~\times 
      \Big[\lambda^4+(mM_c)^4+\lambda'^4-2(mM_c)^2(\lambda^2
      +\lambda'^2)-2\lambda^2\lambda'^2)\Big]^{1/2}\nonumber \\
    & = & \frac{8\mX^4(\wtl^2A_\lambda)^2}{9\pi\lambda^3 M_c^2 M_P^2} 
      \int_{\lambda_0}^\lambda d\lambda' (\wtl'^2A_{\lambda'})^2(\lambda+\lambda')
      \Bigg[(\lambda^2 +\lambda'^2) 
      E\left(\frac{(\lambda-\lambda')^2}{(\lambda+\lambda')^2}\right)
      -2\lambda\lambda'
      K\left(\frac{(\lambda-\lambda')^2}{(\lambda+\lambda')^2}\right)
      \Bigg]~,~~~~~
  \label{eq:GammaaatoGaSum}
\end{eqnarray}
where $K(x)$ and $E(x)$ denote the complete elliptic integrals of the
first and second kind, respectively:
\begin{equation}
   K(x) ~=~ \int_0^{\pi/2}\frac{d\theta}{\sqrt{1-x^2\sin^2\theta}}~,
   ~~~~~~~~
   E(x) ~=~ \int_0^{\pi/2}d\theta\sqrt{1-x^2\sin^2\theta}~.
\end{equation} 

\begin{figure}[b!]
\centerline{
  \epsfxsize 3.0 truein \epsfbox {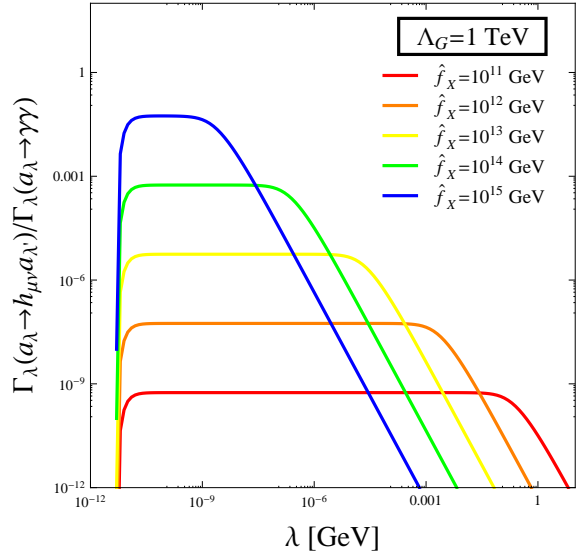} }
\caption{The ratio
$\Gamma(a_\lambda\rightarrow h_{\mu\nu}a)/
\Gamma(a_\lambda\rightarrow \gamma\gamma)$, shown 
as a function of $\lambda$ for several different values
of $\fhatX$.  Here we have set $\Lambda_G = 1$~TeV and
$\xi = g_G = 1$, and we have taken the 
compactification scale to be $M_c = 10^{-11}$~GeV.  It is 
clear from this plot that this ratio is safely below unity
for $\fhatX$ within our preferred region
$10^{14} - 10^{15}$~GeV.
\label{fig:GravitonToPhotonWidthRatio}}
\end{figure}

In order to compare $\Gamma(a_\lambda\rightarrow h_{\mu\nu}a)$ to the
rate for $a_\lambda$ decays to SM fields, we numerically integrate 
Eq.~(\ref{eq:GammaaatoGaSum}) over $\lambda'$ and compare the 
resulting expression to the decay rate 
$\Gamma(a_\lambda\rightarrow \gamma\gamma)$ to photon pairs.  
In Fig.~\ref{fig:GravitonToPhotonWidthRatio}, we plot the ratio
$\Gamma(a_\lambda\rightarrow h_{\mu\nu}a)/
\Gamma(a_\lambda\rightarrow \gamma\gamma)$ as a function of 
$\lambda$ for a variety of different choices of $\fhatX$.  In
each case, we have set $\Lambda_G = 1$~TeV, $M_c = 10^{-11}$~GeV, and
$\xi = g_G = 1$.  It is evident from this plot that only for values of 
$\fhatX$ above the preferred range $\fhatX \sim 10^{14} - 10^{15}$~GeV 
does the decay rate for $a_\lambda\rightarrow h_{\mu\nu}a_{\lambda'}$ 
become similar in magnitude to the rate for axion decays into brane fields.  
Indeed, for values $\fhatX$ within this preferred range, 
$\Gamma(a_\lambda\rightarrow h_{\mu\nu}a)/\Gamma(a_\lambda\rightarrow \gamma\gamma)$ 
never exceeds 0.06, even for the lightest modes in the tower.  Furthermore, 
for values of $\fhatX$ of this magnitude, the lifetimes for all $a_\lambda$ 
light enough to have 
$\Gamma(a_\lambda\rightarrow h_{\mu\nu}a)/\Gamma(a_\lambda\rightarrow \gamma\gamma)$ 
near this maximal value are parametrically larger than the present age 
of the universe, even once the additional contribution to 
$\Gamma_\lambda$ from $a_\lambda\rightarrow h_{\mu\nu}a_{\lambda'}$ 
decays is taken into account.  Consequently, the decays of such
fields are not cosmologically relevant, and since the branching 
fraction for all other, heavier $a_\lambda$ into final states involving 
KK gravitons is utterly negligible.   We therefore conclude that intra-ensemble
decays are not phenomenologically relevant for bulk-axion models of 
dynamical dark matter.     
    
Up to this point, we have focused chiefly on the effect of the tensor
KK modes of the higher-dimensional graviton field on axion production 
in the early universe.  However, we have yet to address the effect 
of graviscalars such as the radion on axion production. 
Since our minimal DDM model involves only a single
extra dimension, only a single physical graviscalar mode (proportional to 
$h_{55}^{(0)}$) appears in the theory.  Furthermore, while the masses
of the $h_{\mu\nu}^{(m)}$ are dictated by the compactification geometry 
alone, the mass of this radion field depends on the details of the
mechanism through which the radius of the extra dimension is stabilized, and
is consequently highly model-dependent.  In this paper, we assume that 
the physical radion field is sufficiently heavy so as not to play a
significant role in the decay phenomenology of the light $a_\lambda$ fields 
which contribute significantly to $\Omegatotnow$.  Nevertheless, we note
that in scenarios which involve multiple extra dimensions of comparable size, or 
scenarios in which a specific model for radius stabilization is invoked,
graviscalars may play a more significant role in the phenomenology of the dark 
sector.


\subsection{Axion Production from Cosmic Strings\label{sec:CosmicStringProd}}

 
A population of cold axions can also be generated by the decays of topological defects. 
In our axion DDM model, this includes decays of the cosmic strings associated with the 
breaking of the global $U(1)_{X}$ symmetry.  Such decays are relevant in situations in 
which this symmetry remains unbroken until after inflation, \ie, $H_I \gtrsim 2\pi\fhatX$.  
By contrast, in situations in which $H_I \lesssim 2\pi\fhatX$ and
the $U(1)_X$ is spontaneously broken prior to the inflationary epoch, cosmic 
strings and other topological defects are washed out by the rapid expansion of 
the universe during cosmic inflation.  Consequently, in this latter case, 
axion production from the decays of cosmic strings can safely be ignored.

In this paper, we are primarily interested in high values of 
$\fhatX \sim 10^{14} - 10^{15}$~GeV, as these values characterize our preferred 
region of parameter space.  Likewise, we will primarily be interested in 
relatively low values of $H_I$, which may be realized naturally in the LTR cosmology.  
For this reason we shall assume that $H_I \lesssim 2\pi\fhatX$ in what follows.  
We see, then, that no significant population of axions is produced by 
cosmic-string decay.


\subsection{Axion Production from the Thermal Bath\label{sec:ThermalProd}}


Another mechanism through which a relic population of axions may be produced
in the early universe is a thermal one: via their interactions with the
SM fields in the radiation bath.  Unlike the axion population generated by
vacuum misalignment, which is characterized by a highly non-thermal velocity 
distribution (essentially that of a Bose-Einstein condensate) and is therefore 
by nature cold, this population is characterized by a thermal velocity distribution.  
Indeed, the properties of a thermal population of axions can differ substantially 
from that of a population of axions generated via misalignment production.   

A number of processes contribute to thermal axion production in the early
universe, and the processes which are the most relevant for the production of
standard axions dominate for each $a_\lambda$ in this scenario as well. 
Among hadronic processes, which play an important role in axion production 
when $c_g$ is non-vanishing, $q \gamma \rightarrow q a_\lambda$ and 
$q g \rightarrow q a_\lambda$ dominate for $T \gtrsim \LambdaQCD$, while
pion-axion conversion off nuclei (including all processes of the form 
$N \pi\rightarrow N' a_\lambda$, where $N,N' = \{n,p\}$ and $\pi$ denotes 
either a charged or neutral pion) and 
the purely pionic process $\pi \pi \rightarrow \pi a_\lambda$ dominate at lower
temperatures.  The rate for the high-temperature process is~\cite{QGPAxionProductionRate}
\begin{equation}
  \Gamma(q \gamma \rightarrow q a_\lambda)~=~\frac{g_s^2T^3\wtl^4 A_\lambda^2}
    {64\pi^5\fhatX^2} \ln\left[\left(\frac{T}{m_g}\right)^2 + 0.406\right],
\end{equation} 
where $m_g$ is the plasma mass
for the gluon, given in terms of the effective number of quark flavors $N_f$ at 
temperature $T$ by
\begin{equation}
  m_g(T) ~=~ \frac{g_s T}{3}\sqrt{3+N_f/2}. 
\end{equation}
Likewise, the rates for pion-conversion off nuclei and pionic production are well
estimated by the 
expressions~\cite{ChangChoiThermalNucleonRate,RaffeltPionAxionCrossSection,LTRAxionsKamionkowski} 
\begin{eqnarray}
  \Gamma(N \pi \rightarrow N' a_\lambda)&=&\frac{T^{7/2}m_N^{3/2}\wtl^4 A_\lambda^2e^{-m_N/T}}
    {6\zeta(3)(2\pi)^{5/2}\fhatX^2f_\pi^2}
    \Big[1.64\big(5C_{an}^2+5C_{ap}^2+2C_{an}C_{ap}\big)+6C_{a\pi N}^2\Big]
    \int_0^\infty  dx_1\frac{x_1 y_1^3}{e^{y_1}-1}\nonumber\\
  \Gamma(\pi \pi \rightarrow \pi a_\lambda)&=&
    \frac{3\zeta(3)T^5 C_{a\pi}^2\wtl^4 A_\lambda^2}
    {1024\pi^7\fhatX^2f_\pi^2}
    \int_0^\infty\int_0^\infty  \frac{dx_1 dx_2 x_1^2x_2^2}{y_1y_2(e^{y_1}-1)(e^{y_2}-1)}
    \int_{-1}^{1} d\mu \frac{(s-m_\pi^2)^3(5s-2m_\pi^2)}{s^2T^4}~,
\end{eqnarray} 
where, once again, $\zeta(x)$ denotes the Riemann zeta function, and the effective 
coupling coefficients $C_{ap}$, $C_{an}$, $C_{a\pi}$, and $C_{a\pi N}$ are given in
Eqs.~(\ref{eq:CapAndCan}) and~(\ref{eq:CaNAndCapi}).  Since these processes are
mediated by strong interactions, they tend to dominate the production rate for a
hadronic axion at temperatures $T\gtrsim 100$~MeV, at which the number densities 
of pions and other hadronic species are unsuppressed.   
  
In addition to these hadronic processes, there are several process
involving the interactions between the $a_\lambda$ and the $e^\pm$ and photon fields
which contribute to the axion production rate, and indeed dominate that rate 
at temperatures $T \ll \LambdaQCD$.  The first of these is the inverse-decay process
$\gamma\gamma \rightarrow a_\lambda$, the rate for which is given by  
\begin{equation}
  \Gamma(\gamma\gamma\rightarrow a_\lambda) ~=~
  2\frac{\lambda^5 G_\gamma(\wtl^2 A_\lambda)^2}{\zeta(3)\fhatX^2 T^2}
  K_1\left(\frac{\lambda}{T}\right)
  \label{eq:InverseDecayRate}
\end{equation}
where $K_1(x)$ and $K_2(x)$ respectively denote the Bessel function of the 
first and second kind, and $G_\gamma = \alpha^2c_\gamma^2/256\pi^2$.  Another 
is the Primakoff process $e^\pm \gamma  \rightarrow e^\pm a$.  For
$T,m_e \gg \lambda$, the rate for this process  
is well approximated by~\cite{QEDPlasmaAxionProductionRate}
\begin{equation}
  \Gamma_{\mathrm{Prim}}(e^\pm \gamma \rightarrow e^\pm a_\lambda) ~=~ 
    \frac{\alpha^3c_\gamma^2 n_e}{192\zeta(3)\fhatX^2}\wtl^4 A_\lambda^2
    \left[\ln\left(\frac{T^2}{m_\gamma^2}\right)+0.8194\right]~,
    \label{eq:ElectronPrimakoff}
\end{equation}
where the plasma mass $m_\gamma$ of the photon is given by $m_\gamma = eT/3$.
In the approximation of vanishing chemical potential, 
the number density of electrons (plus positrons) $n_e$ takes the well-known form
\begin{equation}
  n_e ~=~ \begin{cases}
    \displaystyle \frac{3\zeta(3)}{\pi^2}T^3,  ~~~~~& T \gtrsim  m_e\vspace{0.25cm}\\
    \displaystyle 4\left(\frac{T m_e}{2\pi}\right)^{3/2}e^{-m_e/T}  ~~~~~& T \lesssim m_e~.
    \end{cases}
\end{equation}
Finally, if $c_e\neq 0$ in Eq.~(\ref{eq:HighT4DAxionAction}) and the axion couples 
directly to the electron field, there can be an additional contribution
to the $e^\pm \gamma \rightarrow e^\pm a$ rate from a process akin to Compton
scattering, but with an axion replacing the photon in the final state.  The rate
for this process can be estimated as~\cite{RaffeltSubMeV}:
\begin{equation}
  \Gamma_{\mathrm{Comp}}(e^\pm \gamma \rightarrow e^\pm a_\lambda) ~\sim~
  \frac{4 \alpha c_e^2 n_e}{\fhatX^2}(\wtl^2 A_\lambda)^2 \times 
  \begin{cases}
  \displaystyle\frac{m_e^2}{T^2} ~~& T \gtrsim m_e\vspace{0.25cm}\\
  1 ~~& T \lesssim m_e~.
  \end{cases}
\end{equation}

\begin{figure}[th!]
\begin{center}
  \epsfxsize 3.0 truein \epsfbox {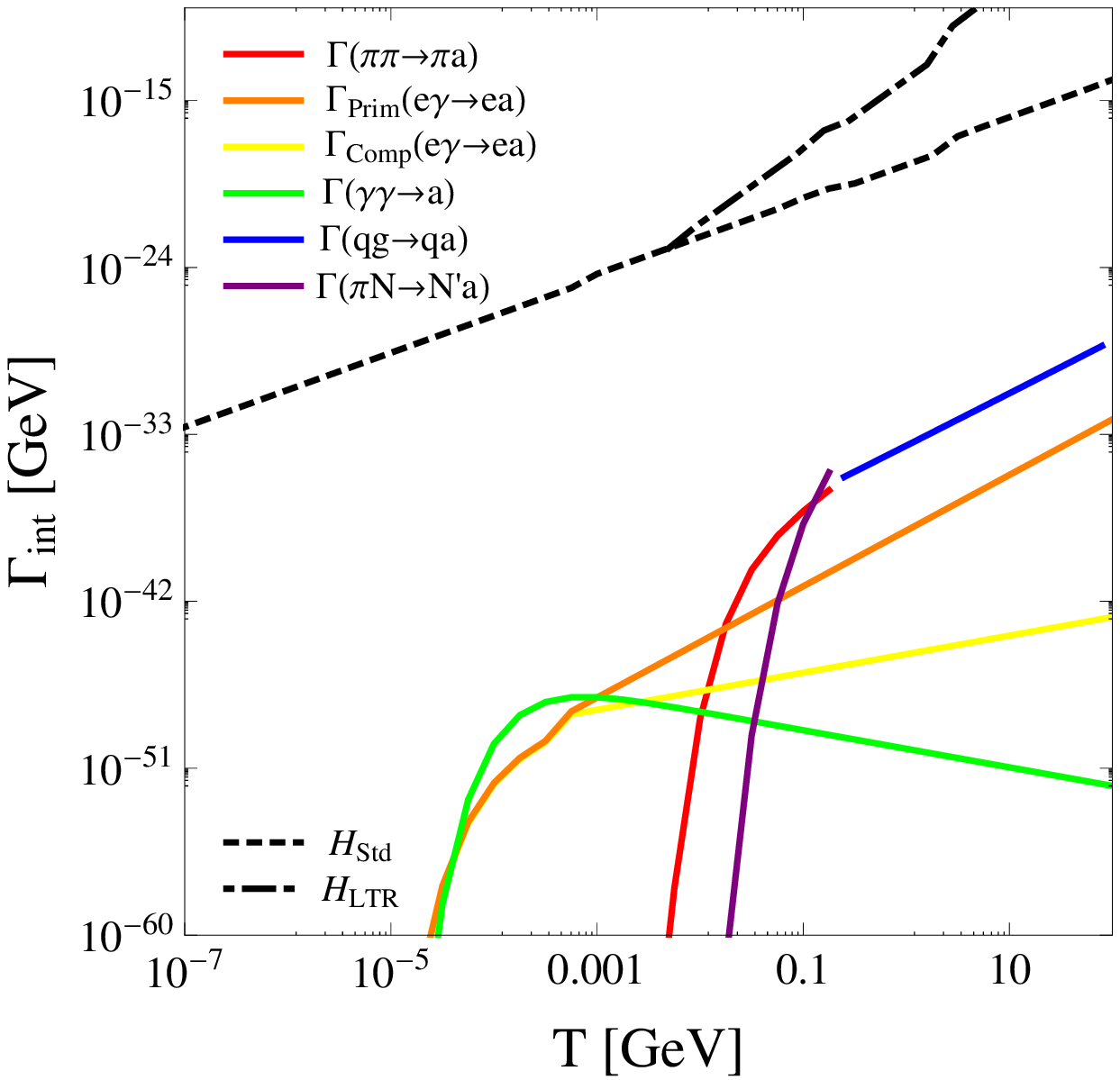} ~~~~
  \epsfxsize 3.0 truein \epsfbox {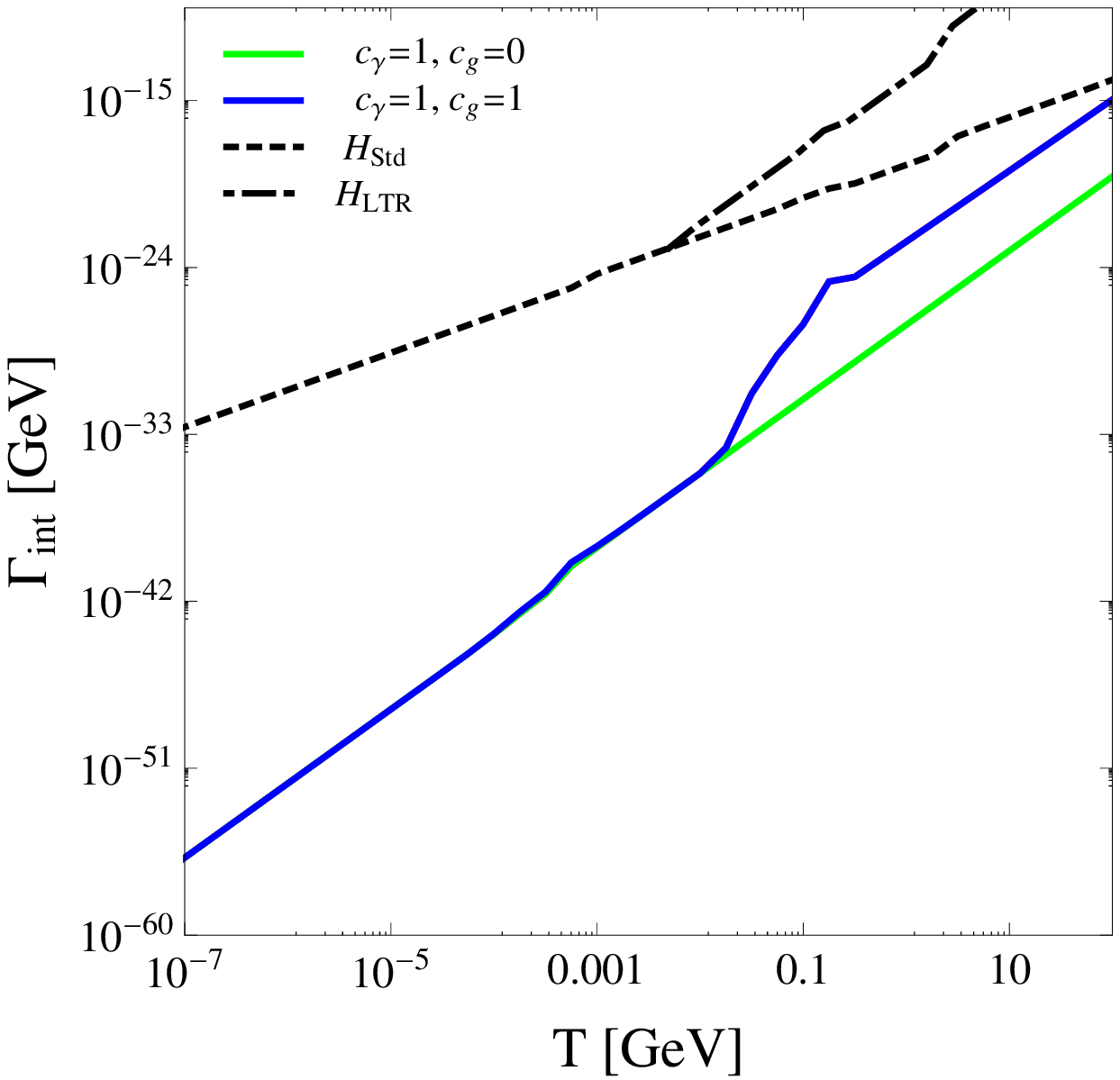} 
\end{center}
\caption{A comparison of the rates associated with different axion-production
processes in the early universe.  Here we have taken $M_c = 10^{-11}$~GeV,
$\fhatX =10^{15}$~GeV, $\Lambda_G = 1$~TeV, $\TRH = 5$~MeV, and  
$\xi = 1$.  The left panel shows the production rate for each
process for an individual axion species $a_\lambda$ with $\lambda = 1$~MeV 
(\ie, a value well within the asymptotic, large-$\lambda$ regime, where the 
rates are the least suppressed).  The right panel shows the integrated 
production rate for each process, including contributions from all modes with
$\lambda < T$.  The most relevant processes for thermal axion production in
this scenario are $\pi \pi\rightarrow \pi  a_\lambda$ production (red curve), 
$e^\pm\gamma\rightarrow e^\pm a_\lambda$ via the Compton process 
(yellow curve), $e^\pm\gamma\rightarrow e^\pm a_\lambda$ production 
via the Primakoff process (orange curve),  inverse decays of the form 
$\gamma\gamma\rightarrow a_\lambda$ (green curve), production via the quark-gluon 
process $qg\rightarrow qa$ (blue curve), and pion-production off nuclei 
(purple curve).  It should be noted that the Compton process requires a non-zero
electron-electron-axion coupling $c_e$, and that the curve shown here corresponds to 
the case in which $c_e=1$.  The value of the Hubble parameter as a function of $T$
in both the Standard (black dashed curve) and LTR (black dash-dotted curve) 
cosmologies are also shown.
\label{fig:RatesVsH}}
\end{figure}

In Fig.~\ref{fig:RatesVsH}, we provide a pictorial comparison of the
rates for the axion-production processes enumerated above as functions
of temperature.  The left panel shows the rates for the production of
a single axion species $a_\lambda$ with $\lambda = 1$~MeV in a scenario
with $\fhatX = 10^{15}$~GeV, $M_c = 10^{-11}$~GeV, and $\Lambda_G = 1$~TeV.
As before, we have taken $\xi=1$ and set $c_g = c_\gamma = c_e = 1$.  The
red curve corresponds to the rate $\Gamma(\pi \pi \rightarrow \pi a_\lambda)$ 
for the pionic process; the orange curve to the rate 
$\Gamma_{\mathrm{Prim}}(e^\pm \gamma \rightarrow e^\pm a_\lambda)$
for the Primakoff process; the green curve to the
rate $\Gamma(\gamma\gamma \rightarrow a)$ for the inverse-decay process;  
the blue curve to the rate $\Gamma(N \pi \rightarrow N' a_\lambda)$ for 
the pion-conversion process off nuclei; and the purple curve to the rate
$\Gamma(q \gamma \rightarrow q a_\lambda)$ for the quark-gluon process.
As the hadron description of the theory is valid only for $T\lesssim \LambdaQCD$, and
likewise, the quark/gluon description is only valid for $T\gtrsim \LambdaQCD$, the rates
$\Gamma(\pi \pi \rightarrow \pi a_\lambda)$, $\Gamma(N \pi \rightarrow N' a_\lambda)$, and 
$\Gamma(q \gamma \rightarrow q a_\lambda)$ are only defined on one side or the other of this
scale.  The yellow curve corresponds to the rate
$\Gamma_{\mathrm{Comp}}(\pi \pi \rightarrow \pi a_\lambda)$ for the Compton-like 
process for $c_e = 1$.  For purposes of comparison, we 
also show the Hubble parameter as a function of $T$ for two different cosmologies:
the standard cosmology (black dashed curve), and an LTR cosmology with 
a reheating temperature $\TRH = 5$~MeV (black dot-dashed curve).  
The value of $\lambda$ we have chosen here is well within the asymptotic
regime for this choice of $M_c$ and $\fhatX$; hence the rates displayed here 
represent take essentially the maximal values possible for any $a_\lambda$ in the
scenario.  In the right panel of Fig.~\ref{fig:RatesVsH}, we show, for the same choice
of $M_c$ and $\fhatX$, the total contribution to the axion-production rate 
obtained by summing the rates for all $a_\lambda$ for which $\lambda \leq T$ ---
\ie, those which will be kinematically accessible at a given temperature. 

The most salient lesson to draw from of Fig.~\ref{fig:RatesVsH} 
is that even after the contributions
from all kinetically accessible $a_\lambda$ states are included in the thermal
axion-production rate, none of the relevant processes by which a thermal population
of axions might be produced comes close to satisfying the $\Gamma \sim H$ criterion.  
This implies that the $a_\lambda$, even when taken together, never attain thermal
equilibrium with the plasma after inflation ends.  Furthermore, these results also  
justify the claims made above, that the electron Primakoff process and inverse 
decays of the form $\gamma\gamma\rightarrow a$ are the most relevant 
axion-production processes for $T\lesssim \LambdaQCD$, while  
hadronic processes dominate the axion-production rate for $T\gtrsim \LambdaQCD$.   

Let us now estimate the contribution to $\Omegatot$ from 
thermal axion production in the context of an LTR cosmology with
a reheating temperature of $\TRH=5$~MeV.  
For concreteness, we focus on the case of a photonic axion with
$c_\gamma = 1$ and $c_g = c_i = 0$ for all $i$; however, the results
for other coupling assignments should not differ drastically from those
obtained here.  We begin by noting that  
any contribution to $\Omega_\lambda$ generated at temperatures $T \gtrsim \TRH$, 
\ie, during the reheating phase, will be substantially diluted 
due to entropy production from inflaton decays.  It is therefore 
legitimate to restrict our attention to axion production within the subsequent 
RD era.  For a photonic axion, the processes which contribute to thermal 
axion production are inverse decays and $e^\pm\gamma\rightarrow e^\pm a$,
the latter of which, since we are assuming $c_e = 0$, is dominated by the
Primakoff process.  The Boltzmann equation for the number density 
$n_\lambda$ of each $a_\lambda$ is therefore effectively given by
\begin{equation}
   \dot{n}_\lambda + (3H + \Gamma_\lambda) n_\lambda ~=~ 
      C_\lambda^{\mathrm{ID}}(T) + C_\lambda^{\mathrm{Prim}}(T)
   \label{eq:ThermProdBoltzEqWithn}
\end{equation}
for $T \lesssim \TRH$, where 
$C_\lambda^{\mathrm{Prim}}(T)$ and $C_\lambda^{\mathrm{ID}}(T)$ are the
contact terms associated with the electron-Primakoff and inverse-decay rates given
in Eqs.~(\ref{eq:ElectronPrimakoff}) and~(\ref{eq:InverseDecayRate}), respectively.
For $T \gg \lambda,m_e$, these contact terms are well-approximated by the expressions
\begin{eqnarray}
  C_\lambda^{\mathrm{Prim}}(T) & \approx & 
     \frac{2\alpha}{3\pi^2}G_\gamma (\wtl^2 A_\lambda)^2
     \frac{T^6}{\fhatX^2}\left[\ln\left(\frac{9}{4\pi\alpha}\right) + 0.8194\right]
    \nonumber \\
  C_\lambda^{\mathrm{ID}}(T) & \approx & 2G_\gamma (\wtl^2 A_\lambda)^2 
     \frac{\lambda^5 T}{\pi^2\fhatX^2} K_1\left(\frac{\lambda}{T}\right)~,
  \label{eq:ContactTermsExplicit}
\end{eqnarray}
where $K_1(x)$ denotes the Bessel function of the first kind.  To obtain a
rough estimate of the relic abundance in situations in which either $m_e$ or $\lambda$
is comparable to or greater than $T$, we 
modify the expression for $C_\lambda^{\mathrm{Prim}}(T)$ given in
Eq.~(\ref{eq:ContactTermsExplicit}) by including an additional exponential factor
$e^{-(\lambda + m_e)/T}$ to model the effect of Boltzmann suppression.  
 
\begin{figure}[t!]
\begin{center}
  \epsfxsize 3.0 truein \epsfbox {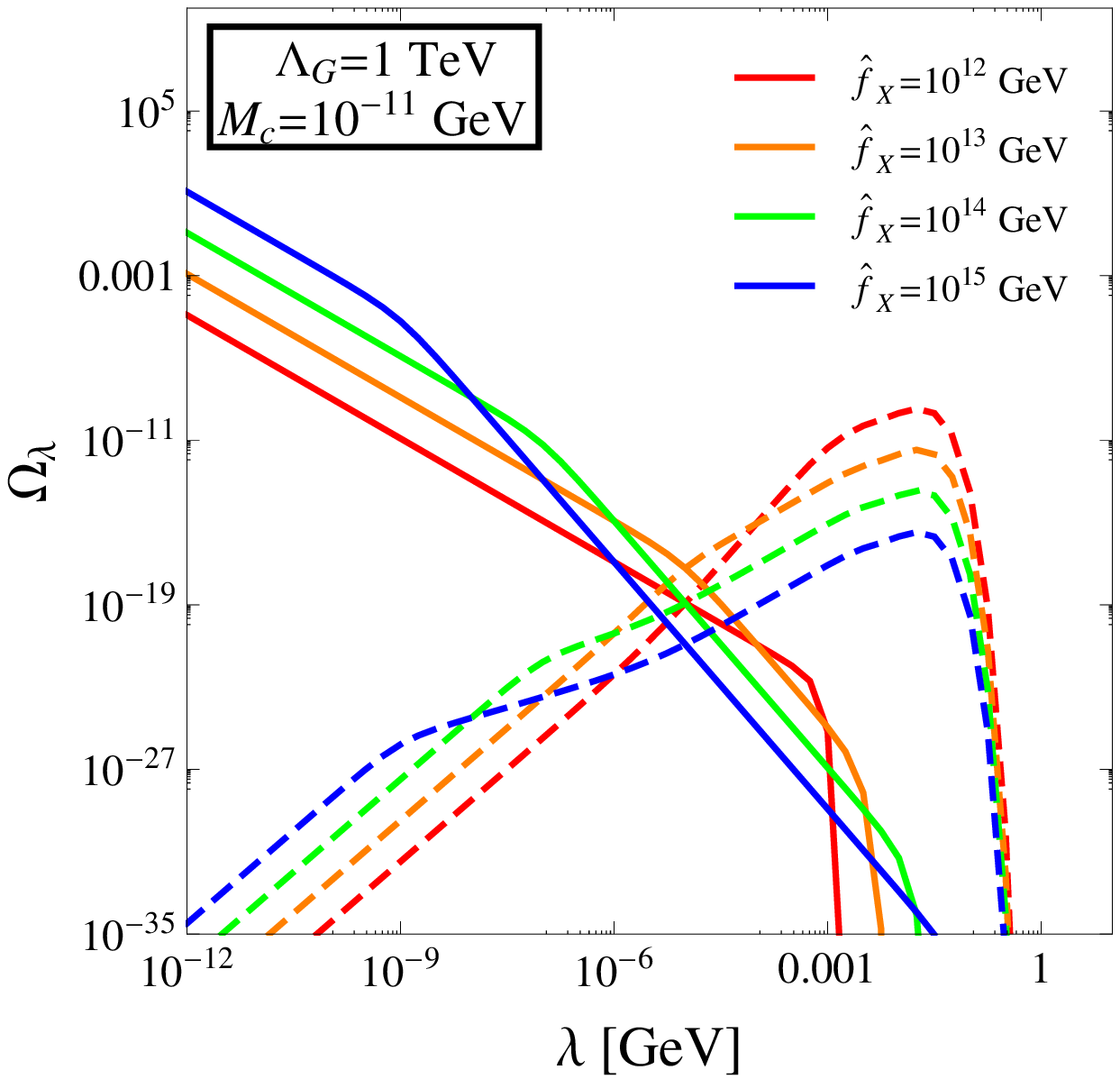} ~~~~
  \epsfxsize 3.0 truein \epsfbox {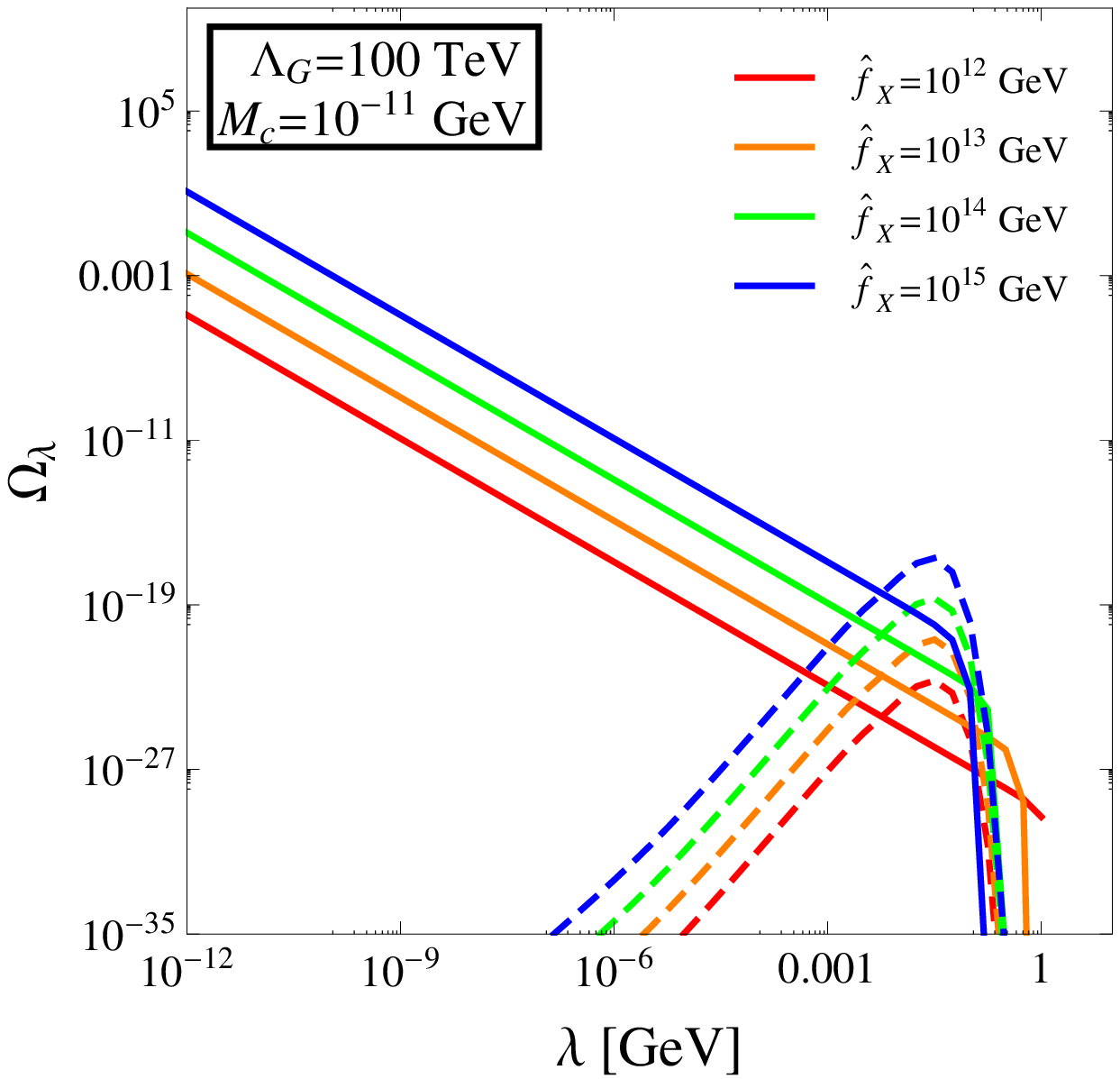} 
\end{center}
\caption{Contributions to the individual mode abundances $\Omega_\lambda$ for 
a photonic axion from thermal production (dashed curves) and misalignment 
production (solid curves), plotted as functions of $\lambda$
for $\fhatX = 10^{12}$~GeV (red curves), 
$\fhatX = 10^{13}$~GeV (orange curves), $\fhatX = 10^{14}$~GeV (green curves), 
$\fhatX = 10^{15}$~GeV (blue curves).  The left panel displays the results for
$\Lambda_G = 1$~TeV, while the right panel displays the results for $\Lambda_G = 100$~TeV.  
The other model parameters have been set to $M_c = 10^{-11}$~GeV, $\TRH = 5$~MeV, 
$\xi = g_G = c_\gamma = 1$.  
\label{fig:OmegaCompFromThemal}}
\end{figure}

From Eq.~(\ref{eq:ThermProdBoltzEqWithn}) we estimate the relic abundance 
of axions produced by interactions with the SM particles in the thermal bath.
To do so, we neglect the decay term and rewrite the resulting equation in
terms of the quantity $Y_\lambda \equiv n_\lambda/s$, 
where $s$ is the entropy density, in order to remove the Hubble term:
\begin{equation}
   s\dot{Y}_\lambda ~\approx~ 
      C_\lambda^{\mathrm{ID}}(T) + C_\lambda^{\mathrm{Prim}}(T)e^{-(\lambda+m_e)/T}~.
   \label{eq:ThermProdBoltzEqWithY}
\end{equation}
By numerically integrating this equation, we obtain an estimate of the thermal contribution 
$\Omega_\lambda^{(\mathrm{therm})}$ to the abundance $\Omega_\lambda$ of each $a_\lambda$ at
present time:
\begin{equation}
  \Omega_\lambda^{(\mathrm{therm})} ~\approx~ \frac{\lambda\Tnow^3\tMRE}{\rhocrit} 
     \int_{\Tnow}^{\TRH}  
  \frac{3}{\kappa(T)}\left(\frac{\TMRE}{T}\right)^{3/\kappa(T)}
  \frac{g_{\ast s}(\Tnow)}{g_{\ast s}(T)}
  \left[C_\lambda^{\mathrm{ID}}(T) + C_\lambda^{\mathrm{Prim}}(T)
     e^{-(\lambda+m_e)/T}\right]dT~,
  \label{eq:OmegaThermNow}
\end{equation}
where $g_{\ast s}(T)$ is the number of interacting degrees of freedom present in the 
thermal bath at temperature $T$, and where $\kappa(T)$ is 
defined in Eq.~(\ref{eq:DefOfkappaForH}).
The results of this integration are displayed in Fig.~\ref{fig:OmegaCompFromThemal}.  In
this figure, we compare the contributions to the relic abundance $\Omega_\lambda$ of a
given $a_\lambda$ from misalignment production and thermal production for a 
variety of different choices of the model parameters.

It is clear from 
Fig.~\ref{fig:OmegaCompFromThemal} that for these parameter assignments,
$\Omega_\lambda^{(\mathrm{therm})}$ only becomes comparable with the relic-density 
contribution $\Omega_\lambda^{(\mathrm{mis})}$ from vacuum misalignment for reasonably 
heavy $a_\lambda$.  Neither $\Omega_\lambda^{(\mathrm{mis})}$ nor
$\Omega_\lambda^{(\mathrm{therm})}$ for such $a_\lambda$ is non-negligible 
compared with the $\Omega_\lambda^{(\mathrm{mis})}$ contribution from the 
lighter modes.  Indeed, summing over $\lambda$ to obtain the total thermal 
contribution $\Omegatot^{(\mathrm{therm})}$ to the axion relic abundance at present time
yields $3.8 \times 10^{-6} \lesssim \Omegatot^{(\mathrm{therm})} \lesssim  3.8 
\times 10^{-4}$.  We may therefore safely conclude that
$\Omegatot^{(\mathrm{therm})} \ll \Omegatot^{(\mathrm{mis})}$
within the preferred region of parameter space for bulk-axion models of 
dynamical dark matter, and that the population of $a_\lambda$ generated by 
the misalignment mechanism dominates the relic density of the DDM ensemble.
  
To summarize the results of this section, we have examined the primary 
mechanisms through which a cosmological population of DDM axions 
may be generated, including misalignment production, thermal production, and 
production by decaying relics.  We have shown that within the 
preferred region of parameter space specified in Ref.~\cite{DynamicalDM2},
the contribution to the total present-day dark-matter relic abundance  
from misalignment production $\Omegatot^{(\mathrm{mis})}$ indeed dominates over the 
contributions from all other production mechanisms.  This justifies the emphasis
placed on misalignment production in Ref.~\cite{DynamicalDM2}.
Still, we note that
although populations of axions produced via those other channels collectively 
represent a negligible fraction of $\Omegatot$, those populations can nevertheless 
play an important role in constraining bulk-axion DDM models.  For example, the 
thermal population of axions discussed above can still leave a significant imprint 
on the diffuse X-ray spectrum despite the small size of $\Omegatot^{(\mathrm{therm})}$, 
because $\Omega_\lambda^{(\mathrm{therm})} \ll \Omega_\lambda^{(\mathrm{mis})}$ when 
$\lambda$ is large.
We shall return to this point in Sect.~\ref{sec:XrayGammaRay}, where we will show
that this imprint is nevertheless consistent with current observational limits.


\section{Phenomenological Constraints on Dark Axion Ensembles\label{sec:Bounds}}


In the previous section we characterized the various mechanisms 
which contribute to the generation of a cosmological population of relic axions in 
axion DDM models and compared the sizes of their contributions to $\Omegatotnow$.  
Given that this population constitutes the dark-matter ensemble in our axion DDM model,
we now turn to examine the relevant phenomenological, astrophysical, and 
cosmological constraints on that population of axions.  
As we shall see, some of these constraints pertain 
generically to any theory of dark matter, or to any theory containing 
late-decaying relics.  Others are particular to models involving light, 
weakly-coupled fields.  Still others pertain to theories with
large extra dimensions in general, regardless of the presence or 
absence of a bulk axion field. 

As we have seen in Refs.~\cite{DynamicalDM1,DynamicalDM2},
the properties of the dark-matter ensemble and its constituent
fields in our bulk-axion DDM model are determined primarily by three parameters:
the compactification scale $M_c$, the $U(1)_X$-breaking scale $\fhatX$, and the 
confinement scale $\Lambda_G$ for the gauge group $G$.  Because these parameters
play such a central role in characterizing the dark sector in our model,
we shall seek to phrase our phenomenological constraints in terms of restrictions
on $M_c$, $\fhatX$, and $\Lambda_G$ whenever possible.  
Of course, in addition to these primary parameters, a number of other ancillary
quantities also have an impact on the phenomenology of 
our model, and thus are also constrained by data.  
These include the scales $H_I$ and $\TRH$ associated with cosmic 
inflation, the coupling coefficients $c_g$, $c_\gamma$, and $c_i$, and so forth.
Generally speaking, these additional parameters play a subordinate role in 
determining the mass spectrum and relic abundances of the $a_\lambda$, 
and the values they take are typically far more model-dependent than $M_c$, 
$\fhatX$, and $\Lambda_G$.  Thus, while certain experimental and observational 
limits serve to constrain the values these additional parameters may take, it 
ultimately turns out to be possible to phrase the majority of constraints on our 
model as bounds on $M_c$, $\fhatX$, and $\Lambda_G$.  Indeed, as we shall see 
in Sect.~\ref{sec:Combined}, most of the critical bounds can be expressed 
conveniently in this manner.
We will also be interested in how these bounds constrain certain derived 
quantities of physical importance, such as the quantity $y$ defined in Eq.~(\ref{eq:DefsOfyandmPQ}), which quantifies the amount of mixing that occurs
across our DDM ensemble.


\subsection{Constraints from Background Geometry\label{sec:ConstraintsLargeED}}


The first set of constraints we consider are those which apply generically 
to theories with extra dimensions, independently of the presence or properties
of the bulk axion field whose KK excitations constitute the DDM ensemble in our
model.  These constraints arise primarily from experimental limits on 
the physical effects to which the tower of KK gravitons necessarily present
in such theories gives rise.  We will primarily focus here on scenarios involving 
$n$ flat extra dimensions in which the fields of the SM are localized on a 3-brane, 
while gravity, as always, necessarily propagates throughout the entirety of the 
$D=(4+n)$-dimensional bulk.

Perhaps the most significant and direct bound on $M_c$
in theories with extra dimensions arises due to
modifications of Newton's law at short distances as a consequence of
KK-graviton exchange.  The lack of evidence for any such effect at 
modified-gravity experiments to date implies constraints on the sizes and 
shapes of those extra dimensions.  For the case of a single large, flat extra 
dimension, the current limit on the compactification scale from
such experiments is~\cite{KapnerModGrav}
\begin{equation}
  M_c ~\gtrsim~ 3.9\times 10^{-12}~\mbox{GeV}~.
  \label{eq:MinimumMc}
\end{equation}
This lower limit on the compactification scale is robust in the sense that
even if there exist additional compact dimensions with radii $r_i \ll 1/M_c$, 
this bound is essentially unaffected.  For this reason, Eq.~(\ref{eq:MinimumMc})
turns out to represent the most significant constraint on the parameter space
of bulk-axion DDM models from considerations which derive solely from 
the presence of extra dimensions. 

There also exist additional constraints on the compactification geometry 
which arise due the relationship
between this scale, the effective four-dimensional Planck scale $M_P$, and 
the fundamental scale of quantum gravity $M_D$.  These constraints are generally
more sensitive to the details of the compactification scenario.  In general, the
fundamental scale $M_D$ is related to $M_P$ by~\cite{ADD}
\begin{equation}
  M_P^2 ~=~ V_n M_D^{2+n}~,
\label{eq:EffPlanckRel}
\end{equation} 
where $V_n$ is the volume of the $n$-dimensional manifold on which the extra dimensions are
compactified.  For the simple case in which this manifold is a flat, rectangular $n$-torus,
the volume $V_n$ is simply the product of $(2\pi r_i)$ for each cycle of the torus.  Assuming
all radii are equal to a common radius $r$, we then have   
\begin{equation}
  r^{-1} ~\geq~ 2\pi M_D^{\mathrm{min}} \left(\frac{M_D^{\mathrm{min}}}{M_P}\right)^{2/n}~.
  \label{eq:McBoundMDmin}
\end{equation}  
Bounds on the scale $M_D$ appearing in the literature are frequently predicated on 
these assumptions.  However, we emphasize that in situations in which the $r_i$ are not
all equal, or in which the compactification geometry differs from that of a flat,
rectangular $n$-torus, those bounds can be considerably modified.  

Under the assumption that the compactification geometry resembles that on 
which Eq.~(\ref{eq:McBoundMDmin}) is predicated, one may derive constraints on $M_D$, 
$r$, or combinations of the two.  For example, one class of
constraints which arise in theories with extra dimensions are those 
implied by the non-observation of effects related to thermal KK graviton 
production in astrophysical sources such as stars~\cite{HannestadRaffeltNeutronStar} 
and supernovae~\cite{KKGravitons1987A,HannestadRaffeltSupernovae}.  A brief 
synopsis of the most relevant bounds in this class is given
in Ref.~\cite{HannestadRaffeltNeutronStar}, all of which depend
crucially on the fundamental quantum-gravity scale $M_D$.  The most stringent of 
these constraints currently derives from limits on photoproduction and stellar heating 
by gravitationally trapped KK gravitons in the halos of neutron stars.  Indeed,
for a theory involving $n$ extra dimensions with equal radii, one finds that for 
$n = 2$, the bound is $r^{-1} \geq 5.8 \times 10^{-7}$~GeV, while for $n = 3$, 
one finds $r^{-1} \geq 3.8 \times 10^{-10}$~GeV~\cite{HannestadRaffeltNeutronStar}.

Collider data also place limits on $r$ and $M_D$ in theories with 
extra dimensions.  Searches for evidence of KK-graviton production in 
the monojet (\ie, $j + \met$) channel have been performed by the 
ATLAS~\cite{ATLASMonojet33pb,ATLASMonojet1fb} and CMS~\cite{CMSMonojet} collaborations.  
The most recent ATLAS analysis~\cite{ATLASMonojet1fb}, 
conducted with $1\mathrm{~fb}^{-1}$ of integrated luminosity, 
constrains $M_D \gtrsim \{3.16,2.50,2.15\}$~TeV at $95\%$~C.L.\ for $n=\{2,3,4\}$ flat 
extra dimensions with equal radii.  The most recent CMS analysis~\cite{CMSMonojet},
conducted at a comparable integrated luminosity, yields the slightly more stringent 
constraint $M_D \gtrsim \{4.03,3.21,2.80\}$~TeV at $95\%$~C.L.\ for 
the corresponding values of $n$.     
Limits from searches for KK-graviton effects in the diphoton~\cite{CMSLargeEDDiphoton} 
and dimuon~\cite{CMSLargeEDDimuon} channels at $36\mathrm{~pb}^{-1}$ and 
$39\mathrm{~pb}^{-1}$ of integrated luminosity, respectively, have also been 
derived by the CMS collaboration, but these are currently less stringent than the 
constraints from the $j + \met$ channel. 

It is important to realize that the aforementioned bounds on $M_D$ as a function 
of the compactification geometry do {\it not}\/ necessarily translate directly into 
analogous bounds on $f_X$ for a given $\fhatX$.  Unlike the graviton 
field, the bulk axion field in our DDM model need not necessarily propagate 
throughout the entirety of the 
extra-dimensional volume, but may in principle also be confined to a
$(4 + n_a)$-dimensional subspace of that volume, where $n_a < n$.  When this is the case, 
$\fhatX$ is related to $f_X$ by the generalization of Eq.~(\ref{eq:fhatInTermsOff}): 
\begin{equation}
  \fhatX^2 ~=~ V_{n_a} f_X^{2+n_a}~.
\end{equation} 
Note that this relationship differs from that which exists between $M_P$ and $M_D$
because $n_a < n$.
In this paper, as in Ref.~\cite{DynamicalDM2}, we focus on the case in which 
the axion field propagates in a single extra dimension of radius $R$, irrespective 
of the size, shape, or number of extra dimensions which compose the totality of the 
bulk.  Accordingly, we define $M_c = 1/R$ to be the compactification scale 
associated with this particular extra dimension, and we shall use this notation 
throughout.  In this paper, we are not aiming to set $M_D$ at or even near the TeV scale, 
since we are not attempting to solve the hierarchy problem, but rather to address 
the dark-matter problem.  We will therefore
assume that the structure of any additional bulk dimensions is such that phenomenological
constraints on $M_D$ and the associated compactification geometry are satisfied. 
Note, however, that the Newton's-law bound in Eq.~(\ref{eq:MinimumMc}) does apply 
to $M_c$, as it applies to the compactification scale associated with any individual 
extra dimension.

Another class of constraints on scenarios involving large extra dimensions
applies to ancillary variables which characterize the cosmological context in
which our model is situated. 
For example, the prediction of the observed abundances of the light elements 
via big-bang nucleosynthesis (BBN) is one of the greatest successes of the 
standard cosmology.  Consistency with these predictions requires that
effects stemming from the presence of these extra dimensions not disrupt BBN.
Successful nucleosynthesis requires that the expansion rate 
of the universe during the BBN epoch, as quantified by the Hubble parameter 
$H(T)$, must not deviate from its usual, four-dimensional value by more than around
10\%~\cite{ADDPhenoBounds}.  In other words, there exists some temperature 
$T_*\geq \TBBN \sim 1$~MeV (usually dubbed the ``normalcy temperature'' in 
the literature) below which the radii of all extra dimensions are effectively fixed
and the bulk is effectively empty of energy density.  A variety of different 
considerations constrain $T_\ast$, most of which are related to the potentially 
observable effects of KK-graviton dynamics in the early universe:

\begin{itemize}
\item Interactions between the SM fields on the brane and the bulk graviton field
result in a transfer of energy from the brane to the bulk, and a consequent cooling
of the radiation bath on the brane.  Substantial energy loss via this 
``evaporative cooling'' mechanism would result in a modification of the expansion
rate of the universe.  At temperatures $T\lesssim T_{\mathrm{BBN}}$, such a modification
would distort the light-element abundances away from those predicted by standard BBN. 
Thus, the strength of the interactions between SM particles and
excitations of the graviton field is constrained.  

\item If the collective energy density associated with the graviton KK modes 
is substantial, that energy density could cause the universe to become 
matter-dominated too early.  In extreme cases, it could even overclose
the universe.   

\item Late decays of KK gravitons could result 
in distortions of the abundances of light elements away from the values predicted 
by BBN~\cite{ADDPhenoBounds}, which accord well with the observed values for 
these abundances.  Such decays could also result in significant entropy production.
    
\item The relationship between the Hubble parameter $H$ and the total energy density
$\rho$ of the universe is modified at early times in higher-dimensional scenarios, 
even when that energy density is overwhelmingly dominated by brane-localized 
states~\cite{EDHubbleCline,EDHubbleBinetruy,EDHubbleShiromizu}.  Such a modification
could have a substantial effect on BBN as well.
\end{itemize}

The constraints on $T_*$ implied by these considerations have been reckoned by a 
number of authors~\cite{ADDPhenoBounds,CosmoConstraintsLargeED}, and while the precise 
values of the bounds so derived again depend on the number, size, and shape of the extra 
dimensions, the value of $M_D$, \etc, the most stringent (which tend to come from 
limits on the late decays of the excited KK modes) generally tend to restrict $T_*$ 
to within the rough range 
$4\mbox{~MeV}\lesssim T_*\lesssim 20\mbox{~MeV}$~\cite{ADDPhenoBounds}.

One possibility for achieving such conditions is to posit that $T_*$ be identified
with the reheating temperature $\TRH$ associated with a period of cosmic inflation
initiated by an inflaton field which is localized on the same 3-brane as the SM 
fields.  During such an 
inflationary epoch, any contributions to the energy density of the universe
from bulk states which existed prior to the inflationary epoch (save for those
which, like the contributions to $\rho_\lambda$ from vacuum misalignment, scale like 
vacuum energy) are inflated away.  Furthermore, if the inflaton field
decays primarily to other brane-localized states, no
substantial population of bulk states is regenerated during the subsequent 
reheating phase.  Thus, by adopting a LTR cosmology with a reheating temperature
$4\mbox{~MeV}\lesssim \TRH\lesssim 20\mbox{~MeV}$, we thereby ensure that 
the relevant constraints related to KK-graviton production in the early universe
are satisfied.  We also note that a reheating temperature of $\TRH\gtrsim 4$~MeV 
is sufficient to ensure that the thermal populations of the SM fields 
(and, in particular, the three neutrino species) required in standard BBN are 
generated by the thermal bath after reheating~\cite{TReheatLimits,KawasakiLTR2}.

In summary, while stringent constraints exist on theories with large extra 
dimensions, these constraints can be satisfied by adopting an LTR cosmology
with $4\mbox{~MeV}\lesssim \TRH\lesssim 20\mbox{~MeV}$ and a compactification 
manifold for which the astrophysical bounds listed above may consistently 
be satisfied for a given choice of $M_D$ and $f_X$.  
Since $\Omegatotnow$ is generated via non-thermal means in our bulk-axion 
model, as discussed in Sect.~\ref{sec:AbundanceConstraints}, such a low 
value of $\TRH$ is not an impediment to obtaining a dark-matter relic abundance 
$\Omegatotnow \approx\OmegaDM$.  In fact, as shown in Ref.~\cite{DynamicalDM2}, 
adopting an LTR cosmology is actually an {\it asset}\/ in terms of generating a  
dark-matter relic abundance of the correct magnitude.  Likewise,
since the relationship between $\fhatX$ and $f_X$ need not be identical
to the relationship between $M_P$ and $M_D$, 
constraints which concern the effects of KK gravitons can be   
satisfied without imposing equally severe restrictions on the parameters 
$\fhatX$, $M_c$, and $\Lambda_G$ which govern the properties of the dark-matter 
ensemble.  Indeed, the only significant model-independent constraint on these 
parameters turns out to be the constraint quoted in Eq.~(\ref{eq:MinimumMc}) 
from tests of Newton's law at short distances.


\subsection{Axion Production with Subsequent Detection: Helioscopes and 
Light Shining Through Walls\label{sec:Helioscopes}}


We now address the constraints which relate directly to the
phenomenological, astrophysical, and cosmological implications associated
with the KK tower of axion fields which constitute the DDM ensemble in our model.
We begin by discussing the limits derived from a wide variety of experiments 
designed to detect axions and axion-like particles via their 
interactions with the photon field. (For extensive reviews of 
these experiments, see Refs.~\cite{KimReview2,JaeckelReview}.)
To date, none of these experiments have seen any conclusive evidence
for such particles, and the null results of these experiments therefore 
imply constraints on the effective couplings between such axion-like particles 
and the photon field.

In order to determine how the results of the experiments listed above 
serve to constrain the parameter space of our bulk-axion DDM model, it 
is useful to divide those experiments into several broad classes, based on
the sort of physical process each probes.  One important class of experiments
comprises those in which axions are produced via their interactions with the 
fields of the SM and then subsequently detected via those same 
interactions.  These include helioscope experiments such as CAST~\cite{CAST} 
and ``light-shining-through-walls'' (LSW) experiments such as BEV, GammaeV, 
and ALPS.  Searches for coherent conversion of solar axions 
to X-ray photons in germanium and sodium-iodide crystals via Bragg diffraction 
which have been performed at experiments such as    
DAMA~\cite{DAMAAxion}, TEXONO~\cite{TEXONOAxion}, SOLAX~\cite{SOLAXAxion}, 
and COSME~\cite{COSMEAxion} also fall into this category.
The characteristic which distinguishes experiments in this class 
from others is that these experiments are affected by decoherence phenomena.
Indeed, it has been observed~\cite{DDGAxions} that in theories with bulk axions,
such phenomena result in a substantial suppression of the rate for any process 
involving the production and subsequent decay of axion modes relative to 
na\"{i}ve expectations.

Let us briefly review the origin of this suppression by 
focusing on the interaction between the photon field and the axion KK modes given
in Eq.~(\ref{eq:HighT4DAxionAction}).  (The results for the coupling between these
modes and the other SM fields are completely analogous.)  We begin by defining a state 
\begin{equation}
  a' ~\equiv~ \frac{1}{\sqrt{N}}\sum_n^N r_n a_n~,
\end{equation}
which represents the particular linear combination of KK eigenstates $a_n$ 
that couples to any physics on the brane, such as $F_{\mu\nu}\widetilde{F}^{\mu\nu}$
or any pair of SM fields.  Here $N \sim f_X/M_c$ denotes the number of modes in the sum.  
Written in terms of $a'$, the relevant term in the interaction
Lagrangian becomes 
\begin{equation}
 L_{\mathrm{int}} ~\ni~
   \frac{\alpha c_\gamma\sqrt{N}}{8\pi^2\fhatX}a' F_{\mu\nu}\widetilde{F}^{\mu\nu}~.
\end{equation} 
In other words, $a'$ couples to the SM fields with a strength proportional to 
$\sqrt{N}/\fhatX \sim 1/f_X$.  Consequently, the cross-section for any 
physical process which involves axions production via interactions with
the SM fields followed by subsequent detection via the same sorts of interactions    
will take the form
\begin{equation}
  \sigma(t) ~\propto~ \frac{N^2}{\fhatX^4} \times P(t)~,
  \label{eq:CouplerToCouplerXSecBasic}
\end{equation} 
where $P(t) = |\langle a'(t)|a'(t_0) \rangle|^2$ is the probability for a
state $a'$ created at time $t_0$ to be in the same state $a'$ at time $t$. 
It can be shown that when $N$ is large, $P(t)$ is given by\begin{equation}
  P(t) ~=~ \frac{1}{N^2} \left[\sum_\lambda \wtl^8 A_\lambda^4 +
     2\sum_\lambda \sum_{\lambda'<\lambda}\wtl^4\wtl'^4A_\lambda^2A_{\lambda'}^2
     \cos \left(\frac{(\lambda^2-\lambda'^2)(t-t_0)}{2p}\right) \right]~,
  \label{eq:Poft}
\end{equation}
where $p$ is the initial momentum of the axion.

At very early times,
when $t\approx t_0$, the cosine factor in $P(t)$ is approximately
unity for all values of $\lambda$ and $\lambda'$.  At such times, all of the 
terms in the sum appearing in the second term on the right side of 
Eq.~(\ref{eq:Poft}) add coherently.  As a result, this term, combined together
with the first term, yields a factor on the order of $N^2$.  However, as the 
system evolves, the cosine terms will no longer sum coherently, and a
random-walk behavior ensues, according to which the two
terms combine to yield a factor of $\mathcal{O}(N)$ rather than of 
$\mathcal{O}(N^2)$.  The
time scale $\tau_D$ associated with this decoherence --- or, more precisely, 
the scale at which $P(t) = 0.1 P(t_0)$ --- is found to be~\cite{DDGAxions}
\begin{equation}
  \tau_D ~ \approx ~ 10^{-5} \left(\frac{2p}{\mPQ^2}\right) \frac{y^2}{N^2}
       ~\approx~
      1.32\times 10^{-29} \left( \frac{p}{\mbox{GeV}}\right)
      \left( \frac{\fhatX}{\mbox{GeV}}\right)^{-2} \mathrm{s}~,       
\end{equation}
where $y$ is defined in Eq.~(\ref{eq:DefsOfyandmPQ}).
Since $\tau_D$ is clearly quite small for any combination of $p$ and $f_X$ values  
of experimental relevance, any method of detecting axions which relies on their
production and subsequent detection will feel the effect of this decoherence. 
By contrast, detection methods which rely on axion production without subsequent
detection (such as missing-energy signals at colliders, energy dissipation
from supernovae, \etc) or which probe for evidence of a cosmic population of 
relic axions (such as microwave-cavity experiments) 
will be unaffected by this phenomenon.  

The consequences of axion decoherence for physical processes in the decoherence
regime are readily apparent.  In this regime, as discussed above, the term in brackets
in $P(t)$ scales like $N$ rather than $N^2$; hence any cross-section
which takes the form given in Eq.~(\ref{eq:CouplerToCouplerXSecBasic})
will scale with $N\sim f_X/M_c$ according to
\begin{equation}
  \sigma(t>\tau_D) ~\propto~ \frac{N}{\fhatX^4} ~\sim~ \frac{1}{N}\frac{1}{f_X^4}~.
  \label{eq:CouplerToCouplerXSecSubbed}
\end{equation}
In other words, such cross-sections are suppressed by an 
additional factor of $N$ relative to the na\"{i}ve expectation obtained by
setting $\fhatX \rightarrow f_X$ in Eq.~(\ref{eq:CouplerToCouplerXSecBasic}).
Thus, due to the decoherence effect, 
any experimental bound on the effective coupling $G_{a\gamma\gamma}$ of 
a single four-dimensional axion to the photon field which takes the form
$G^2_{a\gamma\gamma} < (G_{a\gamma\gamma}^{\mathrm{max}})^2$ translates to a
bound $G^2_{a\gamma\gamma} < (G_{a\gamma\gamma}^{\mathrm{max}})^2/\sqrt{N}$ 
for five-dimensional axion, rather than to 
$G^2_{a\gamma\gamma} < (G_{a\gamma\gamma}^{\mathrm{max}})^2/N$.
Given the parametrization for $G_{a\gamma\gamma}$ given in 
Eq.~(\ref{eq:HighT4DAxionAction}), we can phrase any such constraint as a 
bound on $\fhatX$:
\begin{equation}
  \fhatX ~\gtrsim~ \frac{c_\gamma\alpha}{2\pi G_{a\gamma\gamma}^{\mathrm{max}}}
  \left(\frac{M_c}{f_X}\right)^{1/4}~.
\end{equation}
Using Eq.~(\ref{eq:fhatInTermsOff}), we may rewrite this constraint in the form
\begin{equation}
  \fhatX ~\gtrsim~ \frac{1}{(2\pi)^{13/10}} 
  \left(\frac{c_\gamma\alpha}{G_{a\gamma\gamma}^{\mathrm{max}}}\right)^{6/5}
  \frac{1}{M_c^{1/5}} ~=~
  \big(2.50\times 10^{-4}\big)\, c_\gamma (G_{a\gamma\gamma}^{\mathrm{max}})^{-6/5}M_c^{-1/5}~.
 \label{eq:HelioscopefhatLimit}
\end{equation}

The most stringent limit from the class of experiments categorized above (\ie,
those for which the phenomenon of decoherence is relevant) is currently 
the $G_{a\gamma\gamma} \lesssim 8.8 \times 10^{-11} \mbox{~GeV}^{-1}$ bound
obtained by CAST~\cite{CAST}.  The most stringent limit from crystalline 
detectors is the 
$G_{a\gamma\gamma} \lesssim 1.7 \times 10^{-9} \mathrm{~GeV}^{-1}$
bound from DAMA~\cite{DAMAAxion}, and limits on $G_{a\gamma\gamma}$ from LSW 
experiments are typically roughly three orders of magnitude higher than the
CAST limit.  The corresponding bound on $\fhatX$ from 
Eq.~(\ref{eq:HelioscopefhatLimit}) is
\begin{equation}
  \fhatX ~\gtrsim~ \left(2.92 \times 10^8\right)\, c_\gamma^{6/5} \left(\frac{M_c}{\mbox{GeV}}\right)^{-1/5} 
   \mbox{~GeV}~. 
  \label{eq:CASTConstraint}
\end{equation}
Note that even for $M_c$ at the experimental lower limit given in 
Eq.~(\ref{eq:MinimumMc}), the constraint in Eq.~(\ref{eq:CASTConstraint}) 
is satisfied as long as $\fhatX \gtrsim 5.58 \times 10^{10}$~GeV.   


\subsection{Microwave-Cavity Experiments and Direct Detection of Dark-Matter Axions}


Another class of experiments which place constraints on the couplings of 
axions and axion-like fields to SM particles consists of those which 
involve the direct detection of a cosmological population of axions.  The most
sensitive experiments in this class are those associated with 
dedicated microwave-cavity detectors such as ADMX~\cite{ADMX} and 
CARRACK~\cite{CARRACK}. 
Detectors of this sort are used to search for the resonant conversion of 
dark-matter axions with mass $m_a$ to photons with energies 
$E_\gamma \approx m_a$ in the presence of a strong magnetic field.  
As a result, the observation of a signal at such a detector depends
crucially on whether the mass of the axion in question lies within the
range of photon energies probed.  The axion mass range currently covered by 
ADMX spans only from $1.9\times 10^{-15}$~GeV to 
$3.5\times 10^{-15}$~GeV~\cite{ADMX}, and the projected future mass sensitivity 
extends only as high as $10^{-13}$~GeV.  Likewise, the projected sensitivity for
CARRACK extends only as high as $3.5\times 10^{-14}$~GeV.     

As discussed in Ref.~\cite{DynamicalDM2}, the region of parameter space 
which is the most interesting from a DDM perspective is that within which 
$y \lesssim 1$ and mixing among the light axion KK modes is substantial, for
it is this region within which the full tower contributes meaningfully to 
$\Omegatotnow$.  Within this region of parameter space, the lightest mode in 
the tower has a mass $\lambda_0 \approx M_c/2$.  Taken in conjunction
with the bound on $M_c$ from modified-gravity experiments given in 
Eq.~(\ref{eq:MinimumMc}), this result implies that 
$\lambda_0 \gtrsim 1.5\times 10^{-12}$~GeV in highly-mixed bulk-axion scenarios.
The projected ranges for both ADMX and CARRACK lie well below this threshold 
for $\lambda_0$.  We therefore conclude that no meaningful constraints on 
bulk-axion DDM models can be derived from the results of these experiments.


\subsection{Axion Production without Subsequent Detection: 
Stars and Supernovae\label{sec:StarnAndSN}}


We now turn to examine an additional class of constraints on bulk-axion DDM 
models: those related to astrophysical processes in which the $a_\lambda$ 
are produced through their interactions with the SM field, but never 
directly detected.  Among the constraints in this class are 
limits on axions, moduli, and other light 
scalars derived from the non-observation of their would-be effects on the 
lifetimes, energy-loss rates, \etc, of various astrophysical sources such 
as stars and supernovae.  These effects include the following:

\begin{itemize}
\item Axions and other light fields whose interactions with the particles of the 
SM are extremely weak and whose mean free paths are consequently extremely long 
can dissipate energy from stars extremely efficiently.  Such dissipation can 
accelerate stellar cooling and result in observable alterations in stellar life 
cycles, including the life cycle of our own sun.  

\item Similarly, such light fields can carry away a substantial fraction of the energy 
liberated by supernovae.  Limits may therefore be placed on the strengths of these
interactions from the non-observation of such effects for supernova SN1987A.

\item A diffuse population of long-lived axions or KK gravitons initially produced 
by stars and supernovae could decay at late times, distorting light-element abundances
and producing an observable X-ray or $\gamma$-ray signal in the keV $-$ MeV range or 
higher.  No evidence for such a signal has been seen by EGRET, FERMI, HEAO, Chandra,
COMPTEL, \etc  
\end{itemize}

As is well known, these considerations lead to some of the most 
stringent constraints on standard, four-dimensional QCD axions.  We now turn to 
examine how these limits constrain the parameter space of generalized bulk-axion models.  

The primary distinction between processes in which the presence of the 
$a_\lambda$ is ascertained by direct detection and those in which it 
is only inferred from an energy deficit is that in the latter class of 
processes, the $a_\lambda$ appear as particles in the asymptotic final state.
Thus, the contributions from the individual $a_\lambda$ to the 
overall event rate for any such a process add not at the 
amplitude level, but at the cross-section level.  
The decoherence phenomena discussed in Sect.~\ref{sec:Helioscopes} 
are therefore irrelevant for such processes, and the total cross-section 
$\sigma^{\mathrm{prod}}_{\mathrm{tot}}$ for the production of ``missing energy'' in the
form of $a_\lambda$ fields by any given physical process is simply 
the sum of the individual production cross-sections 
$\sigma^{\mathrm{prod}}_\lambda$ for each axion 
species.  Since the effective coupling between each $a_\lambda$ and any pair of 
SM fields includes a factor $\wtl^2A_\lambda/\fhatX$ from mass mixing, 
as indicated in Eq.~(\ref{eq:ActionInMassEigenbasis}), each of these individual 
production cross-sections scales as
\begin{equation}
  \sigma^{\mathrm{prod}}_\lambda ~\propto~ 
     \frac{1}{\fhatX^2}(\wtl^2A_\lambda)^2~.
  \label{eq:SigmaProdProp}
\end{equation}

When it occurs, axion production will have a characteristic energy scale $\Ech$
determined by the surrounding environment.  This energy scale may be 
associated, for example, with the 
temperature of a star or supernova core, or with the center-of-mass energy 
$\sqrt{s}$ of a collider.  Provided that $\Ech \gg M_c$ (an assumption valid 
for all physical contexts of relevance in bounding the 
large-extra-dimension scenarios considered here), it follows that 
$\lambda \ll \Ech $ for a large number of $a_\lambda$.  Such $a_\lambda$ 
can be considered to be effectively massless as far as production kinematics is 
concerned, implying that to a very good approximation, 
$\sigma^{\mathrm{prod}}_\lambda$ depends on $\lambda$ exclusively 
through the coupling-modification factor appearing in 
Eq.~(\ref{eq:SigmaProdProp}).  (For those modes for which threshold effects 
are important, such an approximation will overestimate 
$\sigma^{\mathrm{prod}}_\lambda$ and result in an overly conservative bound.)  
By contrast,
$\sigma^{\mathrm{prod}}_\lambda$ will be effectively zero for
those $a_\lambda$ with masses $\lambda \gg \Ech$ in any thermal environment 
due to Boltzmann suppression, and will vanish outright in a non-thermal one.
Therefore, it is reasonable to evaluate
$\sigma^{\mathrm{prod}}_{\mathrm{tot}}$ by taking any additional factors in
Eq.~(\ref{eq:SigmaProdProp}) to be essentially independent of $\lambda$ and
by truncating the sum over modes at $\lambda \sim \Ech$.  Thus, we find that
\begin{equation}
  \sigma^{\mathrm{prod}}_{\mathrm{tot}} ~\propto~ \aleph^2(\Ech)/\fhatX^2
  \label{eq:SigmaWithAlephSuppression}
\end{equation}
where the ``effective'' coupling $\aleph(\Ech)$ is given by
\begin{equation}
    \aleph(\Ech) ~\equiv~ 
     \Bigg[\sum_{\lambda=\lambda_0}^{\Ech}(\wtl^2A_\lambda)^2\Bigg]^{1/2}.
   \label{eq:DefOfAleph}
\end{equation}  
Since the number of modes contributing to $\sigma^{\mathrm{prod}}_{\mathrm{tot}}$
is large by assumption, and since their masses are closely spaced, it 
is generally legitimate to approximate $\aleph(\Ech)$ by an integral
\begin{equation}
  \aleph(\Ech) ~\approx~ \Bigg[\frac{1}{M_c}\int_{\lambda_0}^{\Ech} 
    (\wtl^2A_\lambda)^2 d\lambda\Bigg]^{1/2}~.
\end{equation}

\begin{figure}[t!]
\begin{center}
  \epsfxsize 3.0 truein \epsfbox {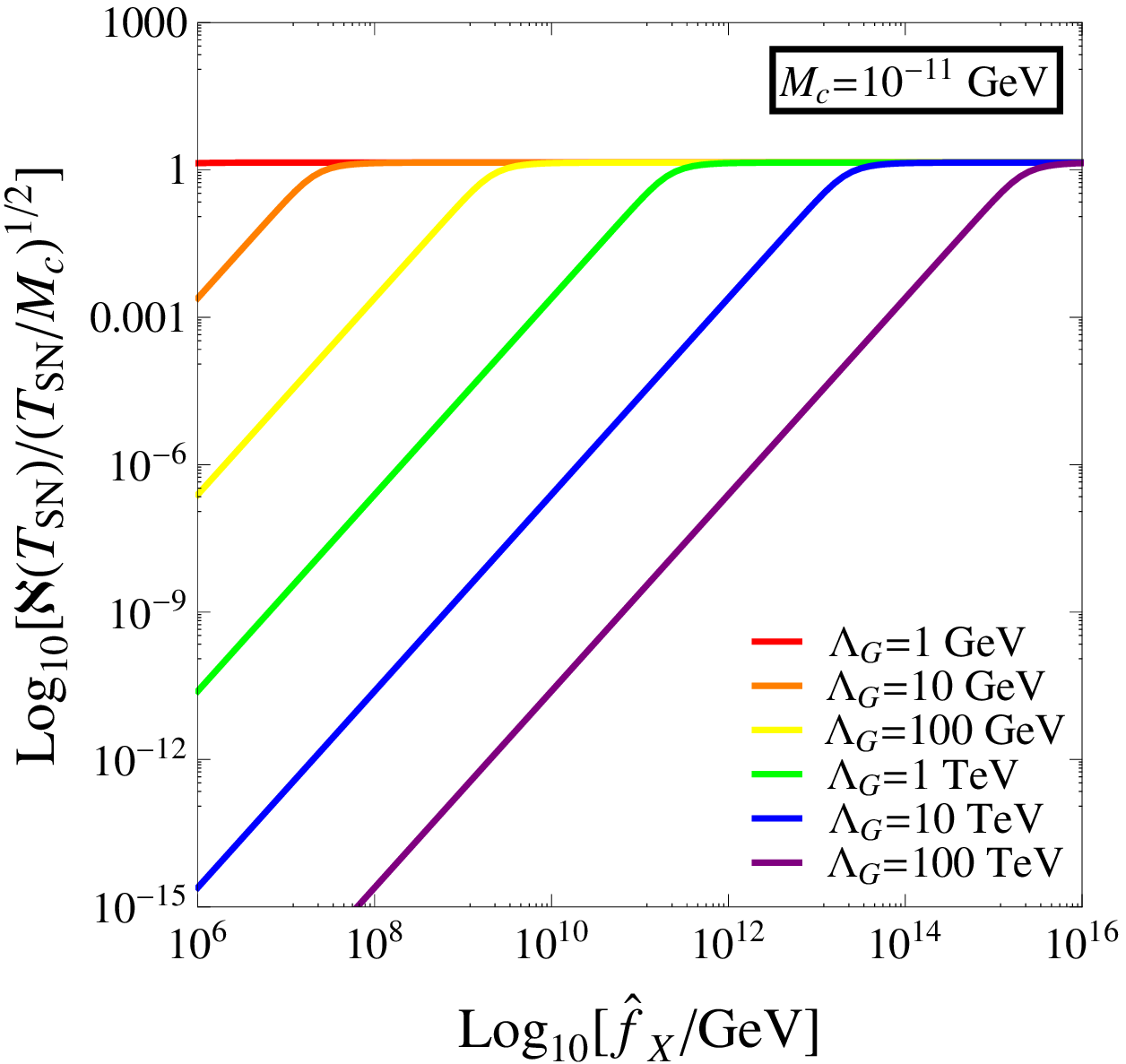} ~~~~
  \epsfxsize 3.0 truein \epsfbox {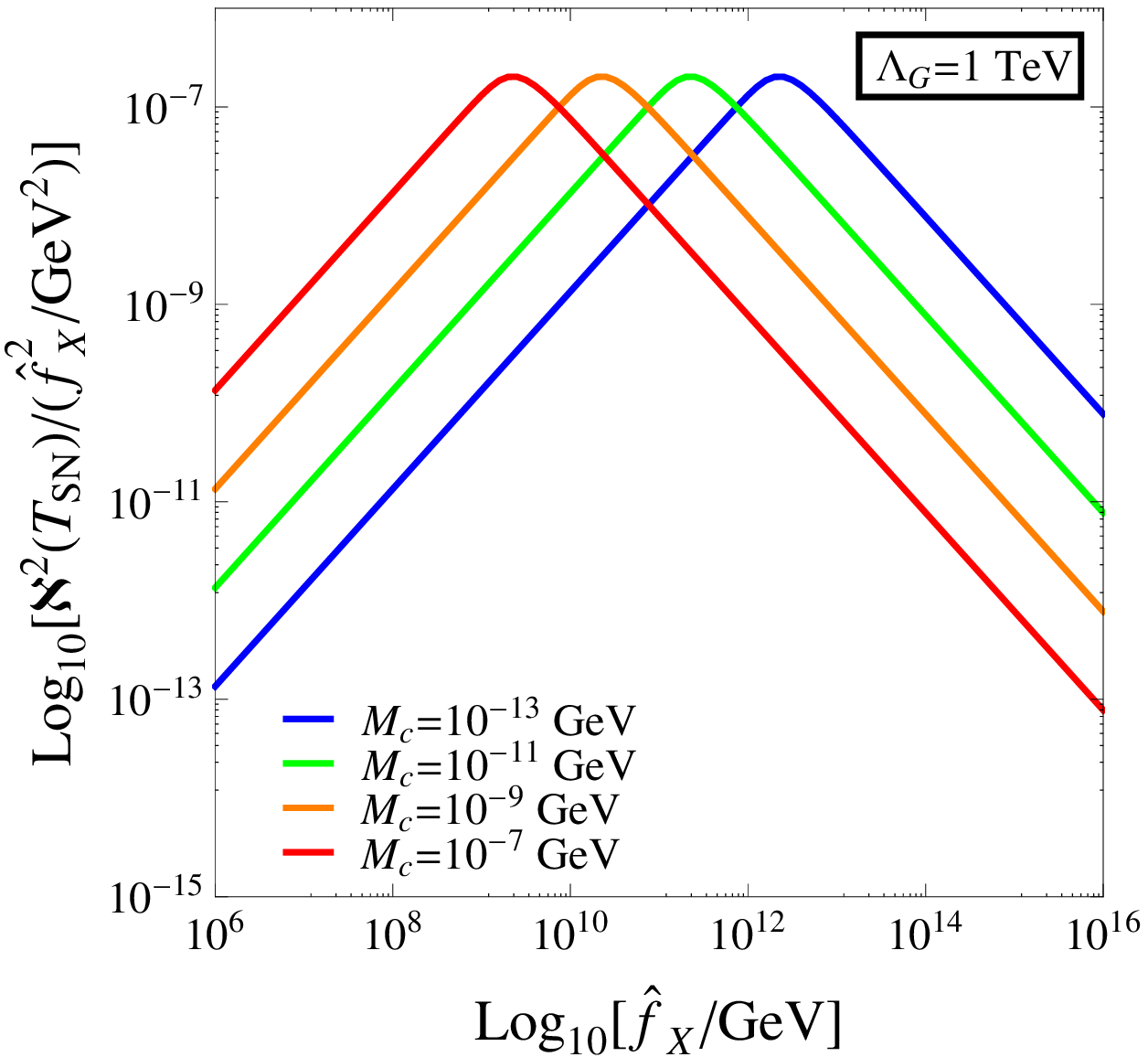} 
\end{center}
\caption{The dimensionless ``effective coupling'' factor 
$\aleph(\Ech)$ defined in Eq.~(\ref{eq:DefOfAleph}), shown as a 
function of the relevant scales $M_c$, $\fhatX$, and $\Lambda_G$.  
In the left panel, we display curves of 
$\aleph(\Ech)$, each corresponding to a particular value of $\Lambda_G$
and normalized to the value $(\Ech/M_c)^{1/2}$ taken by $\aleph(\Ech)$
in the absence of mixing, as a function of $\fhatX$ with fixed 
$M_c = 10^{-11}$~GeV.  It is readily apparent that the net effect of mixing 
within the KK axion tower
is to significantly suppress this effective coupling, thereby loosening
the corresponding production-cross-section constraints.  
In the right panel, we display curves showing the overall 
cross-section-suppression factor $\aleph^2(\Ech)/\fhatX^2$ 
as a function of $\fhatX$ for fixed $\Lambda = 1$~TeV, each corresponding 
to a particular value of $M_c$.  For each set of curves, we have taken 
$\xi = g_G = 1$, and have chosen $\Ech = 30$~MeV, which corresponds roughly to 
the core temperature $T_{\mathrm{SN}}$ of SN1987A.     
\label{fig:AlephPanels}}
\end{figure}

The quantity $\aleph(\Ech)$ clearly plays a crucial role in the phenomenology
of bulk-axion scenarios.  It is therefore worth pausing a moment to examine 
in detail how $\aleph(\Ech)$ depends on the physical scales $\fhatX$, $M_c$, 
and $\Lambda_G$.  A straightforward calculation shows that $\aleph(\Ech)$ 
has the parametric scaling behaviors
\begin{equation}
  \aleph(\Ech) \sim
  \begin{cases}
   \displaystyle
     \vspace{0.25cm}  \frac{\Ech^{3/2} M_c^{1/2}\fhatX^2}{\Lambda_G^4} & 
    ~~~\displaystyle \fhatX \ll \frac{\Lambda_G^2}{M_c}  \\    
   \displaystyle \left(\frac{\Ech}{M_c}\right)^{1/2}& 
    ~~~\displaystyle \fhatX \gg \frac{\Lambda_G^2}{M_c}~.
  \end{cases}
  \label{eq:AlephCases}
\end{equation}  
The first case in Eq.~(\ref{eq:AlephCases}) corresponds to $y \ll 1$, signaling
a highly-mixed axion KK tower for which $\wtl^2 A_\lambda \sim \wtl$.  By contrast,
the second case corresponds to $y\gg 1$, signaling a relatively unmixed axion KK tower
for which $\wtl^2 A_\lambda \sim$~constant.  These results for $\aleph(\Ech)$ are 
illustrated in the left panel of Fig.~\ref{fig:AlephPanels} for 
$\Ech = 30$~MeV, a value which is physically meaningful in that it corresponds 
roughly to the core temperature of supernova SN1987A.  Remarkably, we observe
that $\aleph(\Ech)$ experiences a {\it suppression} for $y \ll 1$.  In other words,
mixing within the axion KK tower acts to suppress the magnitude of the total 
production cross-section for processes in which the $a_\lambda$ appear as 
missing energy.  This is an important result, for it indicates that constraints on the
parameter space of our bulk-axion DDM model derived from limits on axion 
production in stars, supernovae, colliders, \etc, will be considerably 
weaker than one might expect from na\"{i}ve dimensional analysis.  Moreover,
this result applies more generally to any theory involving KK towers of
scalar fields whose squared-mass matrix contains both brane-mass and 
KK-mass terms. 

In order to illustrate more explicitly the physical consequences of $\aleph(\Ech)$ in
our bulk-axion DDM model, we likewise display the behavior of the overall scaling  
factor $\aleph^2(\Ech)/\fhatX^2$ for $\sigma^{\mathrm{prod}}_{\mathrm{tot}}$ in the right 
panel of Fig.~\ref{fig:AlephPanels}.  The results shown
in this panel further illustrate a significant general property of this scaling factor:
namely, that the cross-section is actually suppressed not only for large $\fhatX$, but
also for small $\fhatX$, due to the parametric behavior of $\aleph(\Ech)$ described 
in Eq.~(\ref{eq:AlephCases}).  Thus, for any given choice of $\Lambda_G$ and $M_c$, 
there exists a maximum possible value for $\sigma^{\mathrm{prod}}_{\mathrm{tot}}$, 
which is only attained at some particular value of $\fhatX$.   
These results again illustrate the dramatic effect that $\aleph(\Ech)$ can have in 
suppressing $\sigma^{\mathrm{prod}}_{\mathrm{tot}}$ in our bulk-axion DDM model.    
        
Within the class of constraints from processes in which axions are produced but not 
subsequently detected, use of $\aleph(\Ech)$ allows us to translate experimental 
bounds on four-dimensional axion models into bounds on theories including towers 
of bulk scalars.
The leading such bound is obtained from energy-loss limits from SN1987A.  For
a standard four-dimensional QCD axion, this bound is roughly~\cite{RaffeltSN1987ABound}
\begin{equation}
  f_a ~\gtrsim~ 4\times 10^8\mbox{~GeV}~. 
  \label{eq:SN1987ABoundIn4D}
\end{equation}    
By contrast, in the bulk-axion scenario under consideration here, each 
$a_\lambda$ light enough to be produced within the thermal environment 
of SN1987A can contribute to the overall energy-dissipation rate.  
Since the temperature 
$T_{\mathrm{SN}}$ associated with the supernova core is roughly
$30$~MeV, the appropriate modification of Eq.~(\ref{eq:SN1987ABoundIn4D}) for
a general axion which couples to hadrons 
with a coupling coefficient comparable in magnitude to that of a QCD axion is
\begin{equation}
  \fhatX ~\gtrsim~ \big(4\times 10^8\mbox{~GeV}\big)\, 
     \aleph(T_{\mathrm{SN}})~. 
  \label{eq:SN1987ABoundIn5D}
\end{equation} 
It then follows that in highly-mixed scenarios, this constraint can be 
significantly weaker than the corresponding constraint on KK-graviton 
production derived in Ref.~\cite{ADDPhenoBounds}, due to suppression by 
$\aleph(T_{\mathrm{SN}})$.  Indeed, the corresponding constraint on KK-graviton 
production is directly obtained by replacing $\fhatX\rightarrow M_P$ and 
$\aleph(T_{\mathrm{SN}})\rightarrow T_{\mathrm{SN}}/M_c$ in 
Eq.~(\ref{eq:SN1987ABoundIn5D}).  

While the SN1987A bound is indeed one of the most stringent constraints
on the QCD axion, it is not necessarily applicable for all general axions.  
This is because the bound quoted in Eq.~(\ref{eq:SN1987ABoundIn4D}) is
predicated on the assumption that nucleon bremsstrahlung 
($N + N \rightarrow N + N + a$) and other hadronic processes dominate the
rate for the production of the light scalar in question in the supernova 
core.  This presupposes that the light scalar couples to nuclei with a 
strength comparable to that of a QCD axion.  If this is not the case, however, 
the constraints obtained from SN1987A energy-loss limits can differ considerably 
from the standard QCD-axion bound.  For example, if from among the SM particles, 
the general axion couples only to the photon field, the dominant production 
processes will be $e^-\gamma\rightarrow e a_\lambda$, 
$p^+ \gamma\rightarrow p^+ a_\lambda$, and $p^+n\rightarrow p^+n\gamma a_\lambda$.  
In this case, the considerably weaker bound~\cite{MassoALPsLongPaper}
\begin{equation}
  \fhatX ~\gtrsim~ \big(2.32 \times 10^6  \mbox{~GeV}\big)\, c_\gamma
  \label{eq:SN1987ABoundIn4DPhotonOnly}
\end{equation}
is obtained for a four-dimensional field.  Translating this result
to the case of a KK tower of axions, as above, we find that    
\begin{equation}
  \fhatX ~\gtrsim~ \big(2.32 \times 10^6 \mbox{~GeV}\big)\,
     c_\gamma\, \aleph(T_{\mathrm{SN}})~. 
  \label{eq:SN1987ABoundIn5DPhotonOnly}
\end{equation}
Furthermore, in general axion models, $c_\gamma$ may not necessarily be 
of $\mathcal{O}(1)$.  In other words, the SN1987A constraint is
sensitive to the $U(1)_X$ and $SU(2)\times U(1)_Y$ charges of the fields
in the model, and is thus highly model-dependent. 

An analogous limit on $\fhatX$ can be derived  
from observations of the lifetimes of globular-cluster (GC)
stars.  The ambient temperatures $T_{\mathrm{GC}}$ of such objects are only
$\mathcal{O}(10\mathrm{~keV})$, so axion production primarily proceeds through the 
Primakoff processes $\gamma +e^- \rightarrow a + e^-$ and 
$\gamma + n_Z \rightarrow a + n_Z$, where $n_Z$ denotes a nucleus with atomic 
number $Z$.  (Note that the dominant processes in this environment differ from the 
axion-nucleon-nucleon bremsstrahlung processes which dominate the axion-production 
rate in supernovae.)  Such a bound will therefore arise for any general
axion for which $c_\gamma \neq 1$, regardless of whether or not it couples to
the gluon field.  The observation limit on axion production in 
GC stars is commonly phrased as an upper bound on the effective coupling 
$G_{a\gamma\gamma}$ between a standard, four-dimensional axion (or any other
similar particle) and a pair of photons, and the current bound is  
$G_{a\gamma\gamma} \lesssim 1 \times 10^{-10}\mbox{~GeV}^{-1}$~\cite{PDG}.
Since $T_{\mathrm{GC}}\approx 10$~keV, the corresponding bound on $\fhatX$ is
\begin{equation}
  \fhatX ~\gtrsim~ \big(1.16 \times 10^7 \mbox{~GeV} \big)\,  
     c_\gamma\, \aleph(T_{\mathrm{GC}})~.
  \label{eq:GlobularClusterBound}
\end{equation}   

Note that this constraint is independent of the SN1987A bounds, as it differs from the latter
in two significant ways.  First, because the relevant production process involves 
the coupling of the axion modes to photons rather than to nuclei, it depends on 
$c_\gamma$ alone and not on $c_g$.  Second, since $T_{\mathrm{GC}}\ll T_{\mathrm{SN}}$, 
far fewer of the $a_\lambda$ will be produced with any significant frequency in GC 
stars.  Consequently, the enhancement factor from the sum over kinetically-accessible axion 
modes for GC stars is far smaller.  

Finally, bounds similar to those from SN1987A and GC stars can also be derived from the 
non-observation of effects related to axion production in other astrophysical 
sources, such as our own sun~\cite{SolarLifetimeBounds}.  However, these bounds 
are found to be subleading in comparison with the SN1987A and GC-star constraints, 
essentially because they take place in far cooler environments, where the number of 
kinematically accessible modes is even further suppressed by the cutoff at $\Ech$ 
inherent in $\aleph(\Ech)$.


\subsection{Axion Production at Colliders\label{sec:Colliders}}


We now consider the collider constraints applicable to our bulk-axion 
DDM model.  Due to the highly suppressed couplings between the axion and 
the SM fields in standard four-dimensional axion models, collider data have 
virtually no relevance in constraining the parameter space of such models. 
Nevertheless, because of the huge multiplicity of light modes that arises 
in theories with light bulk fields in large extra dimensions, the net 
contribution to the event rates for certain processes from all of these modes
taken together can potentially yield observable signals.  For example,
modes which are stable on collider time scales all appear as missing energy, and
can lead to signals in channels such as $pp\rightarrow j + \met$ and 
$pp\rightarrow \gamma + \met$.  In addition, the heavier, more unstable modes which 
decay before exiting the detector can potentially give rise to additional 
signature patterns which may include displaced vertices.  Indeed, we have already 
discussed in Sect.~\ref{sec:ConstraintsLargeED} how current limits from LHC data 
constrain the parameter space of one such bulk field --- the higher-dimensional 
graviton --- for which the monojet and monophoton channels mentioned above are of particular 
importance.  Since the $a_\lambda$ in our bulk-axion model couple to the fields of the 
SM in much the same manner as KK gravitons, it is no surprise that the collider 
phenomenology of the $a_\lambda$ turns out to be quite similar to that of KK gravitons.  

We begin by discussing those signals which arise due to the combined
effect of the $a_\lambda$ which are sufficiently long-lived so as 
to manifest themselves in a collider detector as missing energy. 
Collider processes in which the $a_\lambda$ appear as $\met$ are yet further 
examples of the class of processes discussed in the previous section in which 
axions are produced but not subsequently detected.  The net cross-section for any such
process is therefore likewise suppressed by axion mixing in the manner described in
Eq.~(\ref{eq:SigmaWithAlephSuppression}), with $\Ech$ given by the 
center-of-mass energy $\sqrt{s}$ of the collider.

Which specific channels are relevant for the discovery of a bulk axion at hadron colliders
depends crucially on how the five-dimensional axion couples to the SM fields, 
and in particular on 
whether or not it couples appreciably to either light quarks or gluons.  For a 
field with an $\mathcal{O}(1)$ value of either $c_g$ or $c_q$ (where $q=\{u,d,s,c\}$),  
the principal discovery channel at both the Tevatron and 
the LHC is $pp\rightarrow j + \met$, a channel which is also one of the principal discovery
channels for KK gravitons.  Thus, in order to obtain a rough estimate of the 
constraints on the parameter space of our bulk-axion model from the null results
of monojet searches, we translate the bound on the fundamental scale $M_D$ 
established by such searches into a bound on $\fhatX$.   
The cross-section for KK-graviton production in association with a single 
jet at a hadron collider in a theory with $n$ large, flat 
extra dimensions of equal length compactified on an $n$-torus, including 
contributions from all kinematically accessible modes, is 
roughly proportional to~\cite{ADDPhenoBounds}
\begin{equation}
  \sigma_{\mathrm{prod}}(pp\rightarrow j + G) ~\propto ~ \left(\frac{\sqrt{s}}{2\pi}\right)^n
     \frac{1}{M_{D}^{n+2}}~.
  \label{eq:KKGravitonProdSigmaProp}
\end{equation}
This implies that a bound of the form $M_D > M_D^{\mathrm{min}}$ can be translated 
into a rough bound on the parameter space of our bulk-axion model of the form  
\begin{equation}
  \frac{\aleph^2(\sqrt{s})}{\fhatX^2} ~\lesssim~ \left(\frac{\sqrt{s}}{2\pi}\right)^n 
      \frac{1}{(M_D^{\mathrm{min}})^{n+2}}~,
  \label{eq:ColliderLimitBulkAxionTranslated}
\end{equation}
where $\aleph(\Ech)$ is defined in Eq.~(\ref{eq:DefOfAleph}).
While this approximate bound does not take into account the differences in
coupling structure between KK graviton and axion fields or the sum over
polarizations for a massive graviton, it is sufficient to obtain 
parametric estimates of the resulting constraints on our three fundamental
parameters $\fhatX$, $M_c$, and $\Lambda_G$.  

\begin{figure}[ht!]
\begin{center}
  \epsfxsize 2.25 truein \epsfbox {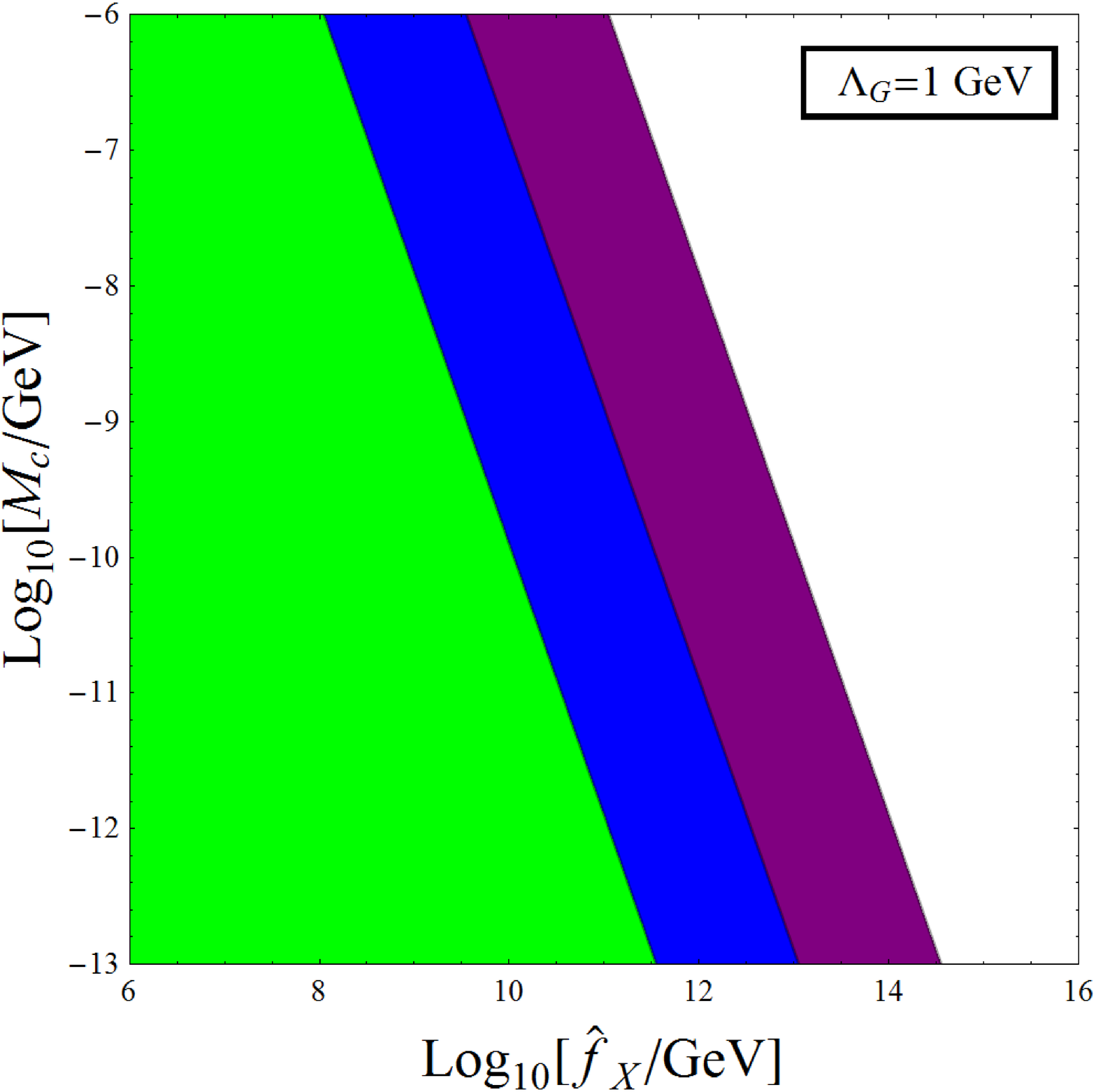} 
  \epsfxsize 2.25 truein \epsfbox {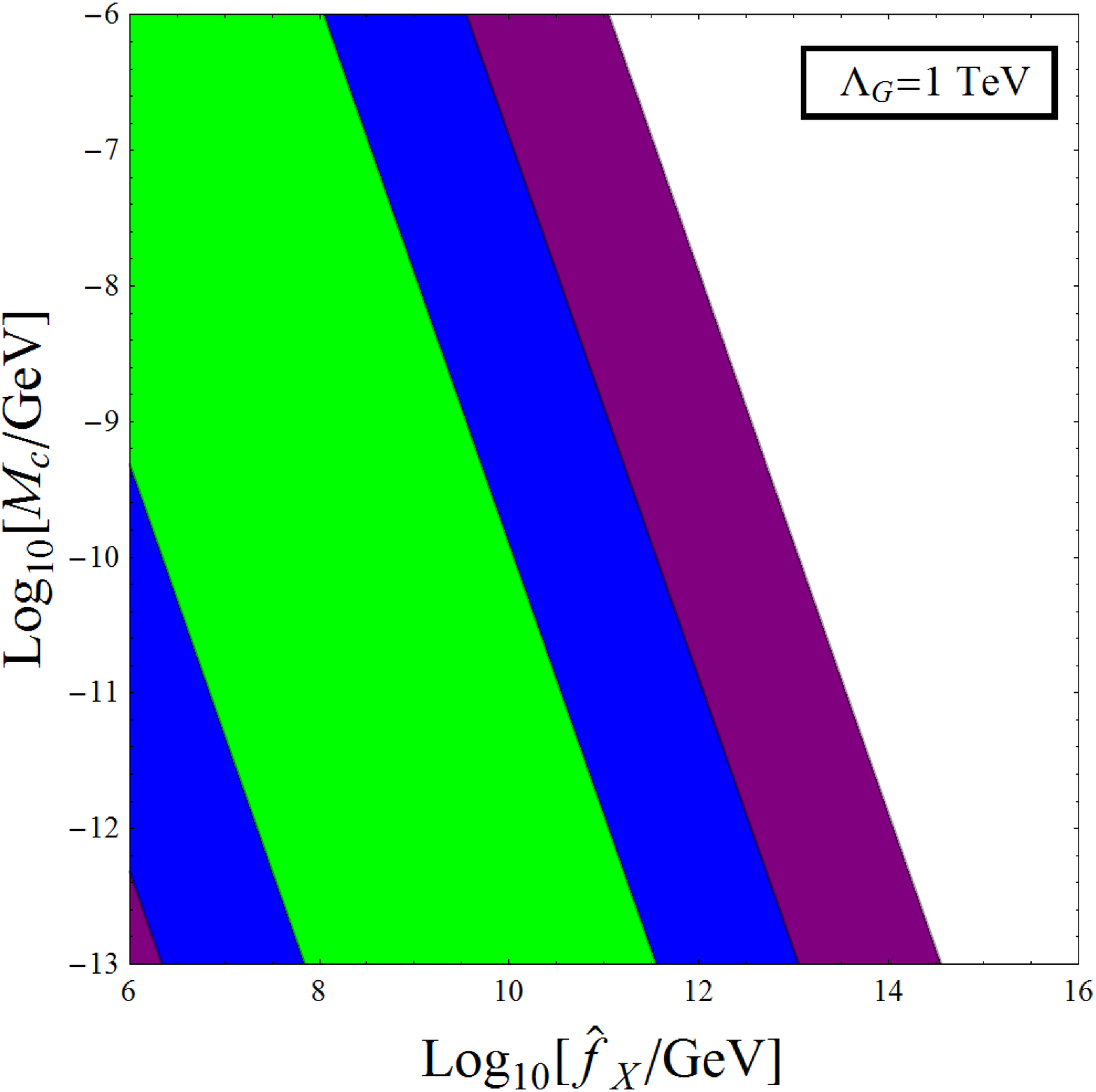} 
  \epsfxsize 2.25 truein \epsfbox {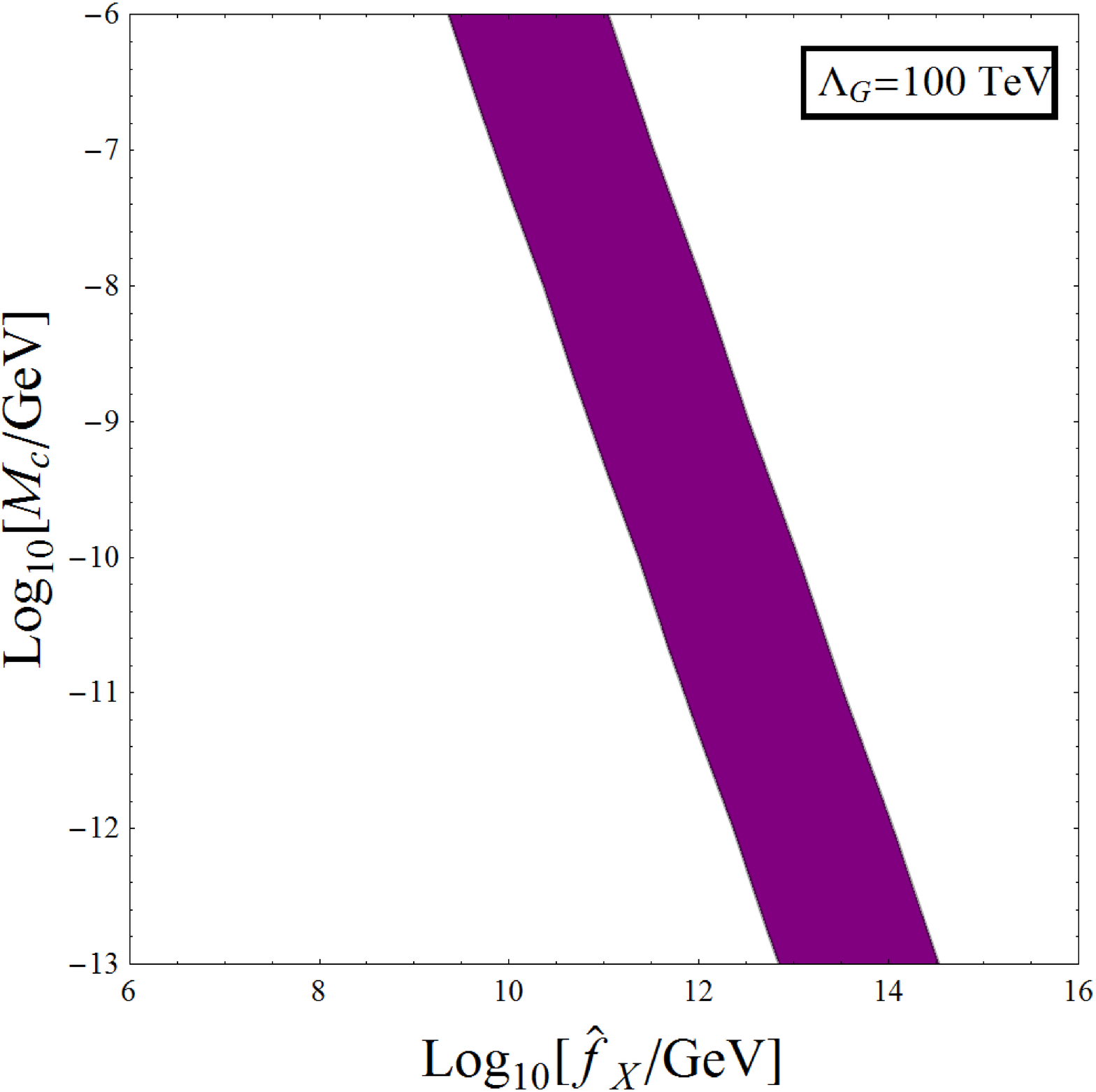}
\end{center}
\caption{Excluded regions of the $(\fhatX, M_c)$ parameter space of our DDM model 
in which the collider constraint in Eq.~(\ref{eq:ColliderLimitBulkAxionTranslated}) 
is violated for $M_D^{\mathrm{min}}=1$~TeV (green); $M_D^{\mathrm{min}}=10$~TeV 
(green and blue); and $M_D^{\mathrm{min}}= 100$~TeV (green, blue, and purple).
As $\Lambda_G$ increases, we see that satisfying the collider constraints becomes 
increasingly easy, particularly for small $\fhatX$.    
In each case, we have taken $\xi=g_c=1$ and assumed
that the axion couples to at least one light, strongly-interacting
SM particle with an $\mathcal{O}(1)$ coupling coefficient $c_g$ or $c_q$.
\label{fig:LHCAExclusion}}
\end{figure}

In Fig.~\ref{fig:LHCAExclusion}, we indicate the rough bounds on the 
parameter space of our bulk-axion DDM model which can be derived in this 
manner, given a chosen value of $M_D^{\mathrm{min}}$.  The contours shown in this
figure correspond to constraints of the form $M_D > M_D^{\mathrm{min}}$ for 
the illustrative values $M_D^{\mathrm{min}} = \{1,10,100\}$~TeV. 

We now compare these results to actual constraints on $M_D$ from current experimental
data and examine the projected LHC reach for our bulk-axion DDM model.  
The most stringent constraints from LHC data (which indeed come from the 
$pp\rightarrow j + \met$ channel) were given in Sect.~\ref{sec:ConstraintsLargeED}.  
Estimates of the future LHC reach for a theory with a single extra dimension are 
$M_D^{\mathrm{min}} \approx \{14,17\}$~TeV at integrated luminosities 
$\mathcal{L}_{\mathrm{int}}=\{10,100\}\mathrm{~fb}^{-1}$, 
respectively~\cite{GiudiceGravitonColliders}.  Likewise, Tevatron data imply a limit 
$M_D^{\mathrm{min}} \approx 2.4$~TeV for a theory with a single extra
dimension~\cite{GiudiceGravitonColliders}.
Comparing these results to those in Fig.~\ref{fig:LHCAExclusion}, we see that current 
collider constraints, while quite stringent, do not significantly impact the preferred 
region of parameter space for our bulk-axion DDM model, even in cases in which the axion 
couples to one or more strongly-interacting SM fields with an $\mathcal{O}(1)$ coupling 
coefficient.
In such cases, since the most stringent current LHC limits imply a bound of roughly 
$M_D^{\mathrm{min}} \approx 1$~TeV, the region of the parameter space of our model 
excluded by these limits roughly corresponds to the green shaded regions shown in
Fig.~\ref{fig:LHCAExclusion}.  
Since the green exclusion regions in this figure embody the most stringent such limits 
applicable to our DDM model, we shall take these to represent our collider constraints 
throughout the rest of this paper.
However, we note that for photonic axions and other axion species which do not couple 
directly to quarks or gluons, the corresponding collider constraints (which arise from 
channels such as $pp\rightarrow \gamma + \met$) are somewhat weaker.  

Before concluding this section, there is one important point which deserves emphasis.
The collider processes we have been discussing thus far are those whose
event rates receive their contributions from
the low-lying modes in the tower --- \ie, those $a_\lambda$ with lifetimes
$\tau_\lambda \gtrsim 10^{-12}$~s.  By contrast, those heavy       
$a_\lambda$ with lifetimes $\tau_\lambda \lesssim 10^{-12}$~s 
tend to decay to pairs of SM fields {\it within} the detector volume. 
The decays of such states can in principle give rise to an entirely different set 
of signature patterns.  For example, a promptly decaying $a_\lambda$ which couples 
to light quarks or gluons as well as photons would in principle contribute to event 
rates in the $pp\rightarrow jjj$ and $pp\rightarrow \gamma\gamma + j$ channels.  
However, since the total event rate in these channels receives contributions 
from a broad spectrum of $a_\lambda$ with different $\lambda$, many
event-selection techniques which are particularly useful in standard searches 
for new physics in these channels cannot be applied to a tower of decaying 
bulk axions.  For example, since the set of decaying axions cannot characterized 
by a single, well-defined mass or cross-section, no identifiable peak can be 
expected to appear in the invariant-mass distribution for the decay products of 
the heavy axions.  Such considerations render the results of standard searches for 
new physics in these channels inapplicable to our bulk-axion model --- and indeed 
to DDM models in general.  Moreover, they also likely render the identification of a 
conclusive signal of non-standard dark-matter physics in these channels
particularly challenging.  Nevertheless, the information that could potentially 
be revealed about the nature of the dark sector via such an identification is of 
sufficient magnitude and importance that an analysis of the discovery potential in 
these channels is an interesting topic for future study.         


\subsection{Axion Decays and Distortions of the Cosmic Microwave 
Background Spectrum\label{sec:CMB}}


Up to this point, we have considered those phenomenological constraints on our 
DDM model which are related to the production of particles which compose our bulk-axion 
ensemble, both with and without their subsequent detection.  By contrast, we now 
turn to discuss an entirely different set of phenomenological constraints, namely 
those which arise due to the potential decays of a {\it pre-existing}\/ 
cosmological population of such particles.  Indeed, such constraints emerge generically in 
all dark-matter scenarios in which the dark sector contains unstable, long-lived 
particles, and can be derived from observational limits on the physical 
consequences of the late decays of those particles.

There are many considerations which can be used to place such limits on scenarios 
involving decaying dark-matter particles.  For example, photons produced via the decays
of such particles 
can yield observable distortions in the CMB spectrum; contribute to the diffuse extragalactic 
X-ray and gamma-ray backgrounds; upset BBN predictions for the primordial abundances of
light elements; and result in unacceptably large entropy production during critical epochs
in the history of the universe.  Constraints on dark-matter candidates from considerations
of this sort depend not only on the decay rate of the particle species in question,
but also on the relic abundance of that species.  For this reason, the constraints 
applicable to single-particle models of dark-matter are generally not directly applicable
to models within the DDM framework.  It is therefore necessary to revisit the observational
limits on dark-matter decays within the context of our bulk-axion model of dynamical dark 
matter and assess how these limits constrain the parameter space of this model. 
 
In this section, we begin our analysis of the constraints on the late decays of the
$a_\lambda$ in our bulk-axion DDM model by examining observational limits on the 
distortions of the CMB which such decays can induce.  The type of CMB distortion to 
which a late-decaying particle contributes depends on the time at which that particle
decays.  In the very early universe, photons produced by particle decays are brought 
into thermal and kinetic equilibrium with CMB photons via a number of processes. 
The dominant processes by which newly-produced photons can equilibrate {\it thermally}\/ 
with CMB photons are double-Compton scattering
($e^- \gamma \rightarrow e^-\gamma\gamma$) and bremsstrahlung
($e^- X^\pm\rightarrow e^- X^\pm \gamma$, where $X^\pm$ is an ion).  However,
once these processes freeze out, photons produced from $a_\lambda$ decays
are unable to thermally equilibrate with the radiation bath, resulting in the 
generation of a non-zero value for the pseudo-degeneracy parameter $\mu$.
The interaction rates for these processes are given by~\cite{HuAndSilkLong}
\begin{eqnarray}
   \Gamma_{\mathrm{DC}}&\approx& 5.73\times 10^{-39} \left(1-\frac{Y_p}{2}\right)
      (\Omega_{\mathrm{B}}h^2)
      \left(\frac{T_{\mathrm{now}}}{2.7 \mbox{~K}}\right)^{3/2}
      \left(\frac{\tMRE}{t}\right)^{9/4} \mbox{GeV}
      \nonumber\\
   \Gamma_{\mathrm{BR}}&\approx& 1.57\times 10^{-36} \left(1-\frac{Y_p}{2}\right)
      (\Omega_{\mathrm{B}}h^2)^{3/2}
      \left(\frac{T_{\mathrm{now}}}{2.7 \mbox{~K}}\right)^{-5/4}
      \left(\frac{\tMRE}{t}\right)^{13/8} \mbox{~GeV}~,
      \label{eq:GammaDCandBR}
 \end{eqnarray}
where $T_{\mathrm{now}} \approx 2.725$~K is the present-day CMB temperature,
$Y_p \approx 0.23$ is the helium mass fraction, $\Omega_\mathrm{B} \approx 0.044$ is the
baryon density of the universe, and $h \approx 0.72$ is the Hubble constant. 
(Note that since $z$ is quite large during the entirety of the relevant 
time frame, we have here approximated $1+z\approx z$.)  Once these
processes freeze out, in the sense that the rates given in Eq.~(\ref{eq:GammaDCandBR})  
drop below the expansion rate $H$ of the universe, photons produced by $a_\lambda$ 
decay will no longer be able to attain thermal equilibrium with the CMB photons.  
Even after double-Compton scattering and bremsstrahlung effectively shut 
off, a number of photon-number-conserving interactions still serve to bring photons 
produced at even later times into {\it kinetic}\/ equilibrium with the radiation bath.
Dominant among these processes is elastic 
Compton scattering ($e^-\gamma\rightarrow e^-\gamma$), 
which efficiently serves to bring photons produced by $a_\lambda$ decays into kinetic 
equilibrium until a much later time $t_{\mathrm{EC}}\sim 9\times 10^9$~s, at which point 
this process too effectively freezes out.  However, since elastic Compton scattering 
conserves photon number, it cannot similarly suffice to 
bring those photons into {\it thermal}\/ 
equilibrium.  As a result, CMB distortions in the form of a non-zero value for the 
pseudo-degeneracy parameter $\mu$ can be generated by $a_\lambda$ decays during this epoch.
In addition, after elastic Compton scattering freezes out, photons produced by $a_\lambda$
decay achieve neither kinetic nor thermal equilibrium with the radiation bath.  
As a result, these photons no longer contribute the generation of $\mu$, but instead 
contribute to the generation of a Compton $y$ parameter (here denoted $y_C$, so as to
distinguish it from the ratio $y=M_c/\mX$).  Finally, at $t\sim 10^{13}$~s,
matter and radiation decouple, and any $a_\lambda$ decays occurring after this
point not affect the CMB, but instead simply persist as a contribution to 
the diffuse photon background.  This last sort of contribution will be dealt 
with separately, in Sect.~\ref{sec:XrayGammaRay}.
  
We thus see that axion decays have the potential to generate both a non-zero $\mu$
and a non-zero $y_C$.  We can therefore establish constraints on our bulk-axion DDM
model by calculating the theoretical predictions for these quantities in our model
and comparing these predictions to observational data.   

We begin our analysis of CMB distortions from $a_\lambda$ decays by addressing those
decays which result in the generation of the pseudo-degeneracy parameter $\mu$.
In general, provided that the additional contribution $\delta\rho_\gamma$ to the 
photon energy density $\rho_\gamma$ from the decay of the $a_\lambda$ fields is small 
compared to the total $\rho_\gamma$, the time-evolution of $\mu$ can
be described by the equation~\cite{HuAndSilkShort,HuAndSilkLong}
\begin{equation}
  \frac{d\mu}{dt} ~=~ \frac{d\mu_a}{dt} 
     - \mu \left(\Gamma_{\mathrm{DC}}+\Gamma_{\mathrm{BR}}\right)~.
  \label{eq:MuCMBBasic}
\end{equation}
Here $\Gamma_{\mathrm{DC}}$ and $\Gamma_{\mathrm{BR}}$ are the 
interaction rates for double-Compton scattering and
bremsstrahlung, respectively, and $d\mu_a/dt$ denotes the differential 
contribution to $\mu$ from axion decay.  For an arbitrary $d\mu_a/dt$,
the solution to this differential equation takes the form 
\begin{equation}
  \mu(t) ~=~ \exp\left[
     \frac{4}{5}\big(2C_{\mathrm{BR}}t^{-5/8}+C_{\mathrm{DC}}t^{-5/4}\big)\right]
     \int_{t_e}^t \left[\frac{d\mu_a}{dt}(t')\right] 
     \exp\left[
     -\frac{4}{5}\big(2C_{\mathrm{BR}}t'^{-5/8}+C_{\mathrm{DC}}t'^{-5/4}\big)\right] dt',
  \label{eq:SolveDiffEqCMBmu}
\end{equation}
where $t_e \approx 1.69\times 10^3$~s is the time scale associated with 
electron-positron annihilation in the early universe, and where the quantities 
$C_{\mathrm{DC}}$ and $C_{\mathrm{BR}}$ are constants related
to the double-Compton-scattering and bremsstrahlung rates $\Gamma_{\mathrm{DC}}$ 
and $\Gamma_{\mathrm{BR}}$ in Eq.~(\ref{eq:GammaDCandBR}) by
$\Gamma_{\mathrm{DC}} \equiv C_{\mathrm{DC}} t^{-9/4}$ and 
$\Gamma_{\mathrm{BR}} \equiv C_{\mathrm{BR}} t^{-13/8}$.  Moreover,
the differential contribution $d\mu_a/dt$ to $\mu$ from axion 
decays is given by the standard expression for 
contributions due to the late injection of photons from a generic source: 
\begin{equation}
  \frac{d\mu_a}{dt} ~=~ \frac{1}{2.143}\left(\frac{3}{\rho_\gamma}
     \frac{d\rho_\gamma}{dt}-
     \frac{4}{n_\gamma}\frac{dn_\gamma}{dt}\right)~.
  \label{eq:dmudtfromadecay}
\end{equation}

In general, the rate of change in the photon energy density is given by 
the Boltzmann equation for the evolution of $\rho_\gamma$.  In our
bulk-axion DDM model, this equation includes a source term from each 
decaying state in the dark-matter ensemble.  Thus, at late times, after 
all of the $a_\lambda$ have already begun oscillating coherently and the 
contribution to $\rho_\gamma$ from inflaton decays can safely be neglected, 
we find that
\begin{equation}
  \frac{d\rho_\gamma}{dt} ~=~ -4H\rho_\gamma + 
     \sum_\lambda \BRgamma\Gamma_\lambda\rho_\lambda~,  
  \label{eq:RhoLambdaAndGammaEvolEqsSGen}
\end{equation}
where $\BRgamma$ is the branching fraction of $a_\lambda$ into a pair of
photons.  Note that the source term in the Boltzmann equation for $\rho_\gamma$ 
is simply a sum of the contributions from the various $a_\lambda$ fields.
Using Eq.~(\ref{eq:RhoLambdaAndGammaEvolEqsSGen}), along with the relations
\begin{equation}
  \frac{1}{(R^4\rho_\gamma)}\frac{d(R^4\rho_\gamma)}{dt} ~=~
    \frac{1}{\rho_\gamma} \left(\frac{d\rho_\gamma}{dt}+4H\rho_\gamma\right),
  ~~~~~~~~
  \frac{1}{(R^3n_\gamma)}\frac{d(R^3n_\gamma)}{dt} ~=~
    \frac{1}{n_\gamma} \left(\frac{dn_\gamma}{dt}+3Hn_\gamma\right)~,
\end{equation}
we can rewrite Eq.~(\ref{eq:MuCMBBasic}) in the form
\begin{equation}
  \frac{d\mu_a}{dt} ~ = ~
     \frac{1}{2.143}\bigg[\frac{3}{\rho_\gamma}
       \sum_\lambda\BRgamma \Gamma_\lambda \rho_\lambda - 
     \frac{8}{n_\gamma} \sum_\lambda\BRgamma\Gamma_\lambda \frac{\rho_\lambda}{\lambda}
     \bigg]~. 
   \label{eq:MuCMBFinal}
\end{equation}
 
For the purpose of establishing a conservative bound, we focus here on the case of
a purely photonic axion.  As we saw in Sect.~\ref{sec:IntraensembleProd},  
the contribution to $\Gamma_\lambda$ from intra-ensemble decays is negligible for 
any $a_\lambda$ which decays on time scales relevant for the generation of CMB 
distortions.  It is therefore justifiable to
approximate $\Gamma_\lambda$ by the expression for $\Gamma(a\rightarrow\gamma\gamma)$
given in Eq.~(\ref{eq:PartialWidthToPhotons}) and thus to take $\BRgamma \approx 1$.
Since the energy density $\rho_\lambda$ associated with each $a_\lambda$ is given 
in Eq.~(\ref{eq:RhoOftEqnWithR}), we find that in this approximation, the first source 
term on the right side of Eq.~(\ref{eq:MuCMBFinal}) takes the form    
\begin{equation}
\sum_\lambda \BRgamma\Gamma_\lambda \rho_\lambda ~\approx~ 
    \frac{1}{2} \theta^2 G_\gamma m_X^4
     \sum_\lambda \lambda \left(\frac{t_\lambda^2}{\tRH^{1/2}}\right)
    (\wtl^2 A_\lambda)^4 e^{-
    \frac{G_\gamma\lambda^3}{\fhatX^2}(\wtl^2A_\lambda)^2 (t-t_G)}
    \times \begin{cases}
     \vspace{0.25cm}
   \tRH^{1/2} t^{-2}  ~~~~ & t \lesssim \tRH \\ 
    \vspace{0.25cm}
   t^{-3/2}  ~~~~ & \tRH \lesssim t \lesssim \tMRE \\
     \tMRE^{1/2}t^{-2}  ~~~~ & t \gtrsim \tMRE~, 
     \end{cases}
    \label{eq:SourceTermRhoPhotExact}
\end{equation}
where we have defined $G_\gamma$ is defined below
Eq.~(\ref{eq:PartialWidthToPhotons}).  The second term takes 
the same form, but with one factor of $\lambda$ fewer in the summand.
 
In principle, one could evaluate this sum numerically at each moment in time,
and then use these results to numerically solve Eq.~(\ref{eq:MuCMBFinal}).
However, we find that by making a few additional well-motivated approximations, 
we can obtain a closed-form, analytical result for $d\mu_a/dt$.  
We begin by dividing the tower into sections, based on the two criteria which 
determine the dependence of $\Gamma_\lambda$ and $\rho_\lambda$ on $\lambda$.  
The first of these is whether the oscillation-onset time for a given 
$a_\lambda$ is within the staggered regime (\ie, $t_\lambda > t_G$), or
the simultaneous turn-on regime (\ie, $t_\lambda = t_G$).  In the former case,  
$t_\lambda$ depends on $\lambda$ according to Eq.~(\ref{eq:tlambdaInBothRegimes});
in the latter case, $t_\lambda$ is independent of $\lambda$.  
The second pertinent criterion concerns the relationship between $\lambda$ and the 
quantity
\begin{equation}
  \lambdatrans ~\equiv~ \pi\mX^2/M_c~,
\end{equation}
introduced in Ref.~\cite{DynamicalDM2}.  This quantity corresponds roughly to 
the transition point between the small-$\lambda$ regime, in which the $a_\lambda$ 
are highly mixed, the large-$\lambda$ regime, in which mixing is negligible.
Indeed, for $\lambda \ll \lambdatrans$, we find that 
$\wtl^2 A_\lambda \approx \sqrt{2}\,\wtl/(1 + \pi^2/y^2)^{1/2}$, while for
$\lambda \gg \lambdatrans$, we find that $\wtl^2 A_\lambda \approx \sqrt{2}$.  Given
these criteria,
our first approximation will be to replace $\wtl^2 A_\lambda$ with its asymptotic
large-$\lambda$ form for all $\lambda > \lambdatrans$, and with its asymptotic 
small-$\lambda$ form for all $\lambda < \lambdatrans$.  Our second will be to 
approximate the sum over $\lambda$ by a set of source-term integrals
$I_i(m,n,\alpha,\beta,\lambda_{\mathrm{min}},\lambda_{\mathrm{max}})$, each 
corresponding to a different regime in the tower of modes characterized by a 
particular dependence of the integrand on $\lambda$.  These 
source-term integrals may be evaluated analytically by making use of 
the identity
\begin{eqnarray}
  I_i(m,n,\alpha,\beta,\lambda_{\mathrm{min}},\lambda_{\mathrm{max}}) 
    &\equiv&
  \alpha\int_{\mathrm{\lambda_{min}}}^{\mathrm{\lambda_{max}}} \lambda^m
    e^{-\beta \lambda^n} d\lambda 
    \nonumber \\ &=&
    \alpha\frac{1}{n} \beta^{-(m+1)/n}\Bigg[
    \Gamma\left(\frac{m+1}{n},\beta \lambda_{\mathrm{min}}^n\right)-
    \Gamma\left(\frac{m+1}{n},\beta \lambda_{\mathrm{max}}^n\right)\Bigg]~,
  \label{eq:IntegralIdentity}
\end{eqnarray}
which is valid for $n>0$ and any real values of $m$, $\alpha$, and $\beta$.
Here $\Gamma(s,x)$ denotes the incomplete gamma function:
\begin{equation}
  \Gamma(s,x) ~\equiv~ \int^\infty_x t^{s-1} e^{-t} dt~. 
  \label{eq:DefOfIncompleteGammaFn}
\end{equation}

Employing the approximations discussed above, we find that the first 
source term on the right side of Eq.~(\ref{eq:MuCMBFinal}) reduces to
\begin{equation}
  \sum_\lambda\BRgamma\Gamma_\lambda \rho_\lambda ~=~ 
     \frac{2G_\gamma\theta^2}{M_c}
     \sum_{i=1}^4 
     I_i\big(m_i,n_i,\alpha_i,\beta_i,\lambda_{i-1}^{\mathrm{CMB}},
     \lambda_i^{\mathrm{CMB}}\big)
         \times \begin{cases}
     \vspace{0.25cm}
   \tRH^{1/2}t^{-2}  ~~ & t \lesssim \tRH \\ 
    \vspace{0.25cm}
    t^{-3/2}  ~~ & \tRH \lesssim t \lesssim \tMRE \\
     \tMRE^{1/2}t^{-2}  ~~ & t \gtrsim \tMRE~. 
     \end{cases}
   \label{eq:SourceTermForPhotons}
\end{equation}
Inserting this result (and the analogous result for the second source term) into Eq.~(\ref{eq:dmudtfromadecay}) and using the fact that
$\mu$-type distortions are generated by decays occurring within the RD era, we 
obtain the result 
\begin{equation}
  \frac{d\mu_a}{dt} ~\approx~ 0.935\times 
     \frac{G_\gamma\theta^2}{M_c t^{3/2}} 
     \Bigg[\frac{3}{\rho_\gamma^{\mathrm{eq}}}\sum_{i=1}^4
       I_i(m_i,n_i,\alpha_i,\beta_i,
       \lambda_{i-1}^{\mathrm{CMB}},\lambda_i^{\mathrm{CMB}})       
       - \frac{8}{n_\gamma^{\mathrm{eq}}}\sum_{i=1}^4
       I_i(m_i-1,n_i,\alpha_i,\beta_i,
       \lambda_{i-1}^{\mathrm{CMB}},\lambda_i^{\mathrm{CMB}})
     \Bigg]~, 
   \label{eq:dmuadtExplicit}
\end{equation}
where the expressions for $\alpha_i$, $\beta_i$, $m_i$, and $n_i$ valid in
each $a_\lambda$ regime are listed in Table~\ref{tab:mnbetaforCMB}.
Note that in obtaining this expression, we have assumed that the additional 
contributions to $n_\gamma$ and $\rho_\gamma$ due to the injection of 
photons from $a_\lambda$ are sufficiently small that these quantities 
can be approximated by the equilibrium expressions 
$n_\gamma^{\mathrm{eq}} = 2\zeta(3)T^3/\pi^2$ and 
$\rho_\gamma^{\mathrm{eq}} ~=~ \pi^2T^4/15$.  
Furthermore, we have used the fact that the time frame during which CMB 
distortions to $\mu$ can arise lies entirely within the RD era. 
Obtaining a final result for the magnitude of $\mu$-type 
distortions to the CMB engendered by the presence of a tower of decaying DDM 
axions is then simply a matter of substituting the result for $d\mu_a/dt$ in 
Eq.~(\ref{eq:dmuadtExplicit}) into Eq.~(\ref{eq:SolveDiffEqCMBmu}) and 
numerically evaluating the integral for a given choice of input parameters.

\begin{table}[t!]
\begin{center}
\begin{tabular}{|c|c|c|c|c|c|c|}\hline
~~~$i$~~~ & ~Oscillation regime~ &
~~Mixing regime~~ &  ~~$m_i$~~ & ~~~$n_i$~~~ &
  $\alpha_{i}$ & $\beta_{i}$ \\\hline 
1& \multirow{2}{*}{\rule[-0.25cm]{0cm}{0.85cm}
     ~~$t_\lambda>t_G$~~} 
  & $\lambda < \lambdatrans$ \rule[-0.25cm]{0cm}{0.7cm}  
     & 3   & 5 & $4\tRH^{-1/2}[1+\pi^2/y^2]^{-2}$ &        
  ~~~$2G_\gamma t(\fhatX\mX)^{-2}[1+\pi^2/y^2]^{-1}$~~~ \\ 
2&  & $\lambda \geq \lambdatrans$ \rule[-0.25cm]{0cm}{0.7cm}
     & $-1$   & 3 & $4\mX^4\tRH^{-1/2}$ & $2G_\gamma t\fhatX^{-2}$ \\\hline
3& \multirow{2}{*}{\rule[-0.25cm]{0cm}{0.85cm}
     ~~$t_\lambda = t_G$~~}
  & $\lambda < \lambdatrans$ \rule[-0.25cm]{0cm}{0.7cm}    
     &  5   & 5 & ~~$t_G^{\kappa_G}\tRH^{3/2-\kappa_G}[1+\pi^2/y^2]^{-2}$~~ &
  ~~~$2G_\gamma t(\fhatX\mX)^{-2}[1+\pi^2/y^2]^{-1}$~~~ \\ 
4&  & $\lambda \geq \lambdatrans$ \rule[-0.25cm]{0cm}{0.7cm}
     &  1   & 3 & $\mX^4t_G^{\kappa_G}\tRH^{3/2-\kappa_G}$ & $2G_\gamma t\fhatX^{-2}$ \\\hline
\end{tabular}
\caption{Values of $m_i$, $n_i$, $\alpha_i$, and 
$\beta_i$ which correspond to
different regimes, labeled by the index $i$, in a generic axion tower, 
for use in Eqs.~(\ref{eq:dmuadtExplicit}) and~(\ref{eq:PhotonSpectrumSum}).  
The symbol $\kappa_G$ denotes the specific value of $\kappa$, as defined in
Eq.~(\ref{eq:DefOfkappaForH}), which corresponds to $t_G$. 
\label{tab:mnbetaforCMB}}
\end{center}
\end{table}

The contribution to $y_C$ from the late decays of the $a_\lambda$ may be 
evaluated in much the same way as the corresponding contribution to $\mu$. 
The decays which contribute to $y_C$ are those which occur during the
window $9\times 10^9 \lesssim t \lesssim 1.2 \times 10^{13}$~s, during which the
rate $\Gamma_{\mathrm{EC}}\sim H$ associated with elastic Compton 
scattering can no longer bring the photons from $a_\lambda$ decay into
kinetic equilibrium even though radiation has yet to decouple from matter.
The evolution of $y_C$ is governed by the 
relation~\cite{DeZotti}
\begin{equation}
  \frac{dy_C}{dt} ~=~ \frac{1}{4\rho_\gamma}\frac{d\rho_\gamma}{dt}.
\end{equation} 
Proceeding with the mode sum as above and adopting the same approximations
as above, we find that
 \begin{equation}
  \frac{dy_C}{dt} ~\approx~
     \frac{2G_\gamma\theta^2}{M_c \rho_\gamma^{\mathrm{eq}}}
     \sum_{i=1}^4 I_i\big(m_i,n_i,\alpha_i,\beta_i,
     \lambda_{i-1}^{\mathrm{CMB}}, \lambda_{i}^{\mathrm{CMB}}\big)
     \times \begin{cases}
     \vspace{0.25cm} t^{-3/2}  ~~~~ & t \lesssim \tMRE \\
     \vspace{0.25cm} \tMRE^{1/2}t^{-2}  ~~~~ & \tMRE \lesssim t \lesssim \tLS \\
     0  ~~~~ & t \gtrsim \tLS~, 
     \end{cases}
   \label{eq:dyCompdtExplicit}
\end{equation}
where $\tLS \sim 1.19 \times 10^{13}$~s is the time of last scattering and 
$I_i(m,n,\alpha,\beta,\lambda_{\mathrm{min}},\lambda_{\mathrm{max}})$ is once 
again given by Eq.~(\ref{eq:IntegralIdentity}).  Note that since matter-radiation 
equality occurs prior to last scattering, at around $\tMRE \sim 10^{11}$~s, the 
epoch during which $a_\lambda$ decays can affect $y_C$ straddles both the RD and 
MD eras.  Numerically evaluating the expression in Eq.~(\ref{eq:dyCompdtExplicit}) 
from $t_{\mathrm{EC}}$ to $\tLS$, we obtain our final results for $y_C$ distortions 
due to late $a_\lambda$ decay.   

 \begin{figure}[ht!]
\begin{center}
  \epsfxsize 3.0 truein \epsfbox {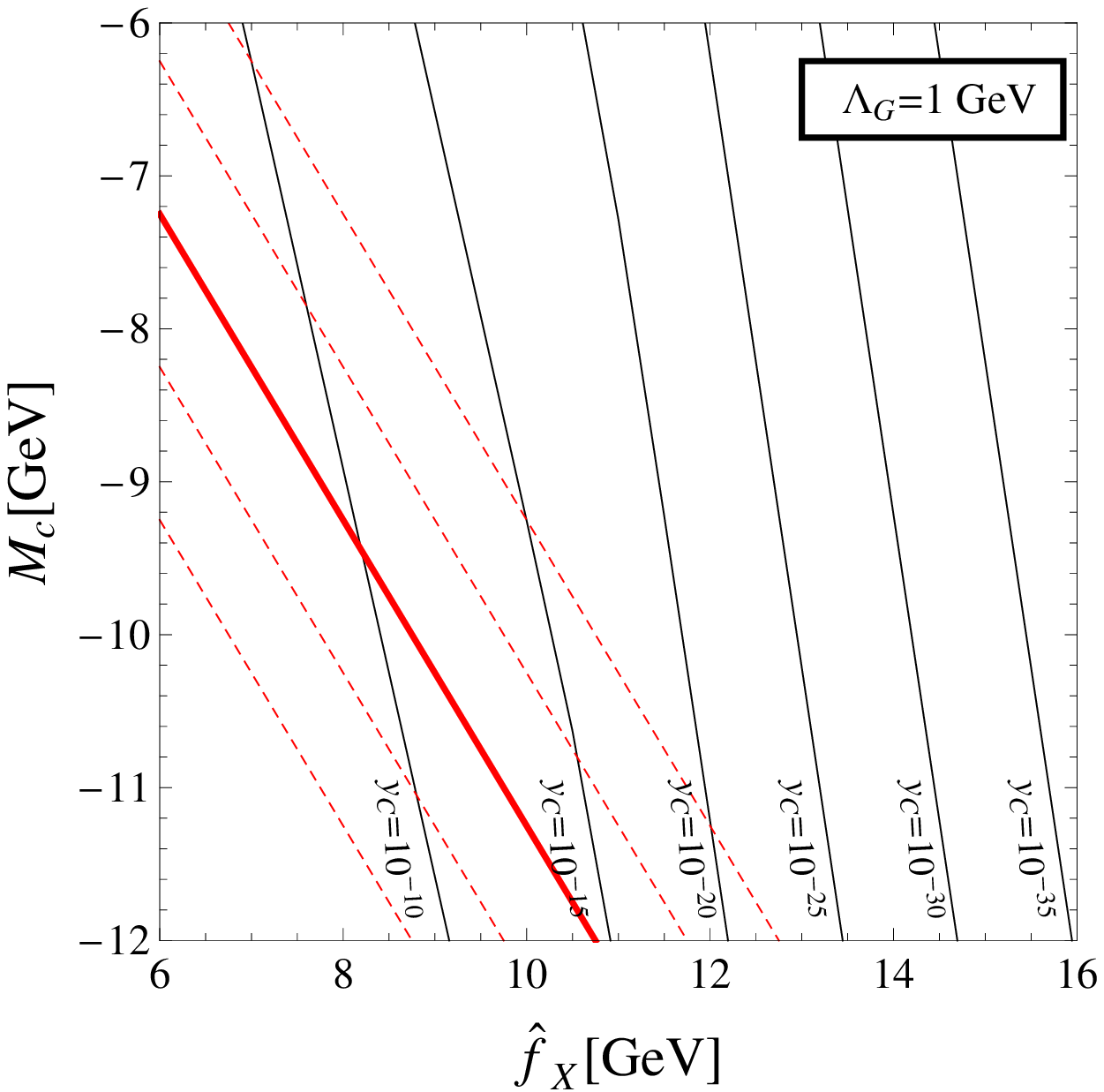} ~~~~
  \epsfxsize 3.0 truein \epsfbox {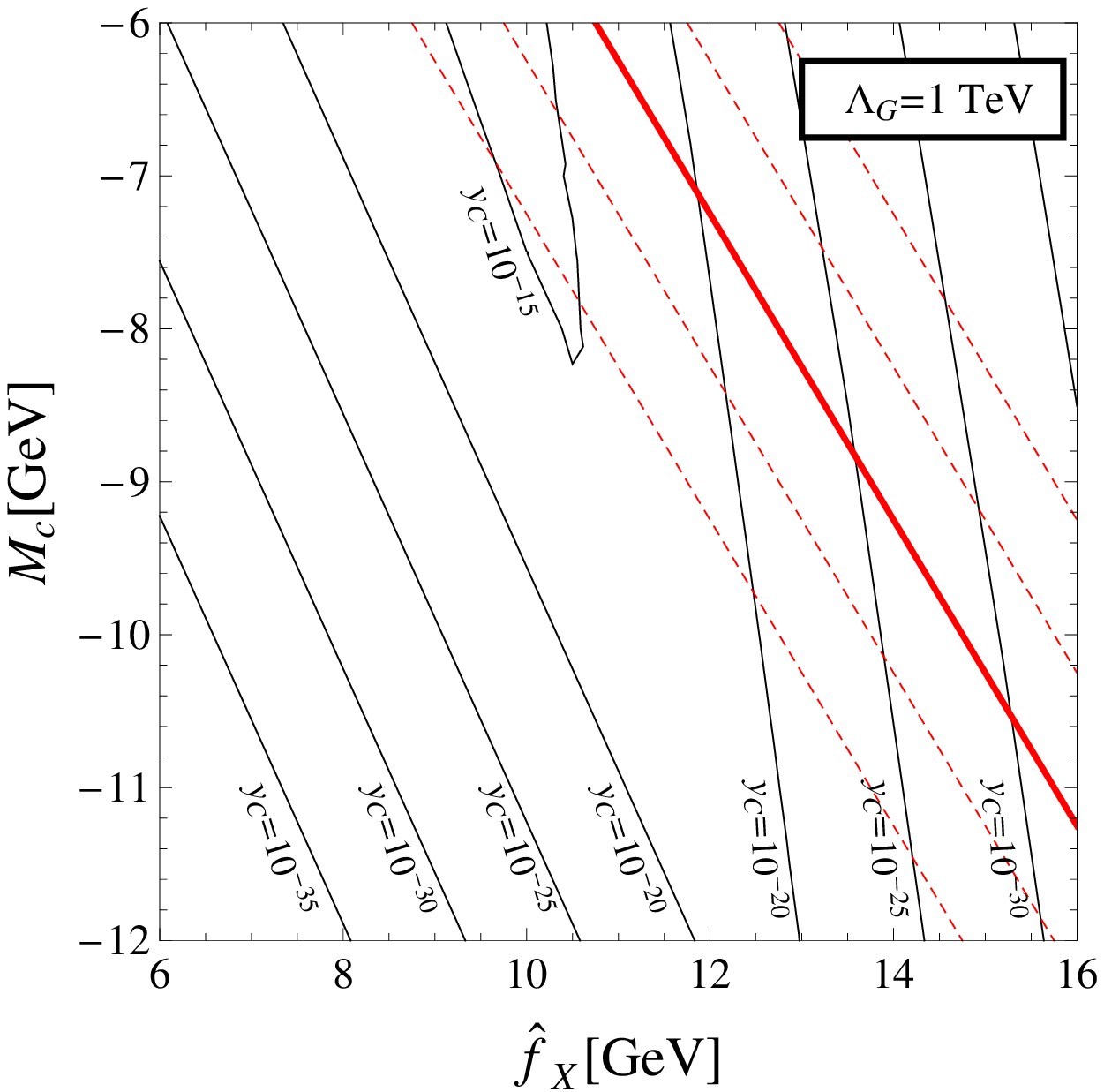} 
\end{center}
\caption{Contours of the CMB Compton-$y$-parameter distortion $y_C$ (black lines) produced 
as a result of axion decays in a bulk-axion DDM model with $\Lambda_G = 1$~GeV 
(left panel) and $\Lambda_G = 1$~TeV (right panel).  In each case, we have assumed a 
photonic axion with $c_\gamma=1$ and have taken
$\xi = g_G = \theta = 1$, with  $H_I = 1$~GeV and $\TRH=5$~MeV.  Contours corresponding
to $y\equiv M_c/\mX = 1$ (solid red line) and to $y = \{0.01,0.1,10,100\}$ (dashed red lines) 
are also shown.  For each panel, it is evident that our bulk-axion DDM model amply 
satisfies the CMB constraints in Eq.~(\protect\ref{eq:yCandmuBounds}) 
for all relevant values of $\fhatX$ and $M_c$, regardless of the value of $y$.
\label{fig:CMByContours}}
\end{figure}

In order to assess the CMB constraints on the parameter space of our bulk-axion 
DDM model, we now compare the results obtained by numerically integrating 
Eqs.~(\ref{eq:SolveDiffEqCMBmu}) and~(\ref{eq:dyCompdtExplicit}) with 
observational limits on $\mu$ and $y_C$.  The current limits 
on these quantities are~\cite{PDG}
\begin{equation}
  |\mu| < 9\times 10^{-5}~, ~~~~~~~~~
  y_C < 1.2 \times 10^{-5}~. 
  \label{eq:yCandmuBounds} 
\end{equation}
The bound on $y_C$ for a photonic axion with $c_\gamma = 1$ yields the 
constraints on $\fhatX$, $M_c$, and $\Lambda_G$ shown in Fig.~\ref{fig:CMByContours}.
In this figure, we display contours of the values of $y_C$ in $(\fhatX, M_c)$ space  
which arise in a bulk-axion model with $\Lambda_G = 1$~GeV (left panel) and with 
$\Lambda_G = 1$~TeV (right panel).  In each case, we have taken 
$\xi = g_G = \theta = 1$, $H_I = 1$~GeV, and $\TRH=5$~MeV.  Contours 
indicating $y \equiv M_c/\mX = 1$ (solid red line) and 
$y = \{0.01,0.1,10,100\}$ (dashed red lines) have also been superimposed. 
For each panel, it is evident that our bulk-axion DDM model amply 
satisfies the CMB constraints for all relevant values of $\fhatX$ and $M_c$, regardless 
of the value of $y$.  

It turns out that the constraints from the corresponding bound on $\mu$ in 
Eq.~(\ref{eq:yCandmuBounds}) are even less stringent than those from the bound 
on $y_C$.  Thus, we conclude that both the $y_C$-type and $\mu$-type distortions 
which result from $a_\lambda$ decays in our bulk-axion DDM model are well below 
present experimental sensitivities.  Indeed, no meaningful constraint  
arises for our bulk-axion DDM model from present limits on distortions in the CMB. 

As we have discussed, neither $\mu$ nor $y_C$ can be affected by
any photons which are produced by $a_\lambda$ decays at times $t \gtrsim \tLS$, 
after radiation and matter decouple.  Such photons do, however, contribute to 
the diffuse photon background.  In the next section, we will discuss the physical 
effects of this diffuse photon background in detail.


\subsection{Axion Decays and Contributions to the Diffuse X-Ray and Gamma-Ray 
Backgrounds\label{sec:XrayGammaRay}}


As mentioned above, the potentially observable effects of late photoproduction 
from axion decays include not only distortions of the CMB, but also imprints on
the diffuse X-ray and gamma-ray backgrounds.  Observational limits on such 
imprints from instruments such as HEAO~\cite{HEAOdiffXRB}, 
COMPTEL~\cite{COMPTELdiffXRB}, XMM, and Chandra~\cite{ChandraDeepFieldData} 
therefore impose additional constraints on the parameter space of our bulk-axion 
DDM model.    
As discussed in Sect.~\ref{sec:AbundanceConstraints}, there are two cosmological 
populations of decaying $a_\lambda$ whose decays to photons can potentially leave 
observable imprints on the diffuse X-ray and gamma-ray backgrounds.  The first is 
the population of cold axions produced by vacuum misalignment, which collectively 
compose the DDM ensemble.  The second is the far smaller
population of axions produced by interactions among the SM fields in the thermal 
bath after inflation.  While the former population provides a far 
greater contribution to $\Omegatot$, the latter population contains a far
larger proportion of heavier, more unstable $a_\lambda$, as indicated in 
Fig.~\ref{fig:OmegaCompFromThemal}.  It is not clear {\it a priori}\/
which population yields the more stringent constraint.  Thus, it is necessary to examine
the contribution to the diffuse photon background from each of these populations 
in turn.

A photon produced at time $t$ with initial energy $E_\gamma(t)$ will only 
contribute to the diffuse photon background if the universe remains transparent
to electromagnetic radiation over the entire range of energies through which 
that photon redshifts as the universe evolves from $t$ to $\tnow$.  A detailed 
analysis of the time scales and photon-energy ranges for which this transparency 
condition is attained is presented in Ref.~\cite{DMDecayChenKamionkowski1}.  
Roughly speaking, the transparency window spans an energy range 
$1 \mathrm{~keV} \lesssim E_\gamma \lesssim 10\mathrm{~TeV}$ and a time range 
$10^{12}-10^{14}\mathrm{~s}\lesssim t \lesssim \tnow$, with the lower 
limit depending on the particular value of $E_\gamma$.  Motivated by these results, 
we approximate the universe to be transparent to all photons with energies which fall 
within this range at all times $t>\tLS$ and opaque to all photons otherwise.  This 
approximation yields a conservative bound.  Moreover, we emphasize that since the 
dominant contribution to the diffuse X-ray and gamma-ray flux in our model
is due to modes which decay at much later times $t\gg \tLS$, our results are 
essentially insensitive to the precise contours chosen for the transparency window.    

The calculation of the photon flux due to late $a_\lambda$ decays proceeds in a 
manner similar to the calculation of the flux from KK-graviton decays outlined in
Ref.~\cite{CosmoConstraintsLargeED}.
The Boltzmann equation for the {\it number}\/ density $n_\gamma$ of photons 
in the presence of a tower of decaying $a_\lambda$ takes the form  
\begin{equation}
  \dot{n}_\gamma + 3 H n_\gamma ~=~  2\sum_\lambda 
      \BRgamma \Gamma_\lambda \frac{\rho_\lambda}{\lambda}~,
\end{equation}   
where once again $\rho_\lambda$ is given by Eq.~(\ref{eq:RhoLambdaInLTRCosmo}).
Solving this equation for $n_\gamma$ as a function of time, we obtain
\begin{equation}
  n_\gamma(t) ~=~  
      2 \frac{s(t)}{\sLS} 
      \sum_\lambda\BRgamma 
      \frac{\rho_\lambda(\tLS)}{\lambda}
      \left[1-e^{-\Gamma_\lambda(t-\tLS)}\right]~, 
  \label{eq:BoltzmannSolXRayPhotons}
\end{equation}
where $s(t)$ is the entropy density of the universe at time $t$, and $\sLS$ is the 
entropy density of the universe at the time of last scattering.
The present-day differential energy spectrum $dn_\gamma/dE_\gamma$ of these 
photons may readily be computed from the relation  
\begin{equation}
  \frac{dn_\gamma}{dE_\gamma} ~=~ \frac{dn_\gamma}{dt}\frac{dt}{dz}
     \frac{dz}{dE_\gamma}~,
  \label{eq:dndEgammaDifferentials}
\end{equation}
where $z$ is the cosmological redshift and $E_\gamma$ is the photon energy at 
redshift $z$.  The first of these factors may be
obtained by explicitly differentiating Eq.~(\ref{eq:BoltzmannSolXRayPhotons}) 
with fixed $s=\snow$, which yields a series of terms of the form  
\begin{equation}
  \left[\frac{dn_\gamma}{dt}\right]_\lambda ~=~  
      2 \left(\frac{\snow}{\sLS}\right) 
      \BRgamma \Gamma_\lambda 
      \frac{\rho_\lambda(\tLS)}{\lambda}
      e^{-\Gamma_\lambda(\tnow-\tLS)}~, 
  \label{eq:dndtGeneralForm}
\end{equation} 
one for each different value of $\lambda$.
The second factor in Eq.~(\ref{eq:dndEgammaDifferentials}) may be obtained by  
noting that the relationship between time and redshift during the present,
matter-dominated era is well-approximated by $t ~=~  \tnow(1+z)^{-3/2}$.
Consequently, for each value of $\lambda$ we have 
\begin{equation}
  \left[\frac{dt}{dz}\right]_\lambda ~=~ 
    -\frac{3}{2}\tnow\left(\frac{2E_\gamma}{\lambda}\right)^{5/2}
\end{equation}
during the epoch of interest.  The third factor in 
Eq.~(\ref{eq:dndEgammaDifferentials}) may be obtained by noting
that each of the photons produced by an axion tower state 
$a_\lambda$ which decays at redshift $z$ will be monochromatic, with 
energy $\lambda/2$, at the moment of decay.  This implies that the present-day 
energies of such photons are given by $E_\gamma(1+z) = \lambda/2$, and hence that
for each value of $\lambda$, we have 
\begin{equation}
  \left[\frac{dz}{dE_\gamma}\right]_\lambda ~=~ 
    -\frac{\lambda}{2E_\gamma^2}~.
\end{equation}
Combining these expressions and summing over $\lambda$, we arrive at a general
formula for the contribution to the diffuse photon flux produced by the tower
of decaying $a_\lambda$:
\begin{equation}
  \left.\frac{dn_\gamma}{dE_\gamma}\right|_{\mathrm{now}} ~=~  
      6\tnow\sqrt{2E_\gamma} \left(\frac{\snow}{\sLS}\right) 
      \sum_\lambda \BRgamma \Gamma_\lambda 
      \frac{\rho_\lambda(\tLS)}{\lambda^{5/2}}
      e^{-\Gamma_\lambda(\tnow-\tLS)}~. 
  \label{eq:dndEgammaGeneralForm}
\end{equation}   
Calculating the contribution to the diffuse X-ray and gamma-ray backgrounds
in our bulk-axion DDM model is then simply a matter of  
applying Eq.~(\ref{eq:dndEgammaGeneralForm}) to the contribution from 
the two relevant populations of decaying axions discussed above.
 
We begin by addressing the contribution from the population of axions 
produced by vacuum misalignment --- \ie, the DDM ensemble itself.
Once again, we focus our attention on the case of a photonic axion, 
for which $\Gamma_\lambda \approx \Gamma(a\rightarrow\gamma\gamma)$ and
$\BRgamma \approx 1$.  In this 
case, we find that the contribution to the present-day diffuse 
photon background from the collective decays of the $a_\lambda$ fields is given by     
\begin{equation}
   \left.\frac{dn_\gamma}{dE_\gamma}\right|_{\mathrm{now}} ~=~ 
   3 \sqrt{2E_\gamma} G_\gamma\theta^2 \mX^4
   \left(\frac{\snow}{\sLS}\right) 
   \sum_\lambda 
   \left(\frac{t_\lambda^{2}\tMRE^{1/2}}{\tLS^{2}\tRH^{1/2}}\right)
   \lambda^{-3/2}(\wtl^2A_\lambda)^4
     e^{\frac{G_\gamma \lambda^3}{\hat{f}_G^2}
     (\wtl^2A_\lambda)^2 (\tnow-t_G)}~. 
   \label{eq:PhotonSpectrumSum}
\end{equation}   
Just as for the contributions to $\mu$ and $y_C$ in Sect.~\ref{sec:CMB}, 
we approximate the sum over axion modes appearing in Eq.~(\ref{eq:PhotonSpectrumSum}) 
as an integral over $\lambda$.  
The lower limit of integration is determined by the requirement that in order
for a photon with redshifted energy $E_\gamma$ to have been produced by 
the decay of the axion species $a_\lambda$ before present day, we must have
$\lambda \geq 2E_\gamma$.  Likewise, photons which decay before 
the processes which equilibrate them with the radiation bath freeze out
will not contribute to features in the diffuse photon background.  Thus, the upper
limit of integration is set by the condition 
$\lambda \lesssim 2E_\gamma (\tnow/\tLS)^{2/3}$.  Furthermore, we must also require 
that $\lambda$ not exceed the cutoff scale $f_G$, or be smaller than the 
lightest mode in the tower.  Once again, we find that the resulting integral 
expressions can be written in terms of the functions
$I_i(m,n,\alpha,\beta,\lambda_{\mathrm{min}},\lambda_{\mathrm{max}})$
defined in Eq.~(\ref{eq:IntegralIdentity}):
\begin{equation}
   \left.\frac{dn_\gamma}{dE_\gamma}\right|_{\mathrm{now}} ~\approx~ 
   12G_\gamma\theta^2  \frac{\sqrt{2E_\gamma}\tnow}{M_c}
   \left(\frac{\snow}{\sLS}\right)
   \left(\frac{\tMRE^{1/2}}{\tLS^{2}}\right)
   \sum_{i=1}^4 
   I_i\big(m_i-5/2,n_i,\alpha_i,\beta_i,\lambda_{i-1}^{\mathrm{XRB}},
   \lambda_i^{\mathrm{XRB}}\big)~,
   \label{eq:FinalXRBFormula}
\end{equation}
where the $\lambda_i^{\mathrm{XRB}}$ are analogous to the 
$\lambda_i^{\mathrm{CMB}}$ appearing in Eq.~(\ref{eq:dmuadtExplicit}).  
Determining the net contribution to the differential photon flux from
decays of the $a_\lambda$ for any particular choice of model parameters
is thus simply a matter of numerically evaluating Eq.~(\ref{eq:FinalXRBFormula}). 

We now turn to consider the observational limits on $dn_\gamma/dE_\gamma$.
The diffuse extragalactic X-ray and gamma-ray background spectra have
been probed by a number of experiments.  In the keV $-$ MeV region, the 
most current data are those from HEAO, COMPTEL, XMM, and Chandra; at
energies above this, the most current data are those from EGRET and FERMI.
Over this entire energy range, the diffuse photon spectrum is well-modeled
by a set of power-law fits, and the non-observation of any discernible, 
sharp features in this spectrum imposes constraints on late relic-particle 
decays to photons.   
For the data from the COMPTEL instrument, the best power-law fit is found to
be~\cite{COMPTELdiffXRB} 
\begin{equation}
  \frac{dn_\gamma}{dE_\gamma} ~ = ~ 
    10.5 \times 10^{-4} \left(\frac{E_\gamma}{5\mbox{~MeV}}\right)^{-2.4} 
    \mbox{~MeV}^{-1}\mbox{cm}^{-1}\mbox{s}^{-1}\mbox{str}^{-1}
    ~~~~~~~~ 800\mbox{~keV} \lesssim E_\gamma \lesssim 30\mbox{~MeV}~,
  \label{eq:COMPTELdiffBG}
\end{equation} 
while the best fit to the HEAO data is found to be~\cite{HEAOdiffXRB}
\begin{equation}
   \frac{dn_\gamma}{dE_\gamma} ~ = ~ \begin{cases}
     \displaystyle
     7.88\times 10^3 \left(\frac{E_\gamma}{\mathrm{~keV}}\right)^{-1.29}
       e^{-(E_\gamma/41.13\mathrm{~keV})}
       \mbox{~MeV}^{-1}\mbox{cm}^{-1}\mbox{s}^{-1}\mbox{str}^{-1}\vspace{0.25cm} 
       &~~~~~~~~  0.1\mbox{~keV}\lesssim E_\gamma \lesssim 60\mbox{~keV} \\
     \displaystyle
     0.43\left(\frac{E_\gamma}{\mathrm{60~keV}}\right)^{-6.5} +\vspace{0.25cm}
     8.4\left(\frac{E_\gamma}{\mathrm{60~keV}}\right)^{-2.58}~~~~~~~~~~~~~~~~~~ \\
     \displaystyle ~~~~~~~~~~ +~
     0.38\left(\frac{E_\gamma}{\mathrm{60~keV}}\right)^{-2.05}
       \mbox{~MeV}^{-1}\mbox{cm}^{-1}\mbox{s}^{-1}\mbox{str}^{-1}
       ~~~~~~~\vspace{0.25cm}
       &~~~~~~~~  60\mbox{~keV}\lesssim E_\gamma \lesssim 160\mbox{~keV} \\
    \displaystyle
    3.8\times 10^5\times \left(\frac{E_\gamma}{\mathrm{keV}}\right)^{-2.6} 
       \mbox{~MeV}^{-1}\mbox{cm}^{-1}\mbox{s}^{-1}\mbox{str}^{-1}
       \vspace{0.25cm}~~~~~~~
       &~~~~~~~~  160\mbox{~keV}\lesssim E_\gamma \lesssim 350\mbox{~keV}\\
   \displaystyle
   2.0\times 10^3 \left(\frac{E_\gamma}{\mathrm{keV}}\right)^{-1.7} 
     \mbox{~MeV}^{-1}\mbox{cm}^{-1}\mbox{s}^{-1}\mbox{str}^{-1}\vspace{0.25cm}  
     &~~~~~~~~ 350\mbox{~keV} \lesssim E_\gamma \lesssim 2\mbox{~MeV}~.
   \end{cases}
\end{equation}
The Chandra satellite has improved upon these diffuse X-ray background 
constraints in the $ 1\mbox{~keV}\lesssim E_\gamma \lesssim 8\mbox{~keV}$ range 
by resolving a large fraction ($\sim 80$\%) of this background into point sources.  
The residual spectrum in this region is well represented by the
power law~\cite{ChandraPowerLawFitHickox}
\begin{equation}
   \frac{dn_\gamma}{dE_\gamma} ~ = ~
      2.6 \times 10^3 \left(\frac{E_\gamma}{\mbox{keV}}\right)^{-1.5} 
      \mbox{~MeV}^{-1}\mbox{cm}^{-1}\mbox{s}^{-1}\mbox{str}^{-1}
      ~~~~~~~~~~~~~ 1\mbox{~keV} \lesssim E_\gamma \lesssim 8\mbox{~keV}~. 
\end{equation}
In the gamma-ray region, the most stringent current limits are those from
EGRET and FERMI.  Data on the diffuse extragalactic gamma-ray background 
from the former instrument~\cite{EGRETdiffGRB} are reliable 
for photon energies within the range 
$1.41\mbox{~GeV}\lesssim E_\gamma \lesssim 30\mbox{~MeV}$, for which we 
find the best fit
\begin{equation}
   \frac{dn_\gamma}{dE_\gamma} ~ = ~
      7.35\times 10^{-3}\left(\frac{E_\gamma}{\mathrm{MeV}}\right)^{-2.35} 
      \mbox{~MeV}^{-1}\mbox{cm}^{-1}\mbox{s}^{-1}\mbox{str}^{-1}
      ~~~~~~~~~~  30\mbox{~MeV}\lesssim E_\gamma \lesssim 1.41\mbox{~GeV}~.
   \label{eq:EGRETdiffBG}
\end{equation}
Note that data exist for
higher photon energies as well, but given that EGRET's energy resolution
is not as good at such high energies, and given that these data
have been superseded by data from FERMI, we do not use them in computing 
this power-law fit.  As for the FERMI data, they are well modeled by the 
power law~\cite{FERMIdiffGRB}
\begin{equation}
   \frac{dn_\gamma}{dE_\gamma} ~ = ~
     9.59\times 10^{-3} \left(\frac{E_\gamma}{\mathrm{MeV}}\right)^{-2.41} 
     \mbox{~MeV}^{-1}\mbox{cm}^{-1}\mbox{s}^{-1}\mbox{str}^{-1}
     ~~~~~~~~~~  274\mbox{~MeV}\lesssim E_\gamma \lesssim  70.7\mbox{~GeV}~.
    \label{eq:FERMIdiffBG}
\end{equation}  

\begin{figure}[b!]
\begin{center}
  \epsfxsize 3.0 truein \epsfbox {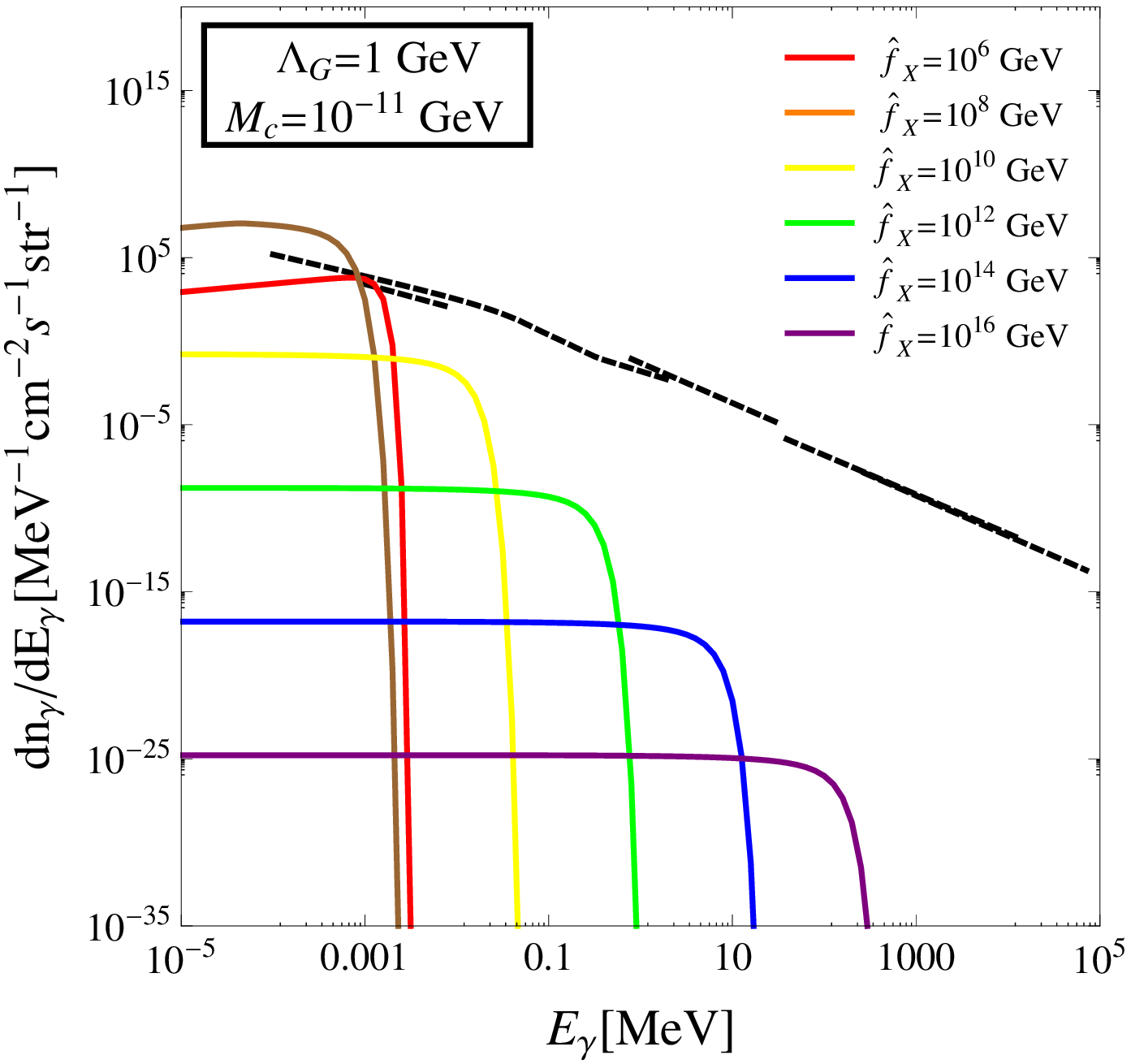} ~~~~
  \epsfxsize 3.0 truein \epsfbox {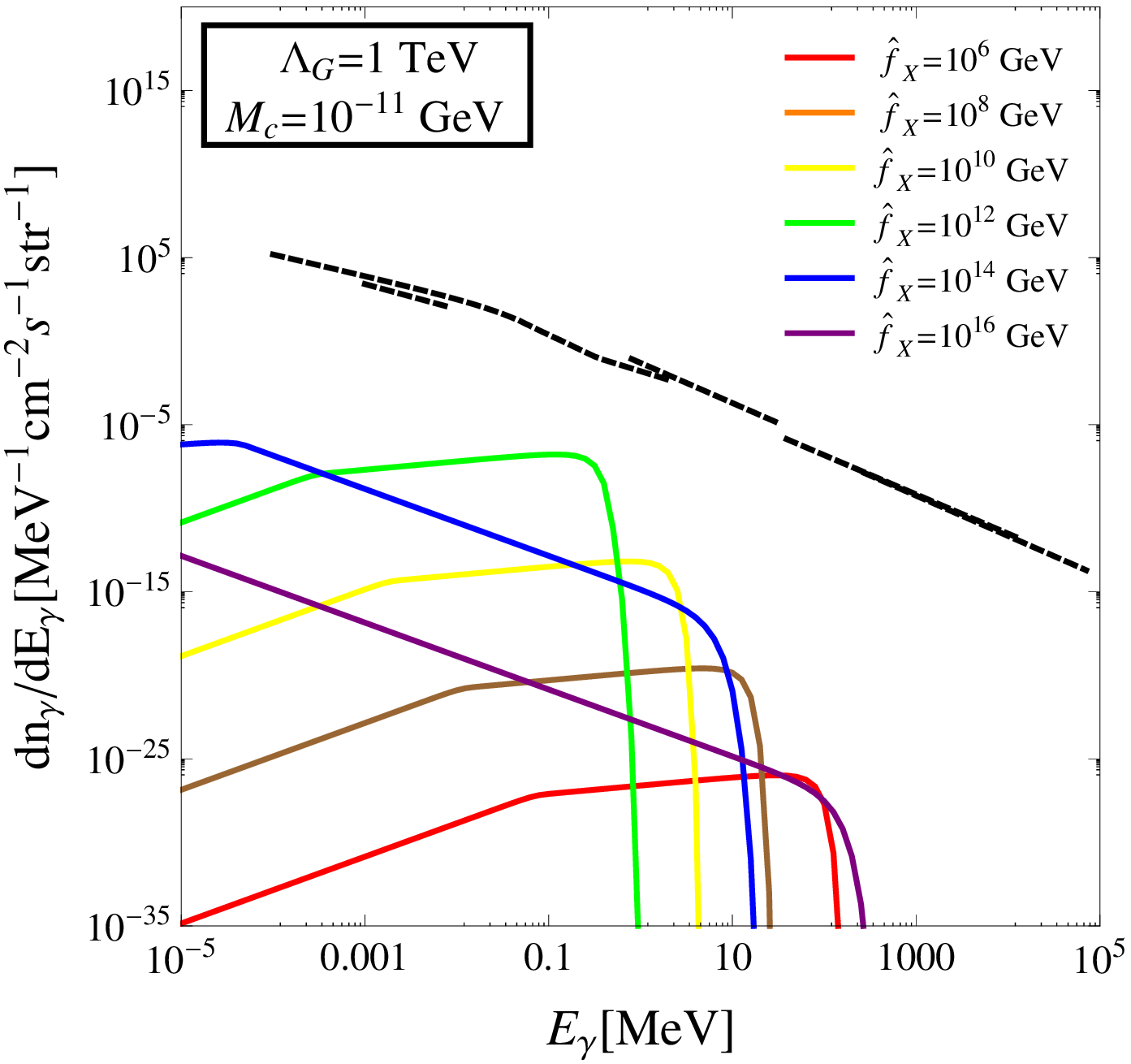} \\ 
  \epsfxsize 3.0 truein \epsfbox {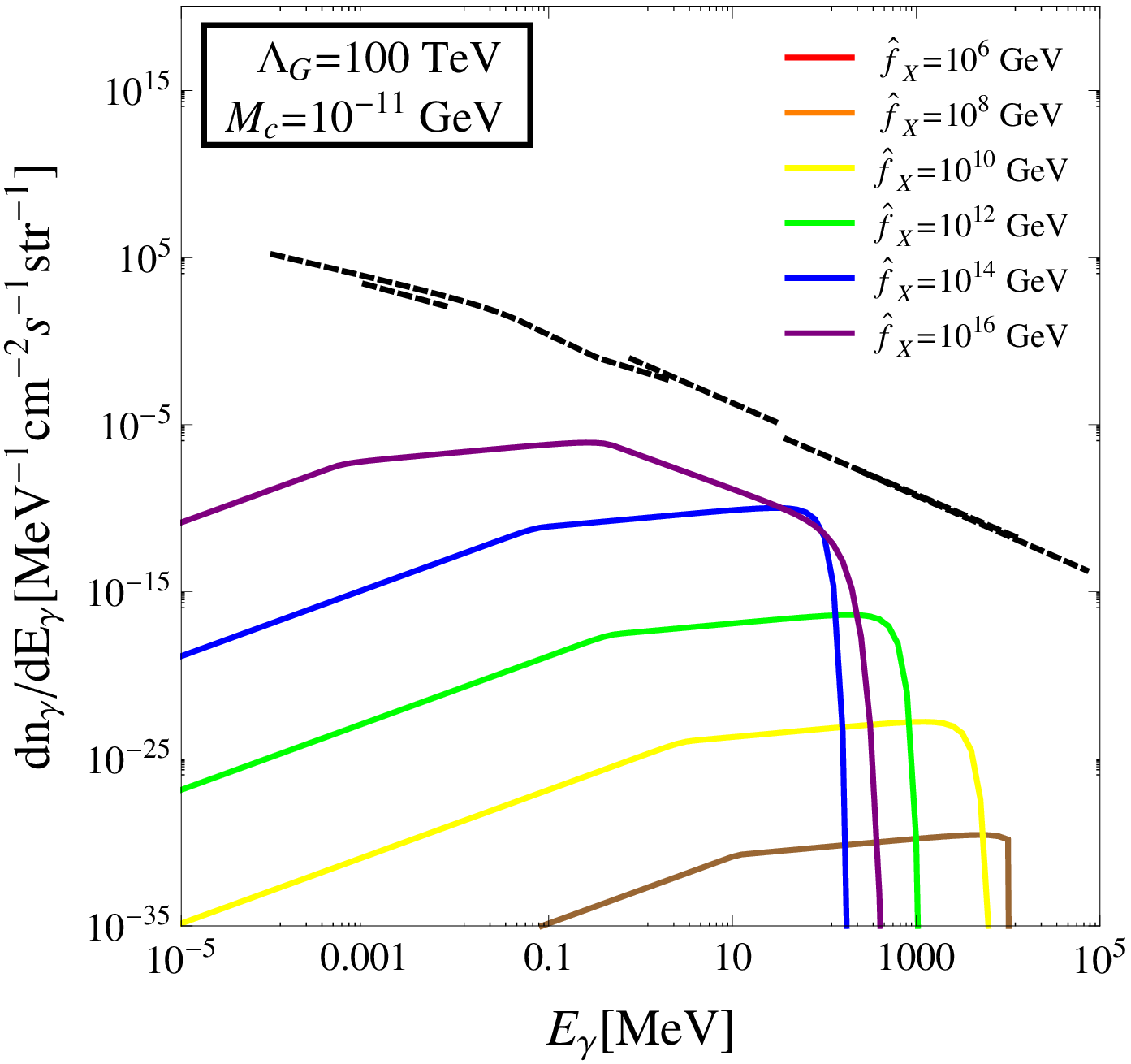}
\end{center}
\caption{The diffuse photon-flux spectrum 
$dn_\gamma/dE_\gamma$ produced from axion decays in our bulk-axion DDM model
with $\Lambda_G = 1$~GeV (upper left panel), $\Lambda_G = 1$~TeV (upper right panel), 
and $\Lambda_G = 100$~TeV (lower panel).
Each solid colored curve corresponds to a different choice of $\fhatX$, within the range
$10^{6} - 10^{16}$~GeV.  In all panels, we have taken 
$M_c = 10^{-11}$~GeV, $H_I = 1$~GeV, $\TRH=5$~MeV, and $\xi = g_G = \theta = 1$.  
By contrast, the dashed black contours represent the upper
bounds on $dn_\gamma/dE_\gamma$ derived from observational limits on the diffuse photon
flux using a number of instruments sensitive in the X-ray and gamma-ray regions.
As evident from these plots, the diffuse-photon-background contribution arising from 
axion decay in our bulk-axion DDM model is consistent with all observational 
limits when $\Lambda_G$ is large.    
\label{fig:XRayGammaRayLimit}}
\end{figure}

In Fig.~\ref{fig:XRayGammaRayLimit}, we show a set of curves (solid colored lines) 
depicting the total contribution to the diffuse gamma-ray background from the decaying 
$a_\lambda$ fields, as given in Eq.~(\ref{eq:FinalXRBFormula}), for several different 
values of $\fhatX$ within the range 
$10^{6} - 10^{16}$~GeV.  Results are shown for $\Lambda_G = 1$~GeV (upper left panel), 
$\Lambda_G = 1$~TeV (upper right panel), and $\Lambda_G = 100$~TeV (lower panel).
In each case, we have taken $M_c = 10^{-11}$~GeV, $\TRH = 5$~MeV, and  
$\xi = g_G = \theta = 1$.  In addition, we have chosen a value for $H_I$ 
sufficiently large that none of the curves
shown is significantly affected by the ``inflating away'' of heavy modes which
begin oscillating before inflation ends.  In addition to these curves, we 
also display contours corresponding to the upper limits on the diffuse X-ray 
and gamma-ray fluxes (black dashed lines) given in 
Eqs.~(\ref{eq:COMPTELdiffBG}) through~(\ref{eq:FERMIdiffBG}).  Any choice of model
parameters for which the differential photon flux $dn_\gamma/dE_\gamma$ exceeds
any one of these observational-limit contours for any value of $E_\gamma$ is excluded.
The results shown in Fig.~\ref{fig:XRayGammaRayLimit} indicate that while it is 
not trivial to satisfy these observational limits in our bulk-axion DDM model, 
the contributions to the diffuse X-ray and gamma-ray fluxes from $a_\lambda$ decay 
are indeed sufficiently small that these limits are satisfied when 
$\Lambda_G$ is large.   

\begin{figure}[h!]
\begin{center}
  \epsfxsize 3.0 truein \epsfbox {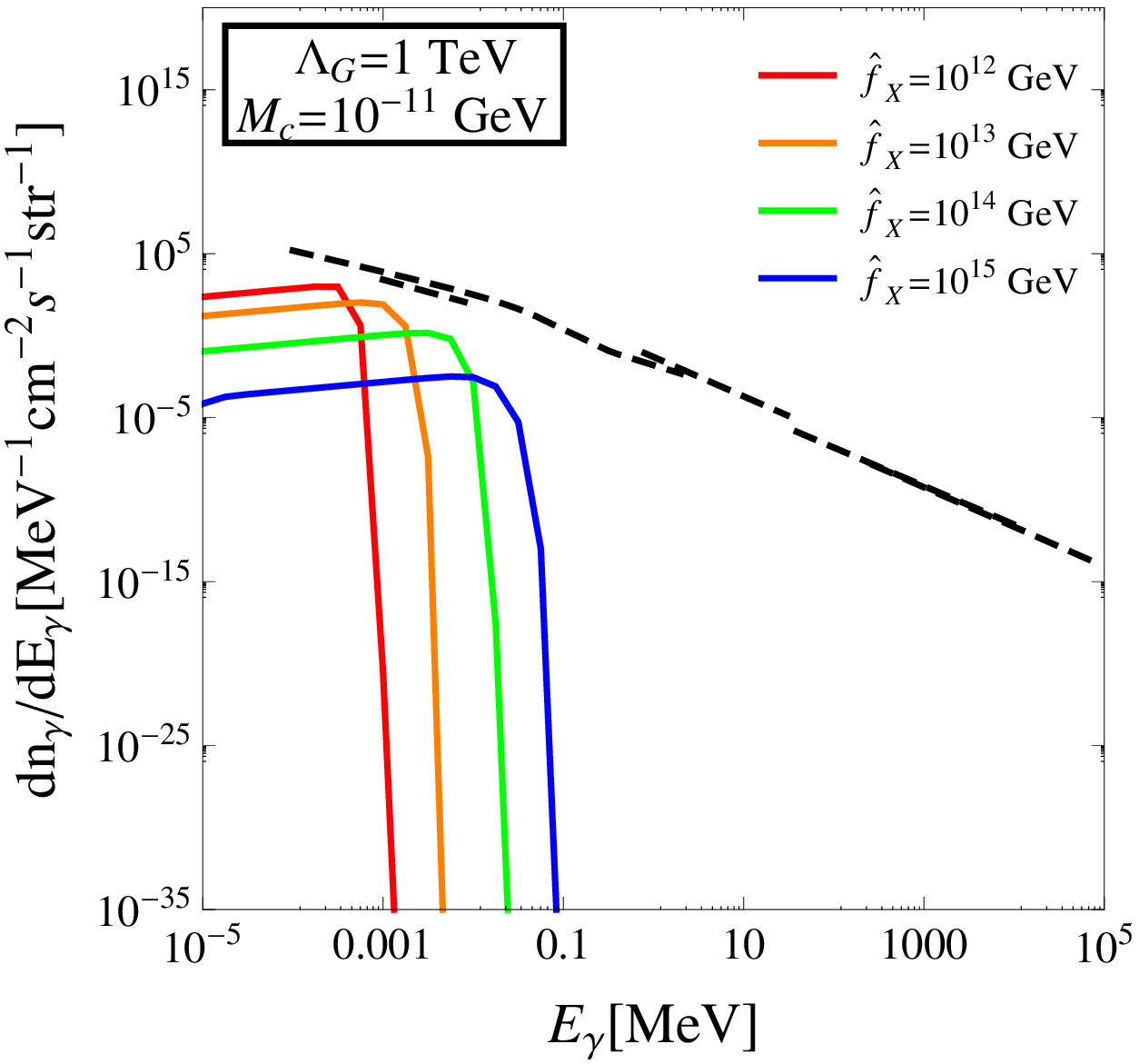} ~~~~
  \epsfxsize 3.0 truein \epsfbox {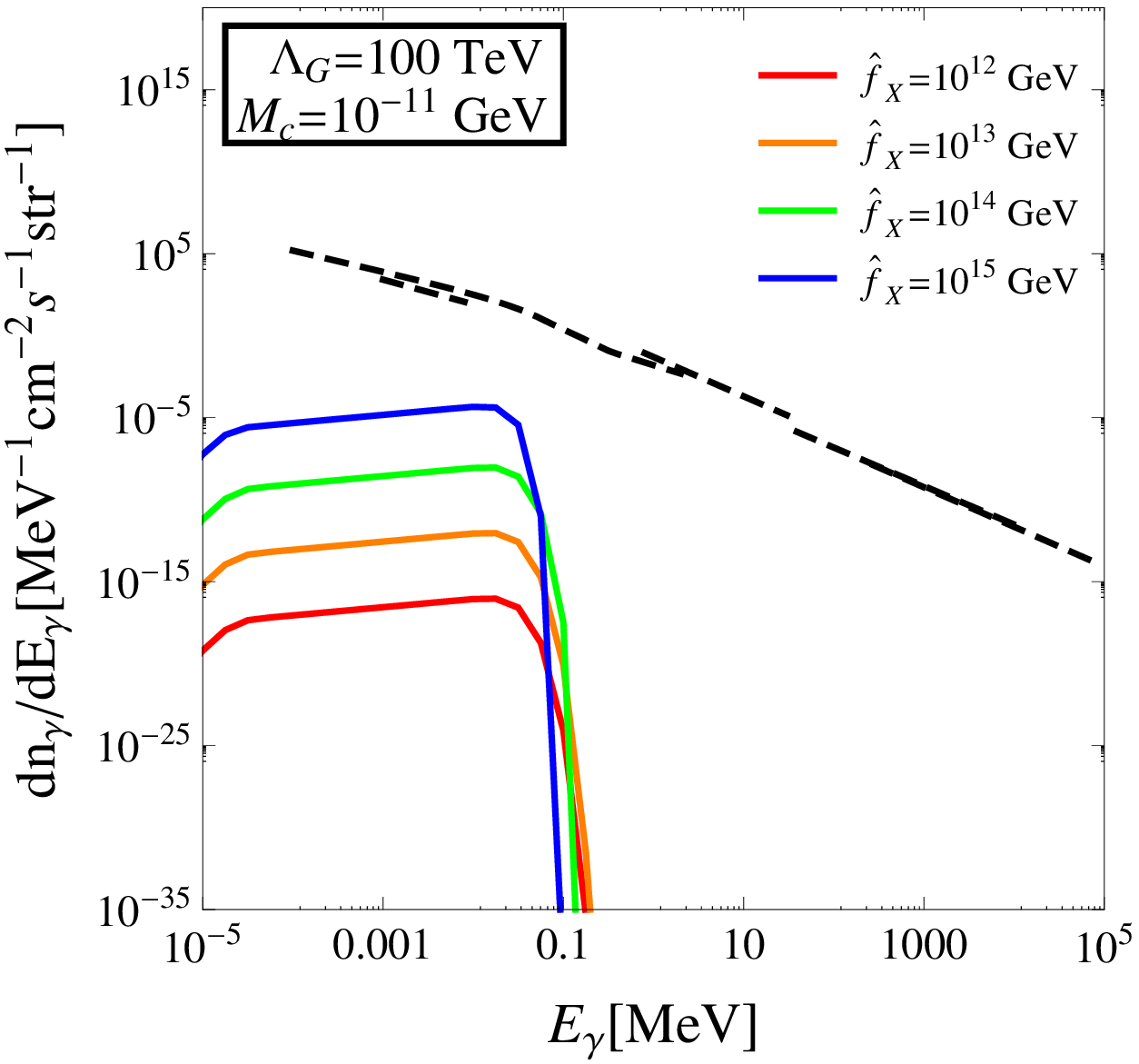} 
\end{center}
\caption{The diffuse photon flux spectrum 
$dn_\gamma/dE_\gamma$ produced from the decays of a 
population of $a_\lambda$ produced by interactions among the SM
particles in the thermal bath after inflation.  The left 
panel shows the results for $\Lambda_G = 1$~TeV, while the right
panel shows the results for $\Lambda_G = 100$~TeV.  In each case, 
we have taken $M_c = 10^{-11}$~GeV, $\TRH=5$~MeV, and $\xi = g_G = 1$.
The solid colored curves indicate the diffuse-photon-flux contributions 
corresponding to different choices of $\fhatX$.
As in Fig.~\ref{fig:XRayGammaRayLimit}, the dashed black contours indicate the 
upper bounds on $dn_\gamma/dE_\gamma$ derived from observational limits on the 
diffuse X-ray and gamma-ray fluxes, and we see that our model is consistent
with these bounds. 
\label{fig:XRayGammaRayLimitTherm}}
\end{figure}

We now consider the contribution to $dn_\gamma/dE_\gamma$ from the population
of axions generated by their interactions with SM fields in the thermal bath
after inflation.    
The contribution to the diffuse photon flux spectrum $dn_\gamma/dE_\gamma$ 
generated by such a population of axions is once again given by  
Eq.~(\ref{eq:dndEgammaGeneralForm}), but with $\rho_\lambda(\tLS)$ in 
Eq.~(\ref{eq:RhoLambdaInLTRCosmo}) now replaced by
\begin{equation}
  \rho_\lambda(\tLS) ~\approx~ \lambda\TLS^3\tLS
     \int_{\TMRE}^{\TRH}  
  \frac{3}{\kappa(T)}\left(\frac{\TLS}{T}\right)^{3/\kappa(T)}
  \frac{g_{\ast s}(\TLS)}{g_{\ast s}(T)}
  \left[C_\lambda^{\mathrm{ID}}(T) + C_\lambda^{\mathrm{Prim}}(T)
     e^{-(\lambda+m_e)/T}\right]dT~,
  \label{eq:RhoThermMRE}
\end{equation}
as follows from Eq.~(\ref{eq:OmegaThermNow}).
To derive an estimate for the expected contribution to 
$dn_\gamma/dE_\gamma$ from the resulting equation, 
we proceed in essentially the same way as we did in calculating 
the contribution from axions produced via vacuum misalignment.  
The results of this calculation are shown in 
Fig.~\ref{fig:XRayGammaRayLimitTherm} for parameter values
within or near the preferred region of parameter space for our 
bulk-axion DDM model.  Specifically, we have taken $M_c = 10^{-11}$~GeV, 
$\Lambda_G = 1$~TeV, $\TRH=5$~MeV, and $\xi = g_G = 1$.  
The solid colored curves shown correspond to several different choices of 
$\fhatX$ ranging from $\fhatX = 10^{12}$~GeV to $\fhatX= 10^{15}$~GeV.  Once 
again, the dashed black lines indicate the observational limits on additional 
contributions to $dn_\gamma/dE_\gamma$.  It is clear from
Fig.~\ref{fig:XRayGammaRayLimitTherm} that while the contribution 
to the diffuse X-ray flux from thermal axions within our preferred 
region of parameter space is certainly not negligible, it is also
consistent with current observational limits.  We therefore conclude that even 
after the contribution from thermal axions is included, our bulk-axion DDM 
model is consistent with X-ray and gamma-ray data.
 
\subsection{Axion Decays and Big-Bang Nucleosynthesis\label{sec:BBN}}


The accord between the primordial abundances of light nuclei inferred from 
observation and the predictions for those abundances within the framework of 
standard BBN has been one of the greatest triumphs of theoretical cosmology.
However, these predictions depend sensitively on the cosmological parameters 
during the nucleosynthesis epoch.
For example, the presence of additional relativistic degrees of freedom in the 
thermal bath during BBN can substantially distort the abundances of the light
elements away from their observed values.  In addition, the decays of unstable 
particles during or after the BBN epoch can also alter these abundances via the
injection of both entropy and energy into the thermal bath.  We must therefore 
ensure that the collective effects of $a_\lambda$ decays in our model 
are sufficiently small so as not to disrupt the successful generation of
light-element abundances via standard BBN.        

Limits on the abundance of a single unstable relic particle $\chi$ from
BBN are typically phrased as bounds on the number density $\hat{n}^\ast_\chi$ 
that $\chi$ would have at present time if it 
were absolutely stable.  In general, the BBN bound on $\hat{n}^\ast_\chi$ for
any given relic particle depends on the
lifetime $\tau_\chi$ of that particle.  The most stringent 
limits are obtained for lifetimes 
$\tau_\chi \sim\mathcal{O}( 10^{9} - 10^{10}$~s),
for which the corresponding constraint is 
roughly~\cite{CyburtEllisBBN,KawasakiHadronicBBN} 
\begin{equation}
  m_\chi\frac{\hat{n}^\ast_\chi}{n_\gamma^\ast}~\lesssim~ 10^{-13}\mbox{~GeV}~,
\end{equation}       
where $n_\gamma^\ast\approx 410.5\mbox{~cm}^{-3}$ denotes the present-day number 
density of photons.  This limit can also be written in the form     
\begin{equation}
  \Omegachinodec ~\lesssim~ 1.7 \times 10^{-5}~,
  \label{eq:OmegathreshChiBBNConstraint}
\end{equation}   
where $\Omegachinodec$ is the relic abundance that $\chi$ would have at present 
time if it were absolutely stable.  

Once again, however, as with other constraints on traditional models of 
decaying dark matter (such as those from the CMB and the diffuse X-ray and 
gamma-ray backgrounds), these constraints are not readily applicable to models 
within the DDM paradigm, since the dark-matter candidate in these models is an 
ensemble with no single, well-defined mass or lifetime.  Thus, we must 
reexamine the derivation of the BBN constraints on decaying relic particles 
in order to determine what restrictions these considerations place 
on the parameter space of our bulk-axion DDM model.  While a detailed calculation
of the precise limits BBN considerations impose on DDM scenarios in general 
is beyond the scope of this paper, it is straightforward to demonstrate that BBN 
constraints do not significantly restrict the parameter space of the particular 
model which concerns us here.  

\begin{figure}[b!]
\begin{center}
  \epsfxsize 2.25 truein \epsfbox {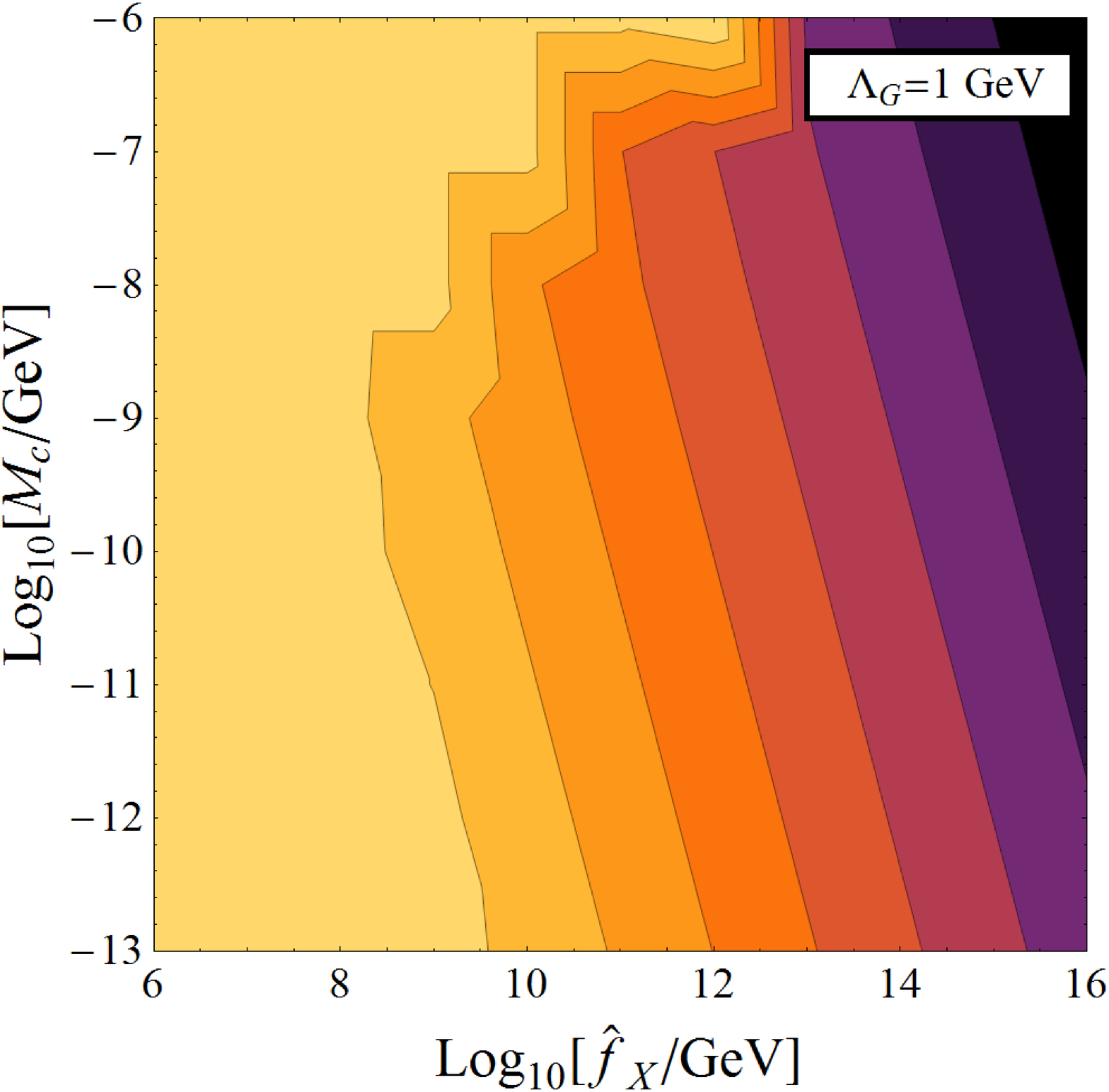} 
  \epsfxsize 2.25 truein \epsfbox {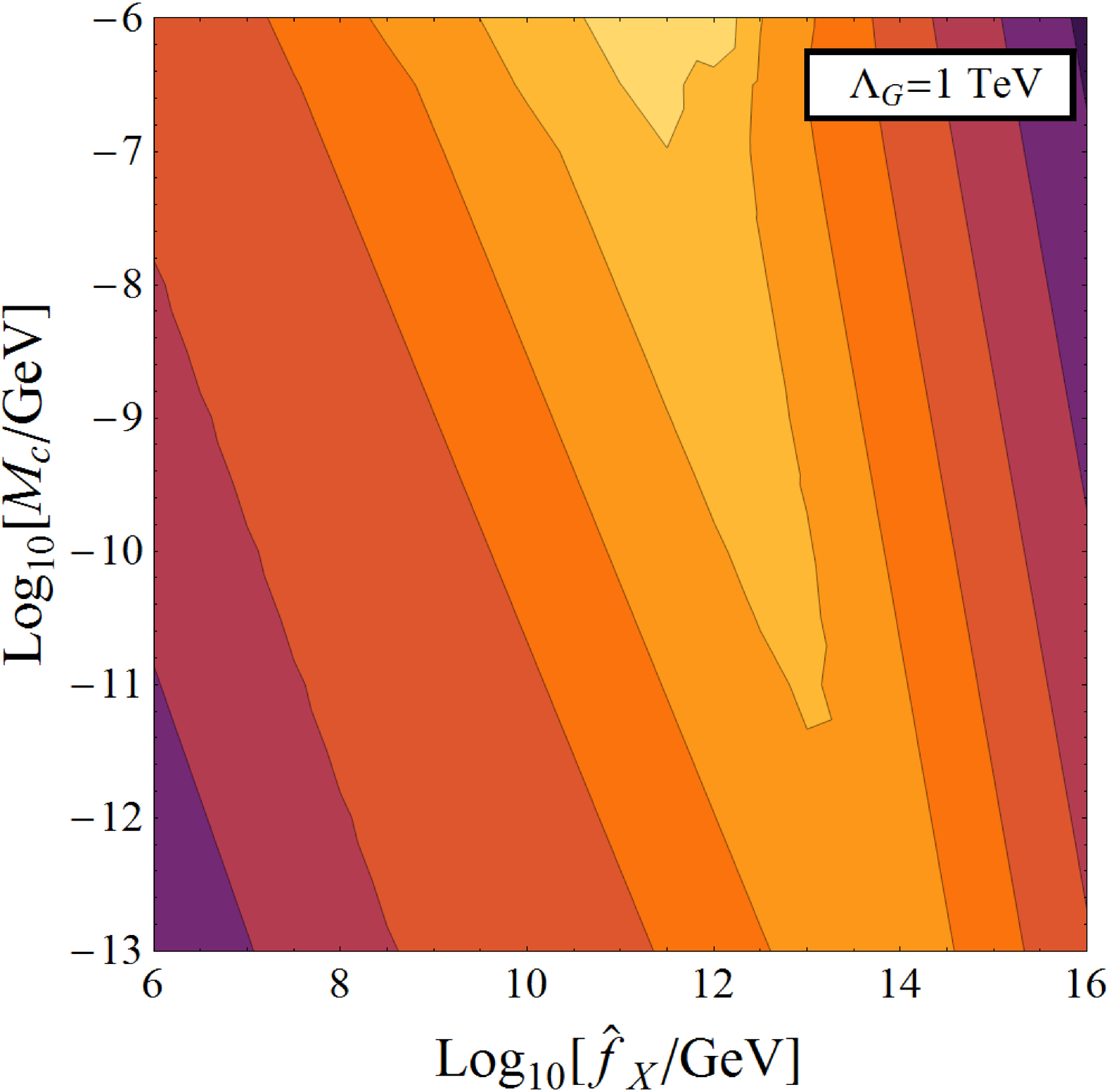}
  \epsfxsize 2.25 truein \epsfbox {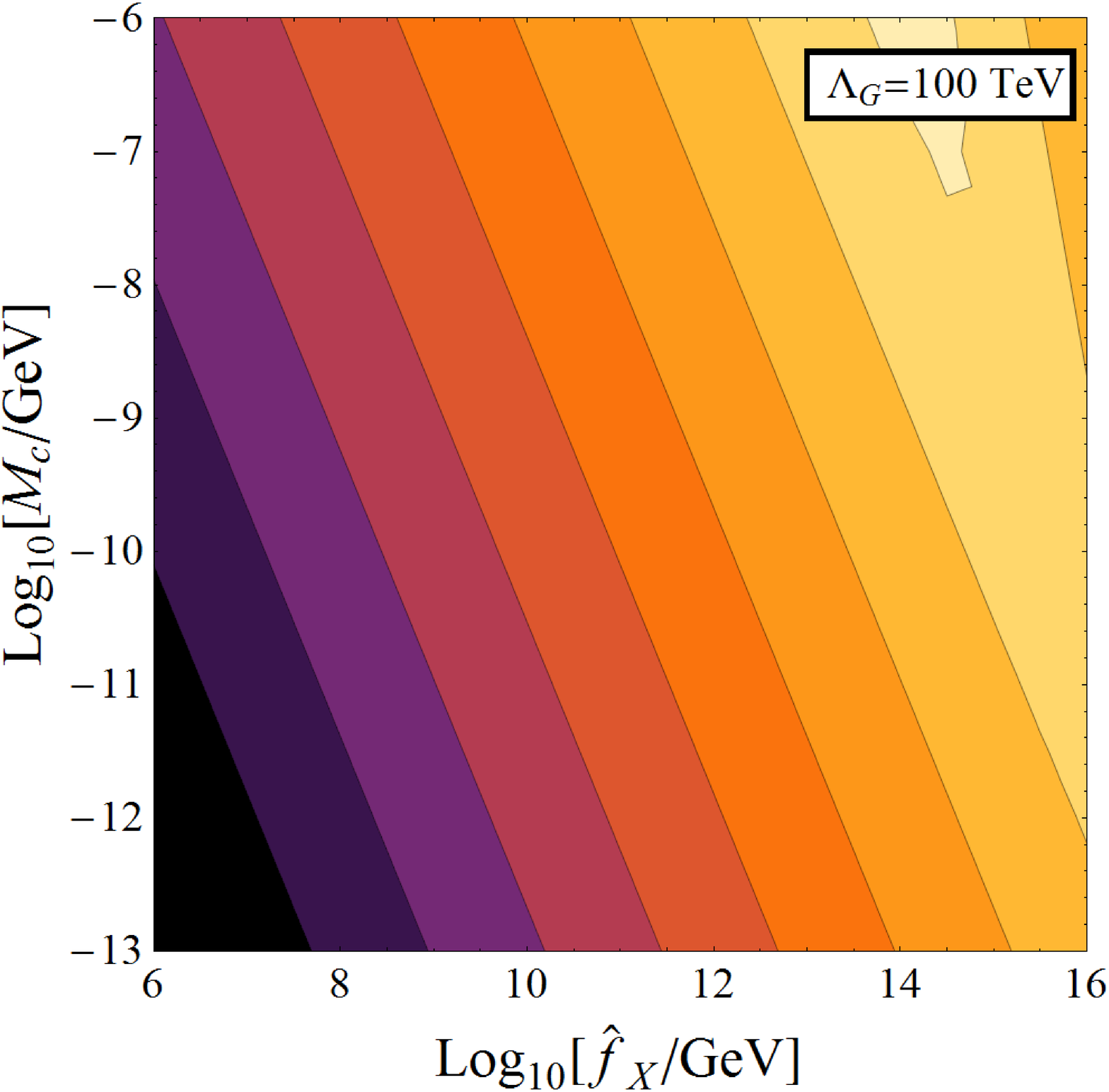}
  \raisebox{0.3cm}{\large$\Omegatotnodec$}\epsfxsize 5.00 truein \epsfbox {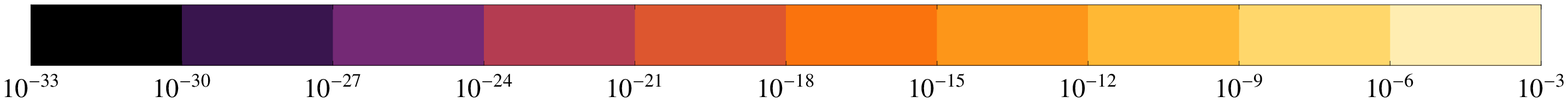}
\end{center}
\caption{Contours of the collective contribution to $\Omegatotnodec$ 
from the set of $a_\lambda$ with lifetimes $\tau_\lambda < \tau_\chi^{\mathrm{min}}$ for 
a DDM ensemble of photonic axions with $c_\gamma=1$. 
The left, center, and right panels display 
the results for $\Lambda_G = 1$~GeV, $\Lambda_G = 1$~TeV, and $\Lambda_G = 100$~TeV,
respectively.  In each case, we have taken $\TRH = 5$~MeV, $H_I = 100$~TeV, and 
$\xi = g_G = \theta = 1$.  In each panel, we see that BBN constraints
are amply satisfied throughout essentially the entire region of parameter space shown.
\label{fig:OmegaThreshPanelsLTR}}
\end{figure}

We begin by noting that in traditional, single-particle dark-matter scenarios, 
an unstable dark-matter candidate $\chi$ with a relic abundance 
$\Omega_\chi \sim \OmegaDM$ is generally consistent with all astrophysical 
and cosmological limits on dark-matter decays, provided that 
$\tau_\chi \gtrsim \tau_\chi^\mathrm{min} \sim 10^{26}$~s~\cite{DMDecayChenKamionkowski1}.    
It therefore follows that any $a_\lambda$ in the DDM ensemble with a lifetime 
$\tau_\lambda \gtrsim \tau_\chi^\mathrm{min}$ will have no impact on BBN within 
regions of parameter space in which the WMAP constraint $\Omegatot \leq \OmegaCDM$ on
the {\it total}\/ dark-matter relic abundance is satisfied.  Thus, we may safely conclude
that our bulk-axion model of dynamical dark matter is consistent with BBN constraints
within such regions of parameter space, provided that  
\begin{equation}
  \Omegatotnodec ~\lesssim~ 1.7 \times 10^{-5}~,
\label{eq:OmegathreshBBNConstraint} 
\end{equation}
where $\Omegatotnodec$ denotes the collective contribution which
the set of $a_\lambda$ with lifetimes $\tau_\lambda < \tau_\chi^{\mathrm{min}}$ 
would have made to the dark-matter relic abundance at present time if they were  
absolutely stable.  In other words, the BBN constraint
we are imposing in Eq.~(\ref{eq:OmegathreshBBNConstraint}) effectively
rests upon the extremely conservative approach of treating all states in the 
DDM ensemble whose lifetimes are less than $\tau_\chi^{\mathrm{min}}$
as if they had lifetimes $\tau_{\lambda}$ which are in the range
which is most dangerous for BBN, namely $\tau_\lambda \sim 10^{9}-10^{10}$~s.
We emphasize that while this criterion is a sufficient condition 
for successful BBN, it does {\it not}\/ represent the true BBN constraint, 
which is always far less stringent.

In Fig.~\ref{fig:OmegaThreshPanelsLTR}, we display contours of 
$\Omegatotnodec$ for a DDM ensemble of photonic axions with $c_\gamma =1$,
as a function of $\fhatX$, $M_c$, and $\Lambda_G$.  
The left panel shows the results for 
$\Lambda_G = 1$~GeV, the center panel for $\Lambda_G = 1$~TeV, and the 
right panel for $\Lambda_G = 100$~TeV.  In each case, we have taken 
$\TRH = 5$~MeV, $H_I = 100$~TeV, and $\xi = g_G = \theta = 1$.
In each panel of Fig.~\ref{fig:OmegaThreshPanelsLTR}, we see that the criterion in
Eq.~(\ref{eq:OmegathreshBBNConstraint}) is amply satisfied throughout
essentially the entire region of parameter space shown.  It therefore follows 
that our bulk axion model is consistent with successful BBN throughout this 
region of parameter space.      


\subsection{Axion Decays and Late Entropy Production\label{sec:Entropy}}


One additional physical consequence of the late decays of unstable relic
particles is the generation of entropy as those particles ``dump''  
their energy density into the radiation bath.  Indeed, a number of 
considerations place constraints on late entropy production 
from decaying particles.  For example, late entropy generation can upset 
the light-element predictions from standard BBN and produce observable 
features in the CMB.  In this section, we examine the effect of 
the late decays of the $a_\lambda$ on the entropy density of the universe 
in our bulk-axion DDM model as a function of time in order to verify
that no perceptible effects can arise which might serve to exclude 
our model.  
     
During any given epoch, the entropy density of the universe is dominated by 
the contribution from radiation and therefore well approximated by 
\begin{equation}
  s ~\approx~ \sum_{i}\frac{4\rho_i(T_i)}{3T_i} ~=~ 
    \frac{\pi^2}{30}g_{\ast s}(T)T^3~,
\end{equation}   
where the index $i$ runs over all relativistic particle species, 
$T_i$ is the temperature associated with any particular such species,
and $g_{\ast s}$ is the number of interacting degrees of freedom at 
temperature $T$.  During the early stages of the history of the universe
(prior to neutrino decoupling), all such species are characterized by a 
common temperature $T_i \approx T$.  During such epochs, 
$g_{\ast s}(T)\approx g_{\ast}(T)$, and the entropy 
density is therefore directly proportional to the total energy density 
$\rhorad$ of radiation.  Indeed, even during subsequent epochs,
$g_{\ast s}$ and $g_\ast$ remain roughly similar, and $\rhorad$ remains a    
good indicator of the entropy density.  Thus, by evaluating the contribution 
to $\rhorad$ from $a_\lambda$ decays in our bulk axion DDM model, we can 
assess the effect of these decays on both the energy and entropy densities 
of the universe.  

In the LTR cosmology, as in the standard cosmology, $\rhorad$ evolves 
according to an equation similar to 
Eq.~(\ref{eq:RhoLambdaAndGammaEvolEqsSGen}):
\begin{equation}
  \frac{d\rhorad}{dt} ~=~ -4H\rhorad + \Gamma_\phi \rho_\phi+
     \sum_\lambda \BRrad\Gamma_\lambda\rho_\lambda~.  
  \label{eq:RhoRadEvolEqGen}
\end{equation}
This equation assumes
the presence of a tower of decaying $a_\lambda$, where $\BRrad$ is the 
total branching fraction of $a_\lambda$ into relativistic particles.  Note,
however, that since we are working within the context of LTR cosmology,  
the effects of inflaton decays on the energy and entropy densities of the 
universe remain relevant until very late times $t\sim \tRH$.
Thus we have explicitly included an additional source term 
$\Gamma_\phi \rho_\phi$ in Eq.~(\ref{eq:RhoRadEvolEqGen}) to account for 
the effect of such inflaton decays, where $\Gamma_\phi$ and $\rho_\phi$ respectively
denote the decay rate and energy density of the inflaton field $\phi$.
  
The contribution to $\rhorad$ from inflaton decays can readily be calculated 
from standard results pertaining to the LTR cosmology (for a review, see, \eg, Ref.~\cite{LTRAxionsKamionkowski}).
As the universe exits the inflationary epoch at a time $t_I \approx 2/(3H_I)$, 
the energy density stored in the inflaton field is initially 
$\rho_{\phi}=\rhocrit =3H_I^2 M_P^2$.  During subsequent epochs, the inflaton 
source term for radiation is approximately given by 
\begin{equation}
  \Gamma_\phi \rho_\phi ~\approx~ \frac{3H_I^2 M_P^2}{2\tRH}
    \left(\frac{t_I}{t}\right)^\kappa e^{-t/2\tRH}~,
  \label{eq:InflatonRadSourceTerm}
\end{equation}
where $\kappa$ is defined as in Eq.~(\ref{eq:DefOfkappaForH}), and
we have used the fact that the inflaton-decay rate is related to the 
reheating time by $\Gamma_\phi \approx 1/(2\tRH)$.  Note that
this source term is negligible at times $t\gg\tRH$, when by definition 
$\rho_\phi \ll \rhorad$, and hence can safely be neglected at such times.
By contrast, at early times $t\lesssim \tRH$, the inflaton source term 
is expected to dominate in Eq.~(\ref{eq:RhoRadEvolEqGen}), in the sense
that 
\begin{equation}
   \Gamma_\phi \rho_\phi ~\gg~ \sum_\lambda \BRrad\Gamma_\lambda \rho_\lambda~.
\end{equation}
Whenever this condition is satisfied,
the contribution to $\rhorad$ from $a_\lambda$ decays is inconsequential 
compared to that from inflaton decays, and the axion source term can therefore 
safely be neglected.  
 
In assessing the contribution from $a_\lambda$ decay, we once again 
choose to focus on the case of a photonic axion with $c_\gamma = 1$; this
implies that the decay mode $a_\lambda \rightarrow \gamma \gamma$ dominates
the contribution to $\rhorad$.   In this
case, the source term for radiation due to $a_\lambda$ decay is just the 
source term for photons given in Eq.~(\ref{eq:SourceTermForPhotons}).  In
this case, solving Eq.~(\ref{eq:RhoRadEvolEqGen}) for $\rhorad$, we find that
\begin{equation}
  \rhorad(t) ~=~ \rhoradbar(t)
     + \int_{t_G}^{t}\left(\frac{t'}{t}\right)^{4\kappa/3} \sum_\lambda 
     \BRgamma \Gamma_\lambda \rho_\lambda(t')dt'~,
\end{equation} 
where $\rhoradbar(t)$ is the solution for $\rhorad(t)$
in the absence of any additional contribution from $a_\lambda$ decays.
Once again making use of the integral functions
$I_i\big(m,n,\alpha,\beta,\lambda_{\mathrm{min}},\lambda_{\mathrm{max}}\big)$ 
defined in Eq.~(\ref{eq:IntegralIdentity}) to approximate the sum over modes,   
we obtain
\begin{equation}
  \rhorad(t) ~=~ \rhoradbar(t)
     + \frac{2G_\gamma\theta^2}{M_c} \int_{t_G}^t dt' 
     \frac{t'^{\kappa/3}}{t^{4\kappa/3}}
     \sum_{i=1}^4 
     I_i\big(m_i,n_i,\alpha_i,\beta_i,\lambda_{i-1}^{\mathrm{CMB}},
     \lambda_i^{\mathrm{CMB}}\big)
         \times \begin{cases}
     \vspace{0.25cm}
   \tRH^{1/2}  ~~ & t \lesssim \tRH \\ 
    \vspace{0.25cm}
    1  ~~ & \tRH \lesssim t \lesssim \tMRE \\
     \tMRE^{1/2}  ~~ & t \gtrsim \tMRE~. 
     \end{cases}
\end{equation}

\begin{figure}[b!]
\begin{center}
  \epsfxsize 2.25 truein \epsfbox {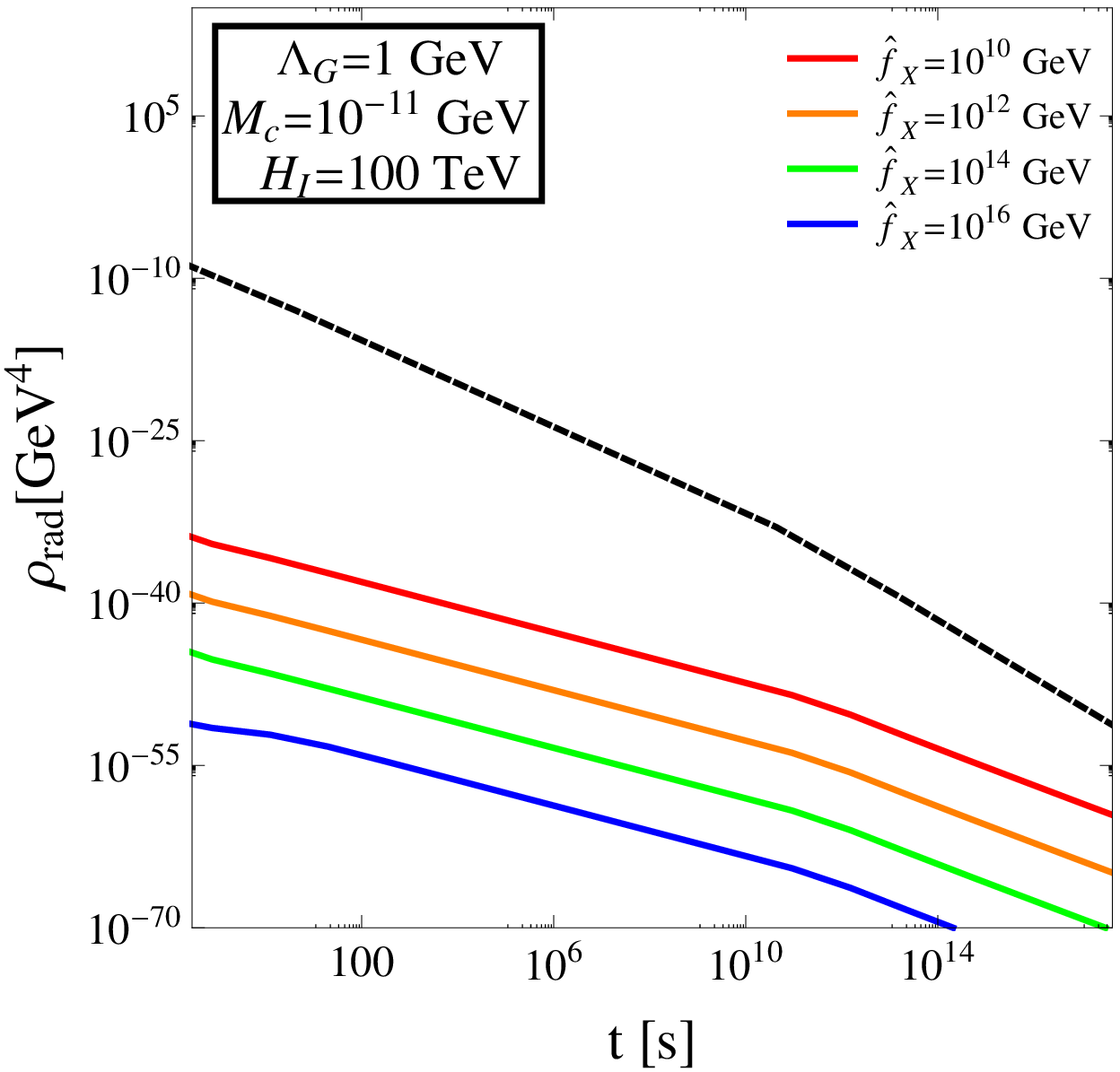}
  \epsfxsize 2.25 truein \epsfbox {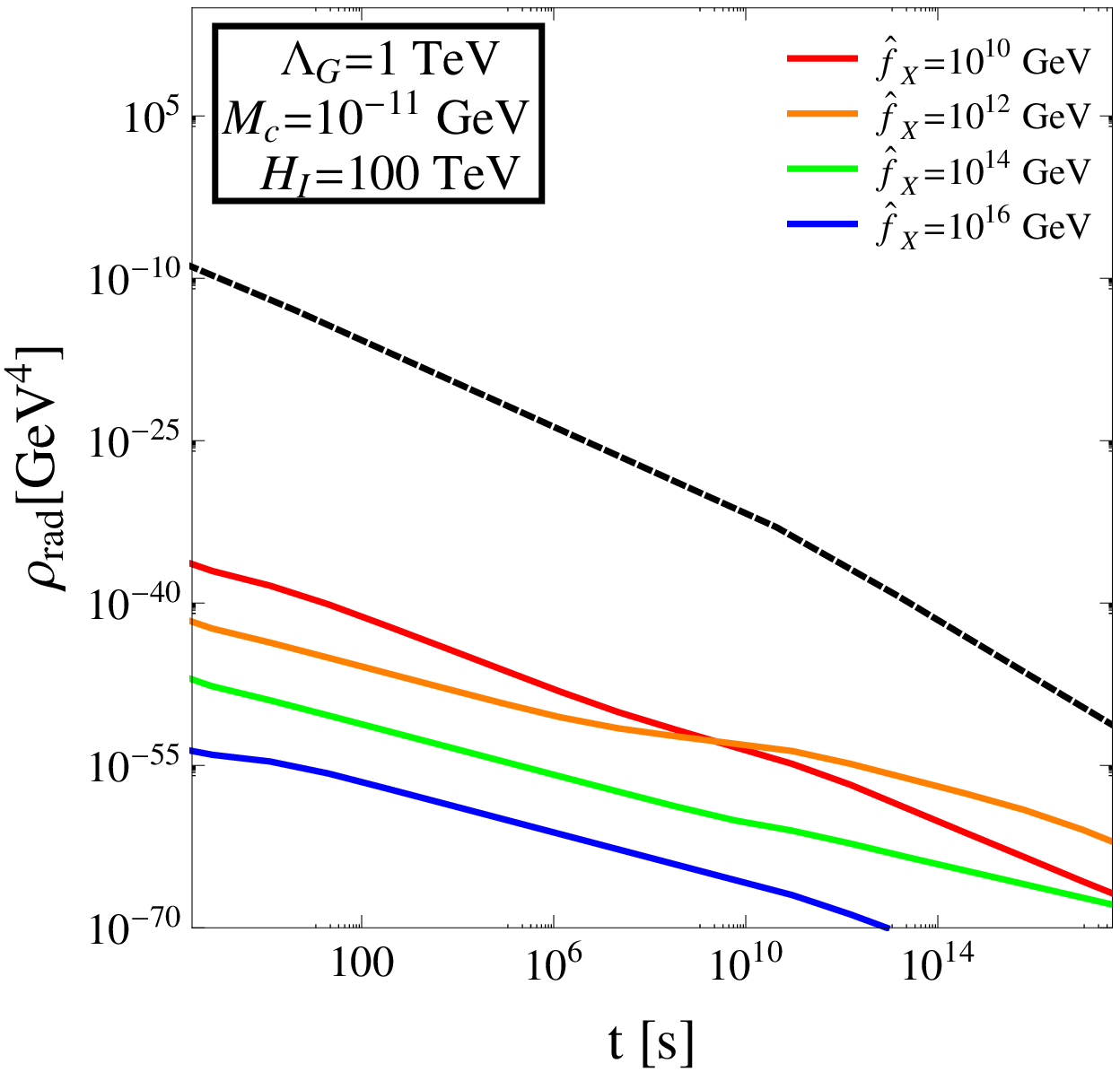}
  \epsfxsize 2.25 truein \epsfbox {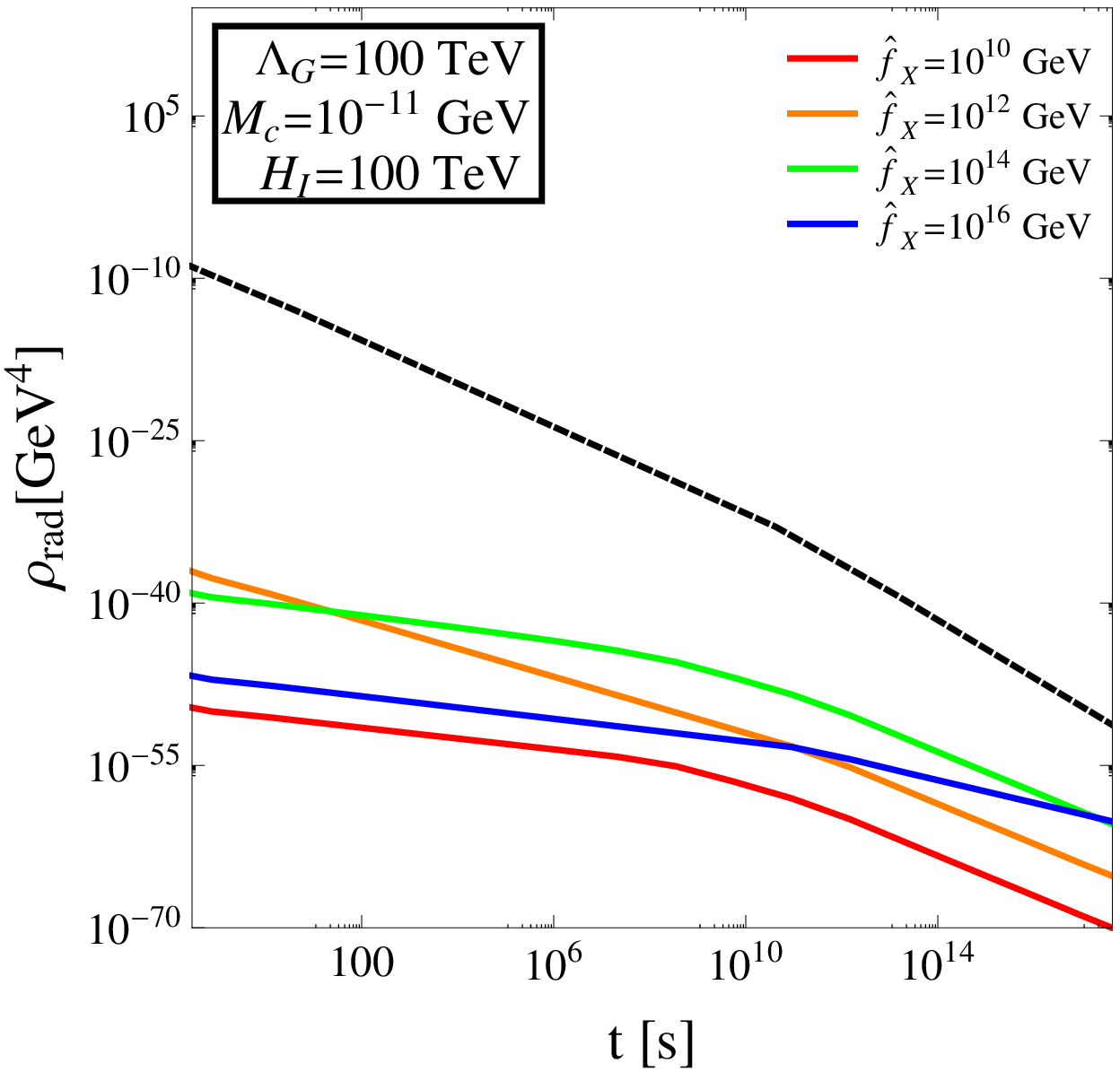}
\end{center}
\caption{The total contribution to the radiation 
energy density $\rhorad$ from photonic $a_\lambda$ decays in our 
bulk-axion DDM model (solid lines), plotted as functions of time 
for a variety of different choices of $\fhatX$.
The left panel shows the results for $\Lambda_G = 1$~GeV, the center panel 
for $\Lambda_G = 1$~TeV, and the right panel for $\Lambda_G = 100$~TeV.   
In each case, we have assumed a photonic axion with $\xi=g_G=\theta=1$, and 
we have taken $M_c = 10^{-11}$~GeV, $\TRH = 5$~MeV, and $H_I = 100$~TeV.
Also shown in each panel is the total value of $\rhorad$ as a function of time 
in the LTR cosmology (black dashed line), which includes the contribution from 
inflaton decay.  In all cases, the collective contribution to
$\rhorad$ from $a_\lambda$ decays at all times $t < \tnow$ remains negligible 
compared to the primordial contribution generated via inflaton decays during
reheating.  Thus our bulk-axion DDM model does not lead to overproduction of either 
radiation-energy density or entropy during any prior cosmological epoch.
\label{fig:RhoRadVsTime}}
\end{figure} 

In Fig.~\ref{fig:RhoRadVsTime}, we show how the contribution to $\rhorad$ from 
$a_\lambda$ decays in our bulk-axion DDM model evolves with time for a variety
of different choices of model parameters.  The left panel
shows results for $\Lambda_G = 1$~GeV, the center panel for $\Lambda_G = 1$~TeV,
and the right panel for $\Lambda_G = 100$~TeV.  The solid colored curves in each panel 
correspond to different choices of $\fhatX$ within the range $10^{10} - 10^{16}$~GeV.
For all curves shown, we have assumed a photonic axion with $c_\gamma=1$, and 
we have taken $M_c = 10^{-11}$~GeV, $\TRH = 5$~MeV, $H_I = 100$~TeV, and 
$\xi=g_G=\theta=1$.  The black dashed curve represents the total value of 
$\rhorad$, which includes the standard contribution from inflaton 
decays during the reheating epoch.  
Since such inflaton decays constitute the dominant source for 
radiation prior to the end of reheating, the range of times shown in
each panel extends from $\tRH$ to present time.  The value of $H_I$ 
has been chosen here to be sufficiently large that the effect of heavier 
$a_\lambda$ with $\lambda \gtrsim 3H_I/2$ being inflated away is unimportant.
Note, however, that for significantly smaller values of $H_I$, the contribution
to $\rhorad$ from axion decays can be further suppressed by this effect.   

The differences among the curves shown in Fig.~\ref{fig:RhoRadVsTime} for different 
choices of $\fhatX$ and $\Lambda_G$ ultimately stem from the effects of axion mixing
on the abundances $\rho_\lambda$ and decay widths $\Gamma_\lambda$ of the
individual axion modes.
The results shown in the left panel correspond to the case in which  
$\Lambda_G$ is sufficiently small that $y\gg 1$ for all choices of $\fhatX$ 
shown.   In this small-mixing regime, $\lambda \gtrsim \lambdatrans$ for all but 
the lowest-lying mode in the axion KK tower, and 
Eqs.~(\ref{eq:DefsOfyandmPQ}) and~(\ref{eq:RhoLambdaInLTRCosmo}) imply that 
$\rho_\lambda \propto\fhatX^{-2}$ and $\Gamma_\lambda \propto\fhatX^{-2}$.
It therefore follows that the photon source term 
$\BRgamma\Gamma_\lambda\rho_\lambda$ associated with each $a_\lambda$ 
within this regime decreases uniformly and substantially with increasing $\fhatX$,
as indicated.  By contrast, as $\Lambda_G$ is increased, 
several competing effects play an increasingly important role in determining the
magnitude of $\BRgamma\Gamma_\lambda\rho_\lambda$ for certain $\lambda$.  
This is because $\lambdatrans$ increases with increasing $\Lambda_G$; hence 
for large $\Lambda_G$ a greater number of the 
$a_\lambda$ are brought into the $\lambda \lesssim \lambdatrans$ regime, in which 
$\rho_\lambda \propto\fhatX^2$ and $\Gamma_\lambda\propto \fhatX^2$.
Increasing $\fhatX$ therefore has the effect of increasing the initial magnitude of 
the photon source terms associated with the $a_\lambda$ in this regime.  However, 
the lifetimes of these modes also increase with increasing $\fhatX$, and hence 
the transfer of their energy density to radiation is deferred until later times, 
when $\rhorad$ is smaller and the contribution from $a_\lambda$ decays 
can have a proportionally greater impact.  The interplay between these effects 
results in the behavior shown in the right two panels of Fig.~\ref{fig:RhoRadVsTime}.

Note that the curves for the total energy density shown in 
Fig.~\ref{fig:RhoRadVsTime}, which are dominated by the 
contribution from inflaton dynamics, drop more rapidly as a function of time than
the contributions from axion dynamics.  This reflects the continuing generation
of new radiation energy density from the ongoing decays of the individual $a_\lambda$
within our DDM ensemble.  In all cases, however, the collective contribution to
$\rhorad$ from $a_\lambda$ decays at all times $t < \tnow$ remains negligible 
compared to the primordial contribution generated via inflaton decays during
reheating.  Thus our bulk-axion DDM model does not lead to overproduction of either 
radiation-energy density or entropy during any prior cosmological epoch.


\subsection{Vacuum Energy and Overclosure\label{sec:Overclosure}}


In traditional dark-matter scenarios involving a single, stable dark-matter 
candidate $\chi$, the dark-matter relic abundance $\Omega_\chi$ increases 
monotonically up to and beyond the present time.  As a result, verifying
that $\Omega_\chi$ satisfies WMAP constraints at the present time is sufficient to
guarantee that $\chi$ does not overclose (or prematurely matter-dominate) 
the universe at all previous times as well.
However, one of the hallmarks of the DDM scenario is that this is no longer true: 
although $\Omegatot$ likewise experiences a Hubble-driven growth during the earliest 
phases of the evolution of the universe, this quantity can nevertheless drop 
during later epochs.  This is possible within the DDM framework because 
the single, stable dark-matter candidate $\chi$ characteristic of most traditional 
dark-matter scenarios is replaced by a complex, multi-component 
dark-matter ensemble whose constituents can have a broad spectrum of lifetimes 
and abundances.  As a result, the decays of certain dark-matter components within the
ensemble can cause $\Omegatot$ to decline --- even prior to the present day.  
Indeed, such behavior for $\Omegatot$ can be quite dramatic, and is illustrated 
in Fig.~6 of Ref.~\cite{DynamicalDM1} for the special case in which the DDM ensemble 
consists of a KK tower of decaying dark fields.  Thus, within the DDM framework,
it is no longer sufficient to verify that $\Omegatot$ satisfies overclosure constraints
at the present time; we must also verify that it has satisfied such overclosure 
constraints (and constraints from premature matter- or vacuum-energy domination) 
at all prior moments during the history of the universe.

It turns out, however, that this is not a problem in our bulk-axion DDM model.  Since our
model already satisfies WMAP constraints at present time within our preferred region of 
parameter space~\cite{DynamicalDM2}, it can run afoul of overclosure constraints in the
past only if the negative rate of change of $\Omegatot$ is sufficiently great that 
$\Omegatot$ might have exceeded unity within the past history of the universe.  However,
as discussed in Refs.~\cite{DynamicalDM1,DynamicalDM2}, this rate of change is described
by an effective equation-of-state parameter $w_{\mathrm{eff}}$, and two things are
already known about the value of this parameter in our model: first, it is extremely 
small at the present day, \ie, 
$10^{-23} \lesssim w_{\mathrm{eff}} \lesssim 10^{-12}$~\cite{DynamicalDM2}, and second,
it was even smaller in the past.  Indeed, this latter assertion follows from the 
generic behavior of $w_{\mathrm{eff}}$ shown in Fig.~8 of Ref.~\cite{DynamicalDM1}:
for a generic KK tower, $w_{\mathrm{eff}}$ reaches its maximum at the present day and 
is exponentially smaller prior to this time.  Thus, working backwards from the present
epoch, and given the finite age of the universe, we see that it is not possible for
$\Omegatot$ to have violated overclosure bounds at any point during the history of the
universe.  

One related concern which arises in our bulk-axion DDM model, due to our reliance on
the misalignment mechanism for the generation of the primordial relic abundances of
the $a_\lambda$ is the risk of premature vacuum domination.  Indeed, any $a_\lambda$ 
for which 
$t_\lambda > t_G$ will contribute to the total dark-energy abundance $\Omegavac$ during 
the period when $t_G \lesssim t \lesssim t_\lambda$, within which its energy density 
$\rho_\lambda$ is non-vanishing but before which it begins oscillating. 
Since $\rho_\lambda$ remains constant during this period, the contribution
to $\Omegavac$ scales like $\Omega_\lambda \propto t^2$ during any MD or RD epoch.  Since 
this represents a rate of increase far faster than that associated with matter or 
radiation, the threat of premature vacuum domination from fields which remain as 
vacuum energy for a long duration is of particular concern.  Indeed, in extreme 
cases, such fields could potentially give rise to an additional period of inflation,
leading to gross inconsistencies with the predictions of BBN, CMB data, and so forth.    

In our bulk-axion model, however, it is straightforward to demonstrate that  
no such inconsistencies with observational data arise.
The masses of all of the $a_\lambda$, with the sole exception of the zero mode $a_0$,
are bounded from below by the Newton's-law-modification constraint in 
Eq.~(\ref{eq:MinimumMc}), since $\lambda_i \geq M_c/2$ for $i>0$.  For all such modes
with $t_\lambda > t_G$, this constraint on $\lambda$ implies a bound 
$t_\lambda > 6.75\times 10^{-14}$~s on the oscillation-onset time of the mode.
(The remaining modes, for which $t_G = t_\lambda$, never contribute to $\Omegavac$.)  
This time scale is sufficiently early that the collective vacuum-energy contribution 
from these $a_\lambda$ poses no threat of overclosure or premature
vacuum-domination.  The $\Omega_\lambda$ contributions from these fields simply do 
not have time to grow to a problematic size.

This leaves only the contribution from $a_0$,whose oscillation time scale
can be substantially longer than the upper limit quoted above for the higher
modes in situations in which $y \gg 1$.  Since $A_{\lambda_{0}} \approx 1$ in this limit,  
Eq.~(\ref{eq:RhoLambdaInLTRCosmo}) implies that prior to the time $t_{\lambda_0}$ at
which it begins oscillating, the relic abundance of $a_0$ is given by 
\begin{equation}
  \Omega_{\lambda_0} ~\approx~ \frac{3}{2}\frac{\mX^2\fhatX^2}{M_P^2}
     \left(\frac{t}{\kappa}\right)^2~.
\end{equation}
Therefore, one finds that by the time of oscillation, which is given by 
$t_{\lambda_0} \approx \kappa_{\lambda_{0}}/2\mX$ 
in this limit, $\Omega_{\lambda_0}$ will have grown to   
\begin{equation}
  \Omega_{\lambda_0}(t_{\lambda_0}) ~\approx~ \frac{3}{8} \frac{\fhatX^2}{M_P^2}~.
\end{equation} 
This result is independent of $\mX$, and implies that the contribution of the 
$a_0$ to $\Omegavac$ is not a cause for concern for sub-Planckian values of $\fhatX$.  
Indeed, this is to be expected: in this regime, $a_0$ functions effectively 
like a four-dimensional axion.  Early vacuum-energy domination is known not to be
a problem for light axions and axion-like particles (see Ref.~\cite{JaeckelReview} and 
references therein) in purely four-dimensional theories. 


\subsection{Misalignment Production and Isocurvature 
Perturbations\label{sec:isocurvature}}

 
In an inflationary cosmology, fluctuations in the energy density of any 
population of particles produced thermally, \ie, via rapid interactions 
in the radiation bath during the reheating phase, stem from the primordial 
perturbations in the energy density of the inflaton field.  Consequently, 
such fluctuations are of the so-called adiabatic type --- that is, they
represent spatial variations in the {\it total} energy density, but not in the 
relative contributions of individual particle species to that total density.
Such variations, in turn, imply fluctuations in the local spacetime curvature
and are therefore sometimes also referred to as curvature perturbations.
By contrast, fluctuations in the energy density of any population of particles
produced via means uncorrelated with the inflaton field (and therefore non-thermal)
can also give rise to fluctuations of the isocurvature 
type --- \ie, perturbations in the relative contributions of different species 
to the total energy density, with that total energy density held fixed. 
Recent WMAP observations of the CMB power spectrum,
taken in combination with baryon acoustic oscillation (BAO) measurements and 
supernova data, place a stringent bound~\cite{WMAP} on any deviations from 
adiabaticity in primordial energy-density fluctuations.  This bound is typically   
expressed in terms of the fractional contribution $\alpha_0$ to the 
CMB power spectrum from axion isocurvature perturbations:   
\begin{equation}
  \alpha_0 ~\equiv~ \frac{\langle (\delta T/T)^2_{\mathrm{iso}}\rangle}
     {\langle(\delta T/T)^2_{\mathrm{tot}}\rangle} < 0.072~, 
  \label{eq:WMAPAlphaIsoBound}
\end{equation}
where $\langle (\delta T/T)^2_{\mathrm{tot}}\rangle$ and 
$\langle (\delta T/T)^2_{\mathrm{iso}}\rangle$ respectively denote the 
total average root-mean-squared fluctuation in the CMB temperature, and the 
average root-mean-squared temperature fluctuation due to isocurvature
perturbations alone.  Since the $a_\lambda$ fields which compose our dynamical 
dark-matter ensemble are presumed to be produced non-thermally, via the 
misalignment mechanism, it is necessary to investigate the implications of this
bound for our model.   

Our discussion of isocurvature perturbations in our bulk-axion DDM model 
in large part parallels the discussion of such
perturbations in traditional QCD axion models presented in 
Ref.~\cite{HertzbergAxionCosmology}, to which we refer the reader for a 
more complete introduction and discussion of the formalism and methodologies used.     
It turns out to be convenient to express the fluctuations of any given $a_\lambda$ in 
terms of the fractional change $S_\lambda$ in the ratio of 
its number density $n_\lambda$ to the entropy density $s$
of the universe.  This quantity can be written in the form
\begin{equation}
  S_\lambda ~\equiv~ \frac{\delta(n_\lambda/s)}{(n_\lambda/s)}
     ~=~ \frac{\delta n_\lambda}{n_\lambda} - 3 \frac{\delta T}{T}~.
  \label{eq:DefOfSlambda}
\end{equation}  
We assume that the production of all other particle species $\psi_i$ 
(\ie, the SM fields) ultimately results from inflaton decay, and that the 
density fluctuations for these species are purely adiabatic, with $S_i = 0$.
Since, by definition, the fluctuation $\delta\rho$ in the total energy density 
vanishes for isocurvature fluctuations, it therefore follows that the sum of
the fluctuations in the energy densities of the various particle species
obeys a constraint which may be written in the form  
\begin{equation}
  \sum_\lambda \rho_\lambda \left(S_\lambda + 3 \frac{\delta T}{T}\right) + 
     3\sum_i \rho_i \frac{\delta T}{T} + 4\rho_{\mathrm{rad}}\frac{\delta T}{T} ~=~ 0~, 
   \label{eq:IsocurvatureNRGConservationEq}
\end{equation} 
where the $\rho_i$ denote the energy densities associated with massive species other 
than the $a_\lambda$, and $\rhorad$ once again denotes the total energy density of 
radiation.  
In our bulk-axion DDM model, the abundances of 
all of the $a_\lambda$ are determined by a single misalignment angle 
$\theta$.  As discussed in Ref.~\cite{DynamicalDM1}, this reflects the
ultimate five-dimensional nature of the axion field.
This in turn implies that the density fluctuations 
$\delta n_\lambda$ for all of these fields are determined by the    
fluctuations $\delta\theta$ in this misalignment angle generated by 
quantum fluctuations during inflation.  
The fact that the fluctuations $\delta n_\lambda$ are all 
determined by $\delta\theta$ implies that the $S_\lambda \equiv S$ are 
essentially equal for all $a_\lambda$; hence 
Eq.~(\ref{eq:IsocurvatureNRGConservationEq}) simplifies to
\begin{equation}
    \Omegatot S ~=~ -3\left(\Omega_{\mathrm{mat}} + 
       \frac{4}{3}\Omega_{\mathrm{rad}}\right) \frac{\delta T}{T}~,
   \label{eq:SlambdaEq}
\end{equation}  
where $\Omega_{\mathrm{mat}}$ denotes the total abundance of matter 
in the universe, including the contributions from baryonic matter, the 
ensemble of dark axions, and any other particles which might contribute to 
the dark-matter relic abundance, and $\Omega_{\mathrm{rad}}$ is
the relic-abundance contribution from radiation.  This expression is 
identical to that which describes the isocurvature perturbations associated 
with a single, four-dimensional 
axion.  Therefore, assuming that the fluctuations in $\theta$ are Gaussian, it
follows that in our axion DDM model, $\alpha_0$ is given by the 
standard expression~\cite{HertzbergAxionCosmology}
\begin{equation}
  \alpha_0 ~=~\frac{8}{25} \left(\frac{\Omegatot^\ast}{\Omega_{\mathrm{mat}}^\ast}\right)^2
           \frac{1}{\langle(\delta T/T)^2_{\mathrm{tot}}\rangle}\,
           \frac{\sigma_\theta^2(2\theta^2 + \sigma_\theta^2)}
           {(\theta^2+\sigma_\theta^2)^2}~,
  \label{eq:IsocurvatureAlpha1}
\end{equation}
where $\Omega_{\mathrm{mat}}^\ast$ denotes the present-day value of 
$\Omega_{\mathrm{mat}}$, and where $\sigma_\theta^2\equiv \langle(\delta\theta)^2\rangle$
denotes the variance associated with fluctuations in $\theta$.

This result makes intuitive sense.  Although our DDM model has essentially partitioned
the total dark-matter abundance amongst a large number of different KK axion fields,
the underlying five-dimensional nature of the KK tower has correlated the individual 
fluctuations of these fields so that they are governed by the fluctuation of a single 
misalignment angle $\theta$.  It is therefore not a surprise that the expected magnitude
for isocurvature fluctuations in our model turns out to be no greater than it is 
standard, four-dimensional axion models.    

All that remains, then, for us to do in order to determine the value of 
$\alpha_0$ in our bulk-axion DDM model, is to assess the magnitude of 
$\sigma^2_\theta$.  Assuming again that the fluctuations in $\theta$ are 
Gaussian, this quantity is given by   
\begin{equation}
  \sigma_\theta^2 ~=~ 
     \frac{H_I^2}{4\pi^2\fhatX^2}~.
\end{equation}  
Since we are operating within the context of an LTR cosmology with 
$\TRH \sim \mathcal{O}(\mathrm{MeV})$, as discussed above, it is by no means
problematic (and in fact quite natural) for $H_I \ll \fhatX$.  Therefore,
as long as $\theta \sim \mathcal{O}(1)$, as might be expected from naturalness 
considerations, it can safely be assumed that $\theta \gg \sigma_\theta$.   
Substituting into Eq.~(\ref{eq:IsocurvatureAlpha1}) the experimentally 
observed~\cite{WMAP} values  
$\langle (\delta T/T)^2_{\mathrm{tot}}\rangle \approx (1.1\times 10^{-5})^2$
and $\Omega_{\mathrm{mat}}^\ast\approx 0.262$ we find that $\alpha_0$ is well 
approximated by  
\begin{equation}
   \alpha_0 ~\approx~ 1.95\times 10^{9}
      \left(\frac{H_I\Omegatotnow}{\fhatX\theta}\right)^2
   \label{eq:IsocurvatureAlpha2}
\end{equation}
in our bulk-axion model.  Combining this result with the upper bound on $\alpha_0$ 
quoted in Eq.~(\ref{eq:WMAPAlphaIsoBound}) yields the constraint
\begin{equation}
   H_I ~\lesssim~ 6.07 \times 10^{-6} 
       \left(\frac{\theta\fhatX}{\Omegatot^\ast}\right)~. 
   \label{eq:HIConstraintIsocurvature}
\end{equation}
We consider the case in which $\Omegatotnow \approx \OmegaDM$ and in which the axion 
ensemble is responsible for essentially the entirety of the observed dark-matter 
relic abundance.  This ocrresponds to $\fhatX \approx 10^{14} - 10^{15}$~GeV.  
We then find that for $\theta \sim \mathcal{O}(1)$,
the resulting constraint $H_I \lesssim 10^9 - 10^{10}$~GeV on the 
Hubble parameter during inflation is 
relatively mild.  Indeed, there is no difficulty in satisfying this constraint in 
either the standard or the LTR cosmology.  We thus conclude that isocurvature 
perturbations do not present any problem for our bulk-axion model of dynamical 
dark matter.  Moreover, a low scale for $H_I$ can be regarded as natural in the context
of an LTR cosmology. 

It is worth remarking, however, that the above results have implications
for the detection of primordial gravitational waves.
Limits on primordial gravitational waves from observations of the CMB can
be conveniently parametrized in terms of the scalar-to-tensor ratio 
$r$.  For example, consider single-field models of inflation, in which 
$r = 16\epsilon$, where $\epsilon = M_P^2(V'/V)^2/(4\pi)$ is 
the inflaton slow-roll parameter, with $V$ and $V'$ denoting 
the inflaton potential and its first derivative with respect to the inflaton field,
respectively~\cite{LiddleAndLyth}.  In the context of our bulk-axion DDM model, the standard 
relation (see, \eg, Ref.~\cite{WMAP}) between $r$ and $\alpha_0$ takes the form
\begin{equation}
  r ~=~ \frac{2\theta^2\fhatX^2}{M_P^2}
     \left(\frac{\OmegaDM}{\Omegatotnow}\right)^2
     \frac{\alpha_0}{1-\alpha_0}~. 
\end{equation}
As discussed above, consistency with the bounds in Eqs.~(\ref{eq:WMAPAlphaIsoBound}) and~(\ref{eq:HIConstraintIsocurvature}) requires that $H_I \ll 2\pi f_X\theta$ 
and $\alpha_0 \ll 1$.  In this regime, one finds that the 
expected tensor-to-scalar ratio is essentially independent of $\Omegatotnow$
and well approximated by
\begin{equation}
  r ~\approx~ 2.7 \times 10^{8} 
    \left(\frac{H_I}{M_P}\right)^2~.
  \label{eq:TensorToScalarTheoretical}
\end{equation}

Current WMAP observations, again in conjunction from BAO and supernova data, 
place an upper bound $r<0.22$ on the tensor-to-scalar ratio~\cite{WMAP}.
Thus, Eq.~(\ref{eq:TensorToScalarTheoretical}) results in a constraint
$H_I \lesssim 6.7 \times 10^{13}$~GeV on the Hubble scale during 
inflation --- a constraint which Eq.~(\ref{eq:HIConstraintIsocurvature})
implies is already automatically satisfied, even for $\mathcal{O}(1)$ 
values of the misalignment angle $\theta$.  The upshot is therefore that 
while there is no conflict between current limits on isocurvature 
perturbations and the predictions of our bulk-axion DDM model, the 
requirement that $H_I$ be relatively small in this model suggests that 
$r$ should likewise be quite small --- at least in the simplest of 
inflationary scenarios.  Constraints on the spectral index $n_s$ from 
WMAP~\cite{WMAP} can simultaneously be satisfied for small $r$ without 
difficulty, for example in negative-curvature models of 
inflation, which tend to predict small $r$~\cite{WMAPInflation}.  

In summary, we conclude that current constraints on isocurvature 
perturbations can be satisfied in our bulk-axion DDM model without 
too much difficulty.  However, we note that any conclusive measurement
of $r$ within the sensitivity range of the Planck satellite would have 
severe ramifications for this model.      


\subsection{Axion Abundances and Quantum Fluctuations During
Inflation\label{sec:InflationScale}}


Thus far in this paper, we have disregarded the effects of the quantum fluctuations 
that naturally arise for any massless or nearly massless field during the 
inflationary epoch.  In particular, the low-momentum modes of any $a_\lambda$ 
in our model with a mass $\lambda \lesssim H_I$ have wavelengths which exceed the 
Hubble length during inflation; excitations of such low-momentum modes are
therefore indistinguishable from a VEV and consequently do not 
inflate away.  These excitations necessarily yield a primordial 
energy-density contribution in our bulk-axion DDM model which cannot be 
avoided in any inflationary cosmology.  Consistency with the relic-abundance
predictions discussed in Sect.~\ref{sec:MisalignmentProd} therefore requires 
that this primordial energy density be small compared to that which results 
from misalignment production.  

In particular, it is possible to formulate a condition that 
ensures that these quantum fluctuations not invalidate our previous analysis. 
Clearly, one criterion that any such condition must enforce is that 
such fluctuations not have a significant effect on the total relic 
abundance of the ensemble.  We may formulate this 
constraint as a requirement that the difference between the full present-day 
relic abundance $\Omegatotjitternow$, which incorporates the effect of these 
fluctuations, and the result $\Omegatotnow$ obtained in the absence of such 
corrections be negligible --- \ie, that  
\begin{equation}
  \big|\,\Omegatotjitternow - \Omegatotnow\big| 
    ~\ll~ \Omegatotnow~.
  \label{eq:HubbleJitterCondit1}
\end{equation}

While the condition in Eq.~(\ref{eq:HubbleJitterCondit1}) is certainly a 
necessary one, it is not by itself sufficient to ensure that
vacuum fluctuations during inflation do not lead to phenomenological 
difficulties for our model.  This is because within DDM framework, 
dark-matter stability is not a requirement, and consistency 
with observational constraints is arranged by balancing decay widths 
against abundances across the entire dark-matter ensemble.  Indeed, 
as we have demonstrated, misalignment production
provides precisely the right relationship between the $\Omega_\lambda$ and 
$\Gamma_\lambda$ to mitigate the deleterious effects of the heavier, more 
unstable states in our ensemble and render our model phenomenologically 
viable.  We must therefore ensure that this delicate balance is not 
disrupted by the effects of vacuum fluctuations during inflation.  

Within the preferred region of parameter space of our bulk-axion DDM model,
as discussed in Sect.~\ref{sec:MisalignmentProd}, the oscillation-onset times 
for the lighter $a_\lambda$ in the tower are staggered in time.  As a result, 
these lighter modes collectively dominate in $\Omegatot$.  It therefore follows
that whether or not the total-relic-abundance constraint in 
Eq.~(\ref{eq:HubbleJitterCondit1}) is satisfied depends primarily on how 
vacuum fluctuations affect the abundances of these most abundant modes alone.   
By contrast, the balancing of lifetimes against abundances depends on the 
properties of the full KK tower, and not merely on the attributes of the 
lighter modes which dominate $\Omegatot$.  The corresponding condition we 
impose on our model therefore represents an even stronger constraint than the 
one appearing in Eq.~(\ref{eq:HubbleJitterCondit1}) and indeed subsumes
it.  To wit, we require 
that the full relic abundance $\widetilde{\Omega}_\lambda$ of {\it each 
axion mode} not differ significantly from the corresponding abundance
$\Omega_\lambda$ obtained in the absence of corrections due to 
vacuum fluctuations during inflation --- \ie, that     
\begin{equation}
  \big|\,\widetilde{\Omega}_\lambda - \Omega_\lambda\big| 
    ~\ll~ \Omega_\lambda   ~~~~~ \mbox{for~all}~\lambda~.
  \label{eq:HubbleJitterCondit2}
\end{equation}
We emphasize that this is an overly conservative constraint, and that
consistency with observational data is certainly possible 
even if vacuum fluctuations do have a significant effect on the 
abundances of certain $a_\lambda$.  However, as we shall demonstrate, 
the restriction that this overly conservative constraint imposes on our model
(which primarily turns out to take the form of an upper bound on $H_I$ for
any allowed choice of $\fhatX$, $M_c$, and $\Lambda_G$) 
is not terribly severe.

In order to determine how this condition restricts
the parameter space of our model, we must first assess what effect  
vacuum fluctuations during inflation have on the individual energy densities
$\rho_\lambda$ and relic abundances $\Omega_\lambda$ of the constituent fields 
in our dark-matter ensemble.  We begin by noting a generic result in
inflationary cosmologies (for a review, see Ref.~\cite{LindeInflationReview}),
namely that the variance $\langle \phi^2\rangle$ in the amplitude of any light scalar 
$\phi$ with a mass $m_\phi \lesssim H_I$ induced by vacuum fluctuations during inflation 
is given by 
\begin{equation}
  \langle\phi^2\rangle ~\sim~ \frac{H_I^3\delta t_I}{4\pi^2}~,
  \label{eq:GenericFlucInPhi}
\end{equation}  
where $\delta t_I$ denotes the duration of inflation.  A fluctuation of this order
will therefore be induced in the amplitude of any axion in our dark-matter ensemble 
with a mass smaller than $H_I$.  Moreover, we note that the 
relationship between $\delta t_I$ and $H_I$ is constrained by the fact
that successful resolution of the smoothness and flatness problems 
requires $N_e \approx H_I \delta t_I \gtrsim 60$, where $N_e$ denotes the 
number of $e$-foldings of inflation.  In typical scenarios, $N_e$ lies only
slightly above this lower bound; hence $\delta t_I$ is typically expected to
be such that $H_I \delta t_I \sim \mathcal{O}(100)$.  We will frequently 
express our results in terms of $N_e$ in what follows. 
 
We begin our discussion the effect of these fluctuations on the abundances 
of the constituent particles in our dark-matter ensemble by examining 
the simple case in which $t_G \lesssim t_I$.  In this case, the axion 
mass-squared matrix attains its 
asymptotic, late-time form before inflation ends, and the $a_\lambda$ are 
consequently already the axion mass eigenstates during the inflationary epoch.  
Thus, we find that the total energy density associated with each $a_\lambda$ 
with $\lambda \lesssim H_I$ at the end of inflation is given by 
\begin{equation}
   \rho_\lambda(t_I) ~\approx~ \frac{1}{2}\lambda^2
    \left(\theta A_\lambda \fhatX + 
    \eta_\lambda\frac{H_I\sqrt{N_e}}{2\pi}\right)^2~,
   \label{eq:RhoLambdaInftIlesssimtG}  
\end{equation} 
where 
\begin{equation}
  \eta_\lambda ~\sim~ 
     \begin{cases}
       \mathcal{O}(1) & \lambda \lesssim H_I \\
       0 & \lambda \gtrsim H_I
     \end{cases}
  \label{eq:DefOfEtaLambdaFluctuation}
\end{equation}
is a random coefficient of which parametrizes the fluctuation in 
the field $a_\lambda$.

Before proceeding further, we remark that the above results depend critically 
on the assumption that $t_G \lesssim t_I$.  In other words, we have assumed that 
the instanton dynamics associated with the gauge group $G$ has already occurred 
and made its contributions to the KK masses prior to the onset of the quantum 
fluctuations that arise due to inflation.  By contrast, if $t_G \gtrsim t_I$,
the quantum fluctuations will occur first, when the axion mass matrix is still 
diagonal and when the KK momentum modes and mass eigenstates coincide.
In such cases, these are the modes which develop quantum fluctuations,
and the mode-mixing induced by the instanton dynamics occurs only later.

This distinction is important, because the resulting energy
density for each $a_\lambda$ takes a somewhat different form
when $t_G\gtrsim t_I$:
\begin{equation}
  \rho_\lambda ~=~ 
   \frac{1}{2} \lambda^2 \,\left[
      \sum_{n=0}^\infty U_{\lambda n} \left(
      \theta \hat f_X \delta_{n,0} + \eta_n 
      \frac{H_I \sqrt{N_e}}{2\pi}\right)\right]^2~. 
\label{eq:RhoLambdaInftIgtrsimtG}
\end{equation}
In this expression, $U_{\lambda n}$ is the mixing matrix in 
Eq.~(\ref{eq:DefOfalambda}) and $\eta_n$ is the analogue of $\eta_\lambda$ 
discussed above, with $\eta_n$ taking non-zero values only when $n\lesssim H_I/M_c$.

{\it A priori}\/, this expression results in a different value for $\rho_\lambda$ 
than that in Eq.~(\ref{eq:RhoLambdaInftIlesssimtG}).  However, it turns out that 
the eventual constraints associated with Eq.~(\ref{eq:RhoLambdaInftIgtrsimtG}) are 
no more stringent than those which we shall eventually calculate for
Eq.~(\ref{eq:RhoLambdaInftIlesssimtG}).  In order to understand why this is the 
case, let us consider an even more dramatic situation in which $\eta_n$ actually 
takes a fixed, positive value $\overline{\eta}$ for all $n$ --- even values of $n$ 
beyond the inflationary cutoff $H_I/M_c$.  In this case, we can make use of 
the identity
\begin{equation}
    \sum_{n=0}^\infty\, U_{\lambda n} ~=~ f(\wtl)\, A_\lambda~,
\end{equation}
where $f(\wtl)\equiv (\wtl^2 + \sqrt{2}-1)/\sqrt{2}$,
in order to rewrite Eq.~(\ref{eq:RhoLambdaInftIgtrsimtG}) in the form
\begin{equation}
   \rho_\lambda ~=~  \frac{1}{2} \lambda^2 \,\left[
    \theta A_\lambda \hat f_X + f(\wtl)\, {\overline{\eta}}\, 
           \frac{H_I\sqrt{N_e}}{2\pi}\right]^2~.
\end{equation} 
Remarkably, this is essentially the same expression as we would have obtained 
from Eq.~(\ref{eq:RhoLambdaInftIlesssimtG}) when $\eta_\lambda =\overline{\eta}$ 
for all $\lambda$, except that the fluctuation contribution now comes multiplied 
by an extra ``scaling'' factor $f(\wtl)$.  It is easy to verify that 
$f(\wtl)\to 1$ as $\wtl\rightarrow\infty$, whereas for small $\wtl$ 
we find that $f(\wtl)\ll 1$.  This indicates that the effects of the 
inflation-related quantum fluctuations are actually {\it suppressed}\/ for the 
lighter modes, relative to what occurs in the case with $t_G\lesssim t_I$.  
The magnitude of this suppression depends on $y$, and is more severe when
$y\ll 1$ ({\it i.e.}\/, when the axion modes are more fully mixed).  We thus 
conclude that the contributions from the quantum fluctuations that arise during 
inflation are greater when they occur {\it after}\/ the instanton dynamics turns 
on (and after the KK mode-mixing), rather than before.  We shall therefore 
concentrate on the $t_G\lsim t_I$ case in what follows. 

Given the result in Eq.~(\ref{eq:RhoLambdaInftIlesssimtG}), we see that 
the effect of vacuum fluctuations on the $\rho_\lambda$ will be small 
for values of $\lambda$ which satisfy the condition         
\begin{equation}
  \theta A_\lambda \fhatX ~\gtrsim~ \frac{H_I\sqrt{N_e}}{2\pi}~.
  \label{eq:VacMisSurpassesHubbleFluc}
\end{equation}
Since $A_\lambda$ is a monotonically decreasing function of $\lambda$, it follows 
that within any given tower of $a_\lambda$, there exists a critical mass value        
\begin{equation}
  \lambdafluc ~\equiv~ \frac{\mX}{\sqrt{2}} \left[
     \sqrt{\left(1+\frac{\pi^2}{y^2}\right)^2
     +\frac{32\pi^2\theta^2\fhatX^2}{N_e H_I^2} }
     -\left(1+\frac{\pi^2}{y^2}\right)\right]^{1/2}
   \label{eq:Lambdafluc}
\end{equation}
below which the effect of vacuum fluctuations on the corresponding energy density
$\rho_\lambda$ is negligible.  These $\rho_\lambda$ are therefore well approximated
by Eq.~(\ref{eq:RhoLambdaInLTRCosmo}), and the corresponding abundances 
$\widetilde{\Omega}_\lambda$ are given by Eq.~(\ref{eq:OmegaLambdaOftEqnLTRtG})
or Eq.~(\ref{eq:OmegaLambdaOftEqnLTRtlambda}), depending on the value of $t_\lambda$.
By contrast, for $\lambdafluc \lesssim \lambda \lesssim H_I$, the effect of vacuum 
fluctuations overwhelms the effect of vacuum misalignment.  
The initial energy density of each $a_\lambda$ in this regime is therefore  
effectively set at $t_I$ and is approximately given by
\begin{equation}
  \rho_\lambda(t_I) ~\approx~ \frac{N_e}{8\pi^2} \lambda^2 H_I^2~.
\end{equation}      
Since the Newton's-law-modification bound on $M_c$ in Eq.~(\ref{eq:MinimumMc}) 
implies that $t_\lambda \lesssim \tRH$ for each such field, it therefore follows 
that at all subsequent times, the corresponding relic abundance is given by  
\begin{equation}
  \widetilde{\Omega}_\lambda ~\approx~
    \frac{3N_e H_I^2}{4\pi^2 M_P^2}
      e^{-\Gamma_\lambda(t-t_I)}\times
      \begin{cases}
     \displaystyle\frac{1}{4} \vspace{0.25cm}~~
      & 1/\lambda ~\lesssim ~t ~\lesssim~ \tRH \\
      \displaystyle\frac{4}{9}\left(\frac{t}{\tRH}\right)^{1/2} \vspace{0.25cm}~~
      & \tRH ~\lesssim~ t ~\lesssim~ \tMRE \\ 
      \displaystyle\frac{1}{4}\left(\frac{\tMRE}{\tRH}\right)^{1/2}~~
      & t~\gtrsim ~\tMRE~.    
    \end{cases}
  \label{eq:OmegaLambdaInf}
\end{equation} 

\begin{figure}[t!]
\begin{center} 
  \epsfxsize 2.25 truein \epsfbox {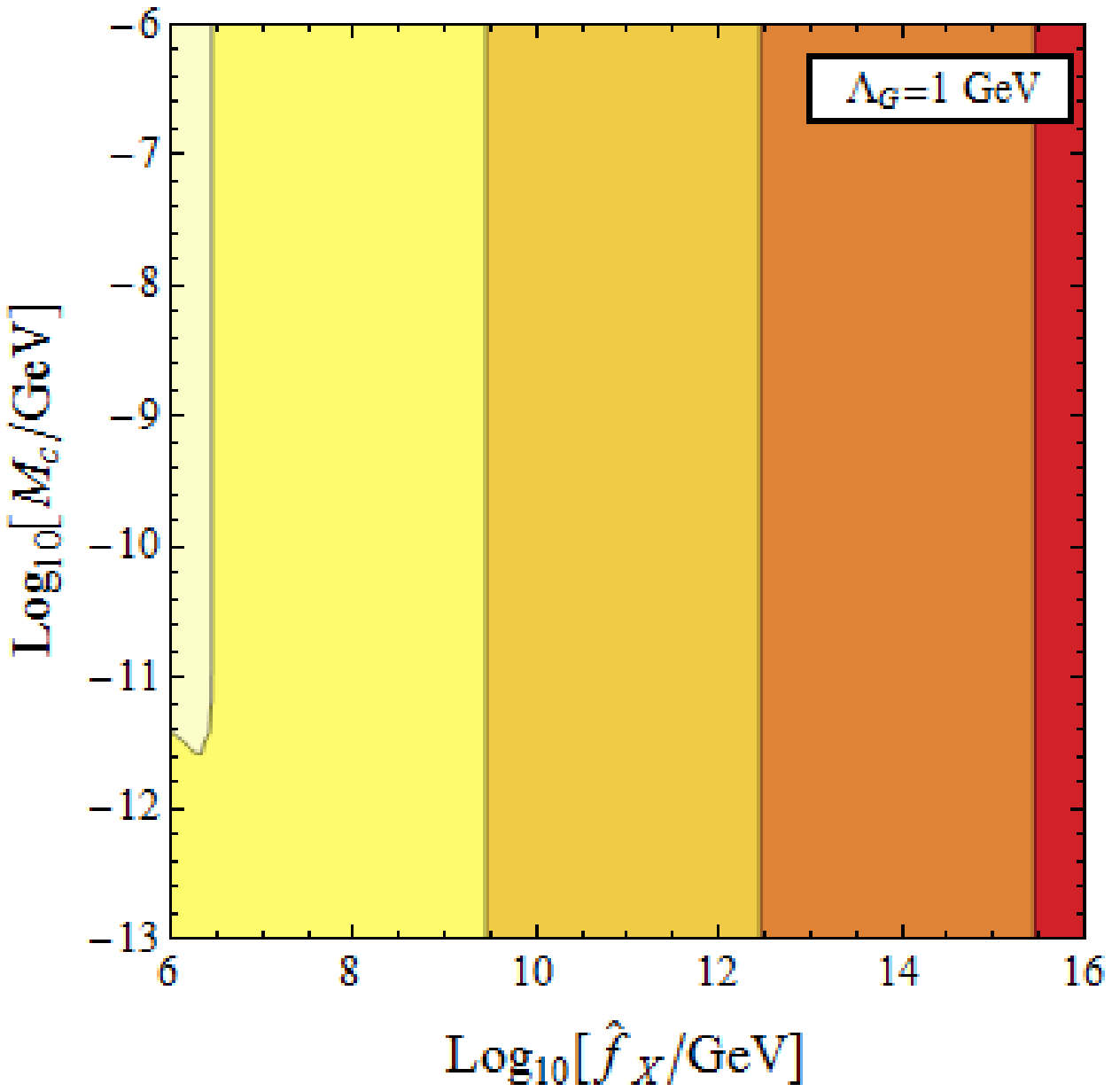}
  \epsfxsize 2.25 truein \epsfbox {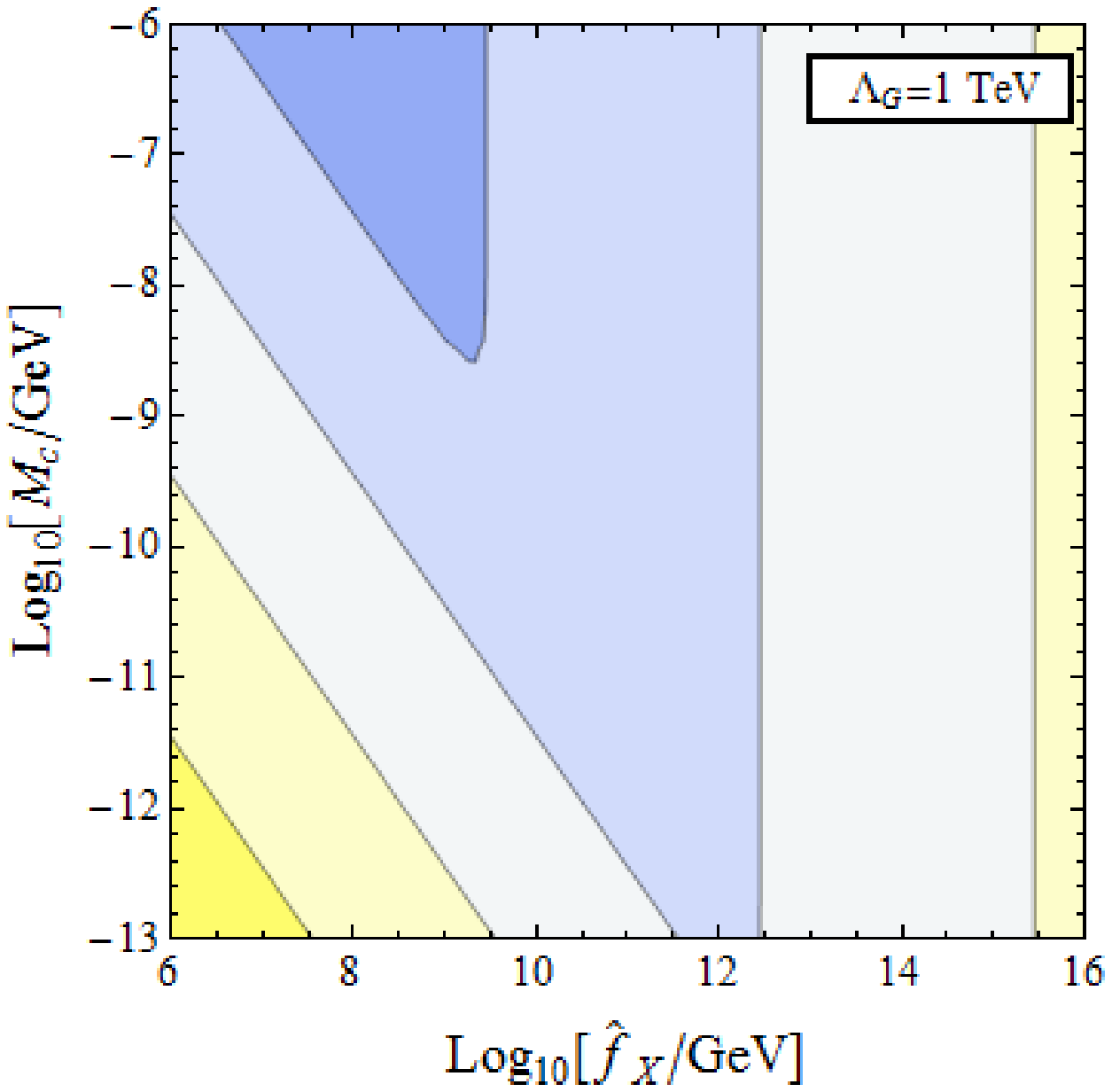}
  \epsfxsize 2.25 truein \epsfbox {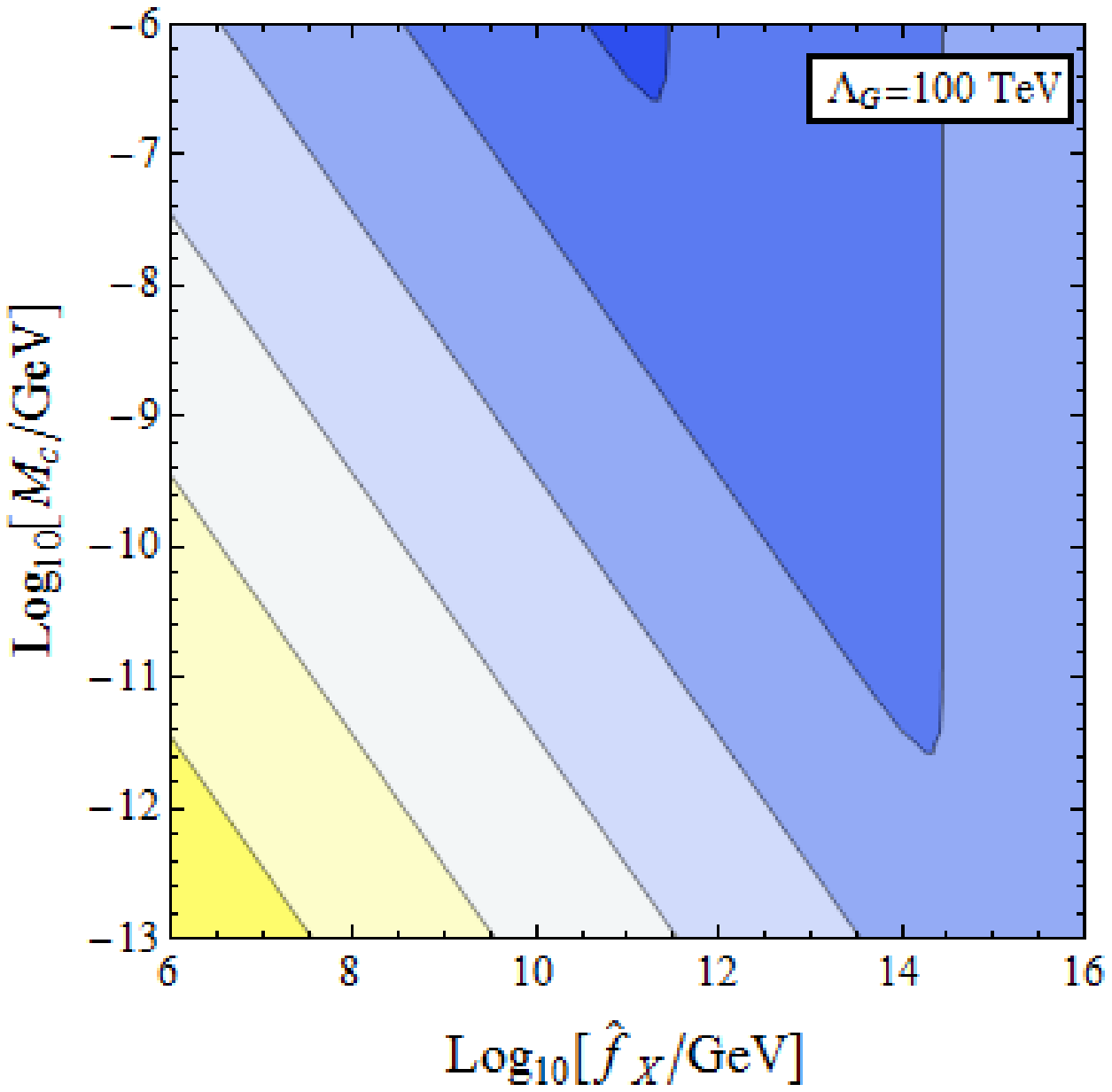}
  \raisebox{0.3cm}{\large$H_I^{\rm crit}$}
     \epsfxsize 5.00 truein \epsfbox {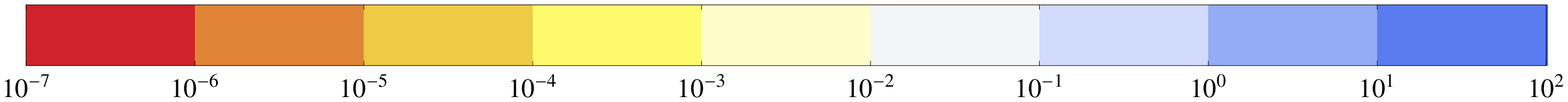}
\end{center}
\caption{Contours of the critical value $H_I^{\rm crit}$ in $(\fhatX, M_c)$
parameter space.  As discussed in the text, choosing $H_I \ll H_I^{\rm crit}$
guarantees that misalignment production dominates over 
vacuum fluctuations in determining the relic abundance 
$\widetilde{\Omega}_\lambda$ of all $a_\lambda$ in our DDM ensemble, 
as desired.  Here, we have taken $\TRH = 5$~MeV, 
$N_e = 100$, and $\xi = g_G = \theta = 1$ and assumed a photonic axion 
with $c_\gamma=1$.  The left, center, and right panels display 
the results for $\Lambda_G = 1$~GeV, $\Lambda_G = 1$~TeV, and 
$\Lambda_G = 100$~TeV, respectively.
\label{fig:HIMaxFlucPanelsLTR}}
\end{figure} 

To summarize, we see that the axion KK tower separates into three distinct regimes
within each of which different physics plays a principal role in determining 
$\widetilde{\Omega}_\lambda$.  In the $\lambda \lesssim \lambdafluc$ regime, 
the effect of vacuum fluctuations on 
$\widetilde{\Omega}_\lambda$ is negligible and the results in 
Sect.~\ref{sec:MisalignmentProd} continue to hold.  In the 
$\lambdafluc \lesssim \lambda \lesssim H_I$ regime, the opposite is true: vacuum 
fluctuations dominate and the abundances of the $a_\lambda$ are given by
Eq.~(\ref{eq:OmegaLambdaInf}).  Finally, in the 
$\lambda \gtrsim H_I$ regime, the wavelengths of even the lowest-lying momentum
modes of each $a_\lambda$ fall short of the Hubble length during the 
inflationary epoch.  Such modes therefore behave unambiguously 
like particles, and are consequently inflated away.
 
We are now ready to address the constraint 
we have imposed on the individual abundances $\widetilde{\Omega}_\lambda$ 
in Eq.~(\ref{eq:HubbleJitterCondit2}).  Since the effect of vacuum fluctuations 
is negligible both for $\lambda \gtrsim H_I$ and for $\lambda \lesssim \lambdafluc$, 
it follows that this constraint will be satisfied whenever $H_I \ll \lambdafluc$.  
Moreover, since $\lambdafluc$ itself decreases with increasing $H_I$, as indicated 
in Eq.~(\ref{eq:Lambdafluc}), we find that our constraint may be expressed in 
the form $H_I \ll H_I^{\rm crit}$, where $H_I^{\rm crit}$ is the value of the 
Hubble parameter during inflation for which $H_I = \lambdafluc$.  
In Fig.~\ref{fig:HIMaxFlucPanelsLTR}, we display contours of
$H_I^{\rm crit}$ as a function of the model parameters $\fhatX$, $M_c$, and $\Lambda_G$.
For the large values of $\Lambda_G$ characteristic of our preferred region of 
parameter space, we observe that the constraint in Eq.~(\ref{eq:HubbleJitterCondit2})
is satisfied for 
$H_I \ll H_I^{\rm crit} \sim \mathcal{O}(10 - 100\mathrm{~GeV})$.  For smaller 
values of $\Lambda_G$, although the constraint is certainly more severe, we 
nevertheless observe that the bound can be satisfied for 
$H_I \ll H_I^{\rm crit} \sim \mathcal{O}(10 - 100\mathrm{~keV})$.
This condition on $H_I$ has non-trivial implications for the construction of 
explicit inflationary models, since values of $H_I$ of this magnitude tend to be 
rather non-generic~\cite{TensorToScalarNonGeneric} among typical classes of 
inflationary potentials.  However, as discussed in Ref.~\cite{DynamicalDM2}, 
such a scale for $H_I$ is certainly not excluded 
(see, \eg, Refs.~\cite{LTRAxionsKamionkowski,lowscaleinflation}).  Moreover, a 
small value for $H_I$ fits naturally within the context of the LTR cosmology.


\subsection{Other Astrophysical Constraints on Light Axions\label{sec:AdditionalConstraints}}


In addition to the constraints we have discussed above, there exist 
a number of additional astrophysical and cosmological bounds on theories 
involving light axions and axion-like particles.  
Indeed, particles of this sort can give rise to a number of 
potentially observable effects~\cite{StringAxiverse}, such as a rotation of the CMB
polarization, modifications of the matter power spectrum, and the enhanced spindown 
of rotating black holes.  However, in order to give rise to 
observable effects of this sort, the particle in question must be exceedingly light, 
with a mass $m\lesssim 10^{-10}$~eV.  In the extra-dimensional scenario we are 
discussing here, the Newton's-law-modification constraint on the 
compactification scale $M_c$ stated in Eq.~(\ref{eq:MinimumMc}) implies 
that all $a_\lambda$ in the tower have masses $\lambda\gtrsim 10^{-3}$~eV in 
any scenario in which $y\lesssim 1$, \ie, in which the full tower of $a_\lambda$ 
contributes significantly to $\Omegatot$. 
Consequently, the additional constraints on ultra-light axions and axion-like fields
discussed in Ref.~\cite{StringAxiverse} are not relevant for our bulk-axion DDM model. 


\section{Synthesis:~ Combined Phenomenological Constraints 
on Axion Models of Dynamical Dark Matter\label{sec:Combined}}


In the previous section, we enumerated the individual astrophysical, phenomenological, 
and cosmological considerations which potentially constrain our bulk-axion DDM
model, and we evaluated the restrictions that each placed 
on the parameter space of this model.  In this section, we summarize how these
individual results, taken together, serve to constrain that parameter space.
Our particular interest concerns the preferred 
region of parameter space outlined in Ref.~\cite{DynamicalDM2}, namely 
$\fhatX \sim 10^{14} - 10^{15}$~GeV, $\Lambda_G \sim 10^2 - 10^5$~GeV, and $M_c$
chosen sufficiently small that $y \lesssim 1$.  Indeed, this is the region within
which the full KK tower contributes non-trivially to the total dark-matter 
relic abundance. 
          
\begin{figure}[b!]
\begin{center}
  \epsfxsize 2.25 truein \epsfbox {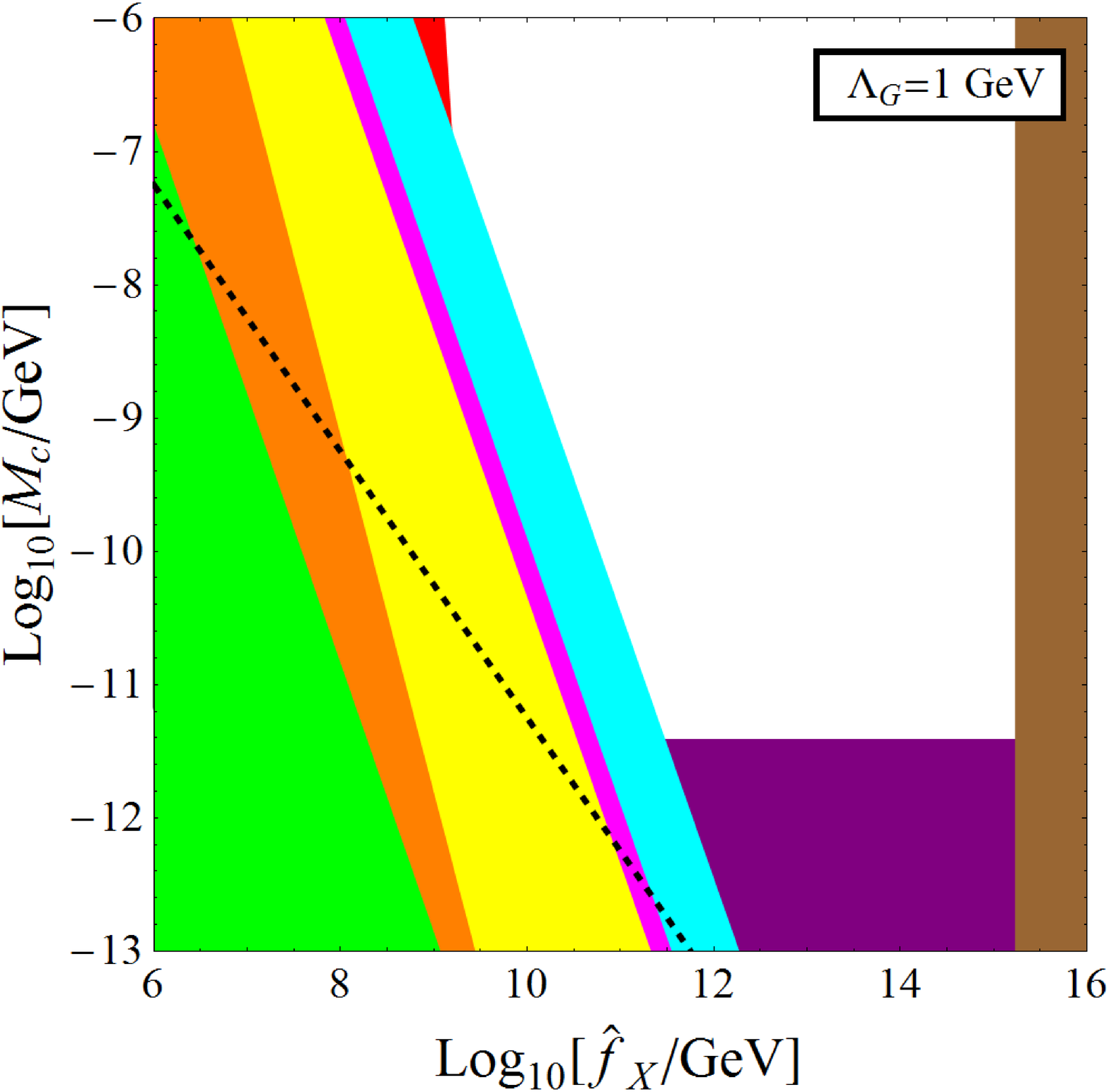}
  \epsfxsize 2.25 truein \epsfbox {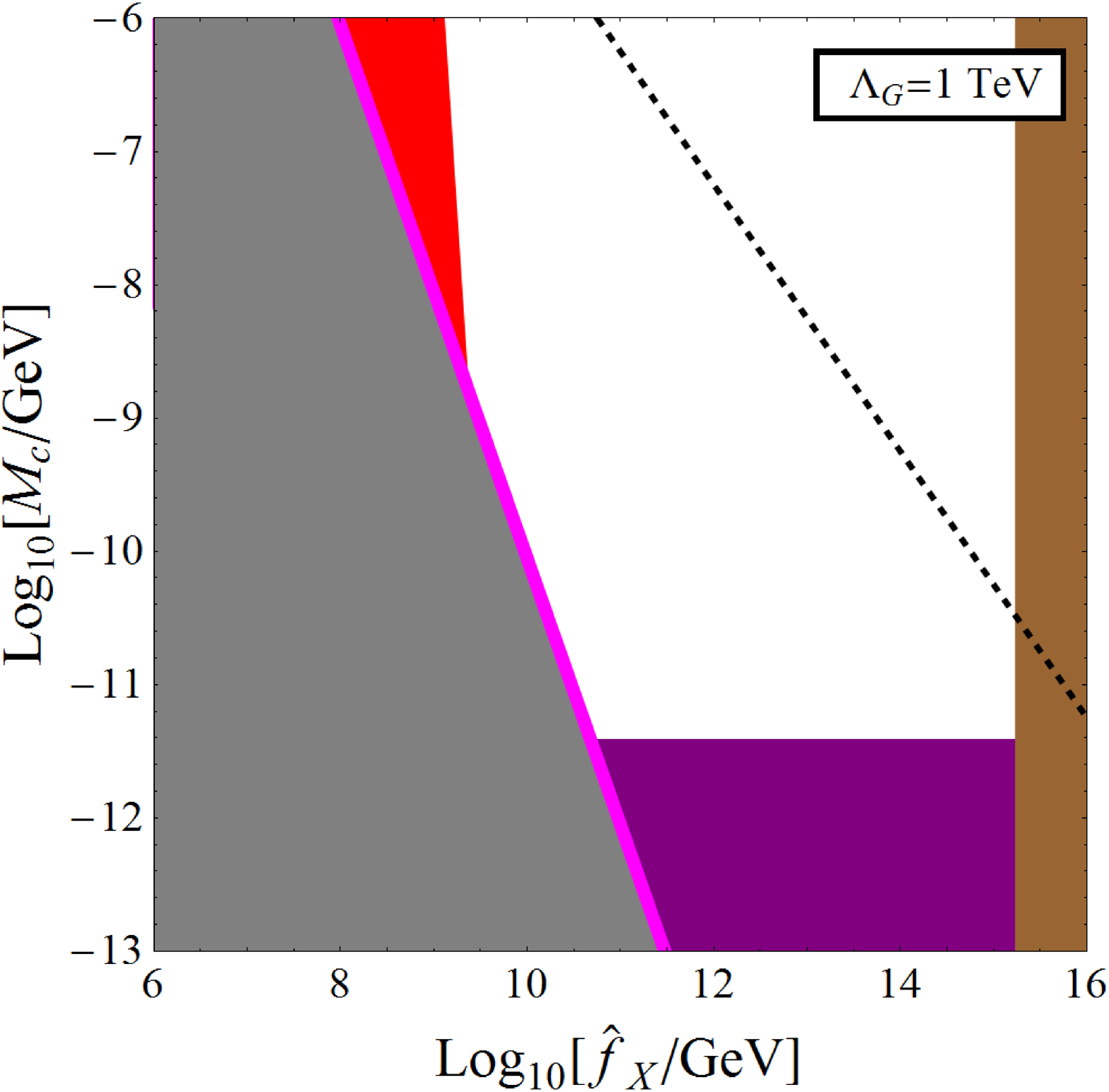}
  \epsfxsize 2.25 truein \epsfbox {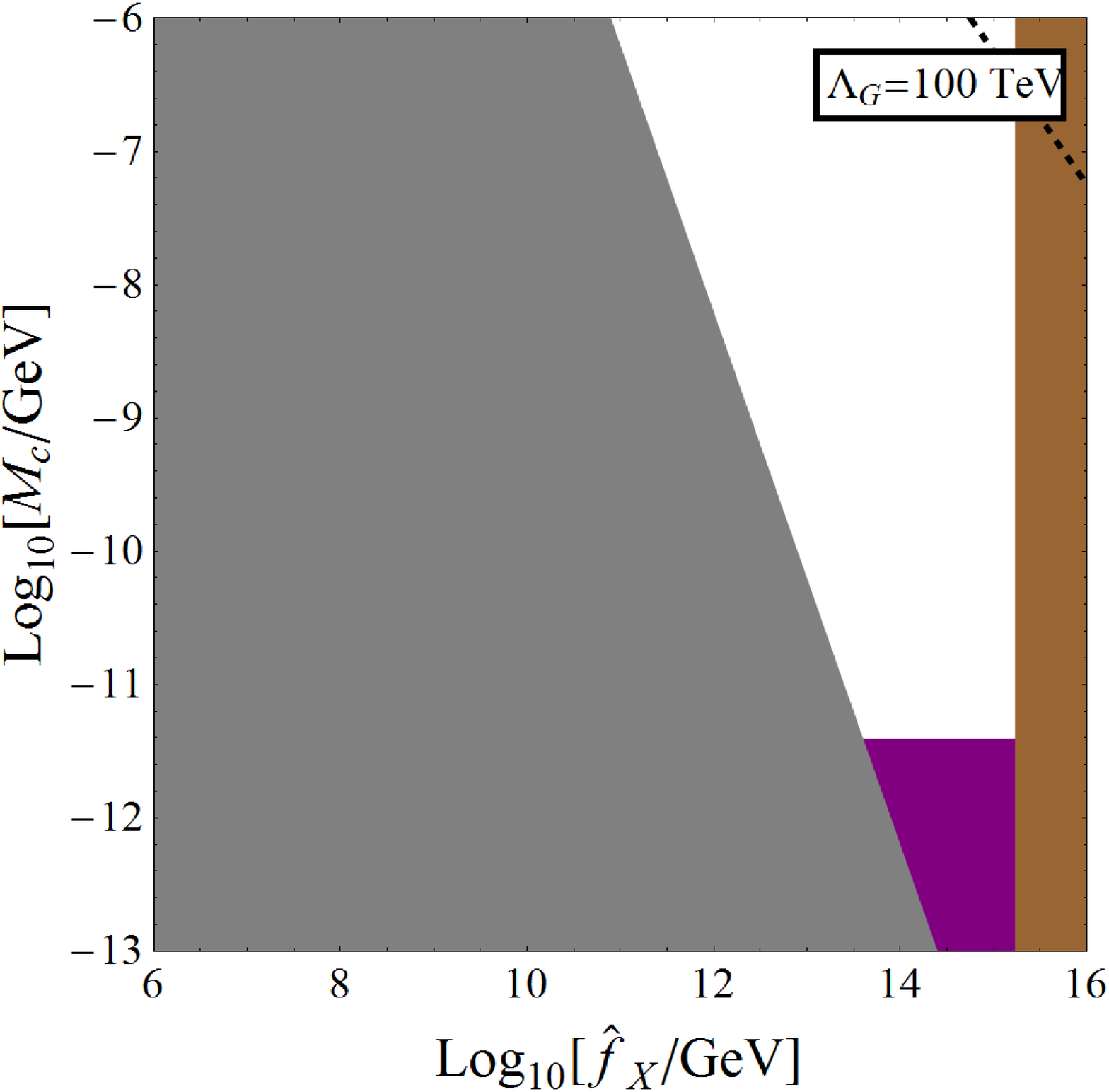}
\end{center}
\caption{Exclusion regions associated with all applicable phenomenological constraints 
discussed in this paper
for our bulk-axion DDM model with $\Lambda_G = 1$~GeV (left panel), 
$\Lambda_G = 1$~TeV (center panel), and $\Lambda_G = 100$~TeV (right panel).
In each case, we have taken $\xi=g_G=1$, $\TRH = 5$~MeV, and $H_I = 10^{-3}$~GeV,
and we have assumed that the axion only couples to the photon field with 
$c_\gamma=1$.  The shaded regions are respectively excluded by
data from helioscope measurements (red), collider considerations (magenta), 
tests of Newton's-law modifications via E\"{o}tv\"{o}s-type experiments (purple), 
measurements of the diffuse extragalactic X-ray and gamma-ray spectra (orange), 
observations of the lifetimes of globular-cluster stars (yellow), energy-loss
limits from supernova SN1987A (cyan), the model-consistency requirement 
$\Lambda_G < f_X$ (gray), overproduction of thermal axions (green), 
and the upper bound on the dark-matter relic abundance from WMAP (brown).  
The dashed black line corresponds to $y=\pi$; smaller values of $y$ correspond
to the region below and to the left of this line. 
\label{fig:MasterConstraintPlotPhotonic}}    
\begin{center}
  \epsfxsize 2.25 truein \epsfbox {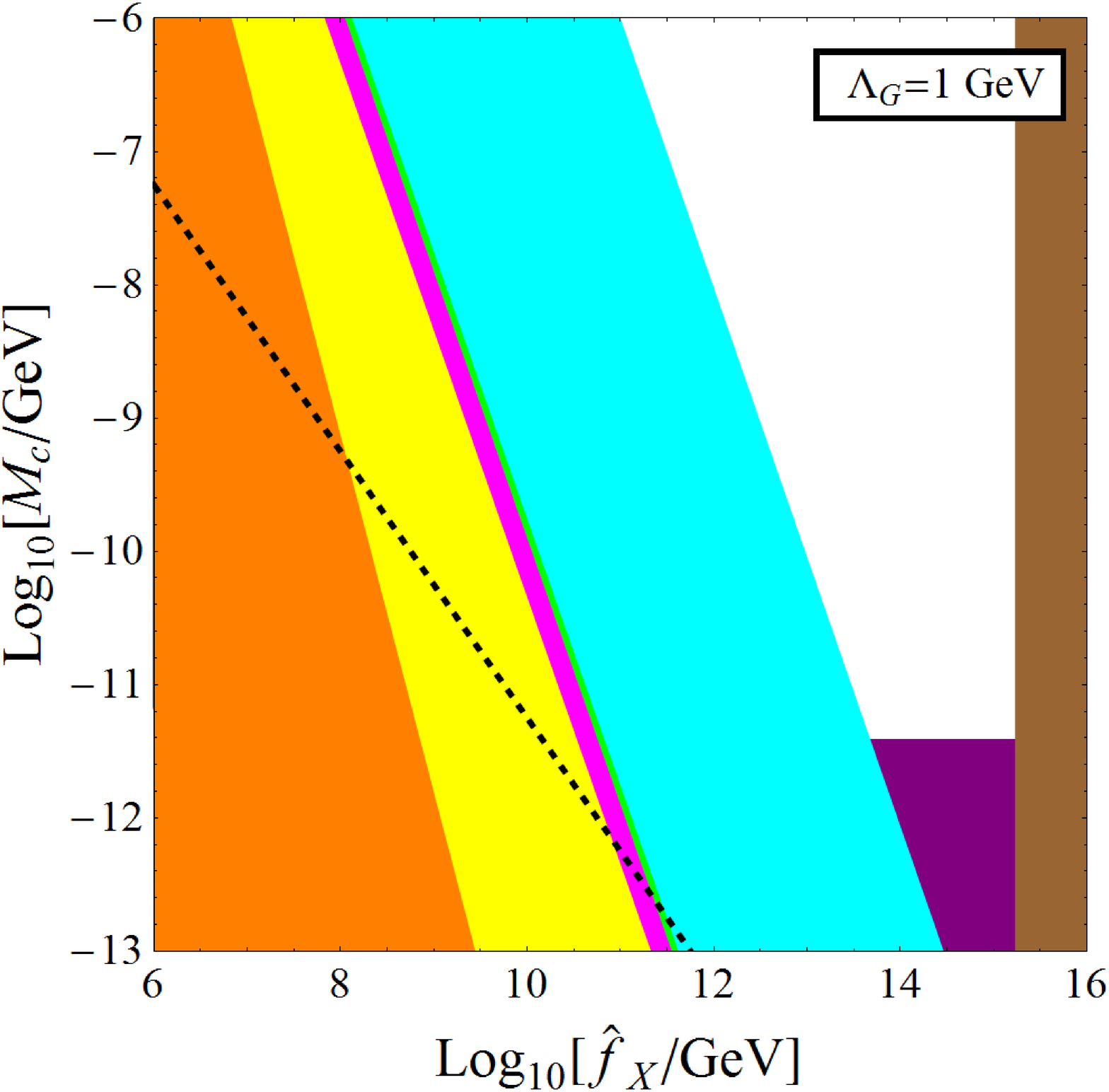}
  \epsfxsize 2.25 truein \epsfbox {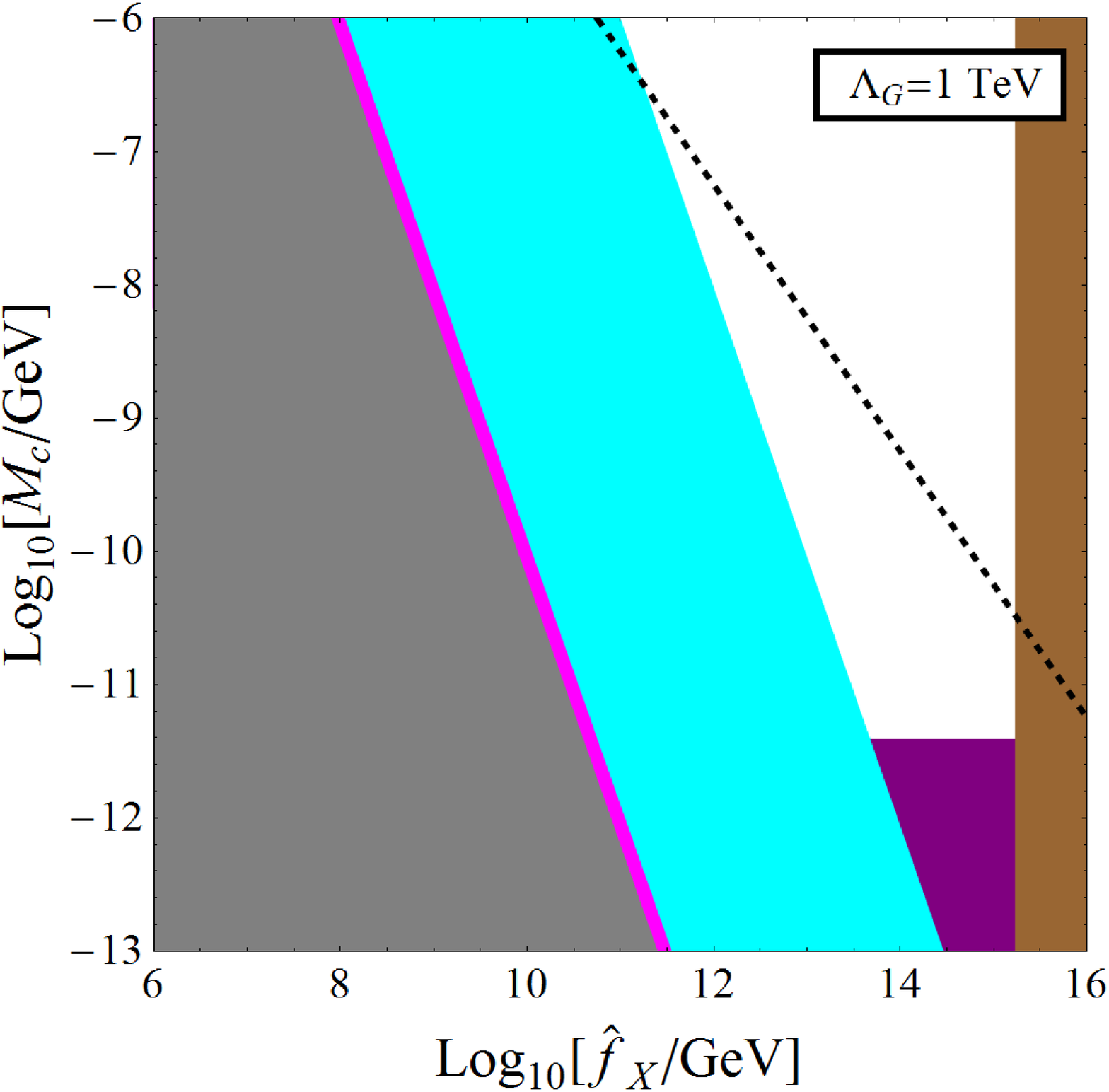}
  \epsfxsize 2.25 truein \epsfbox {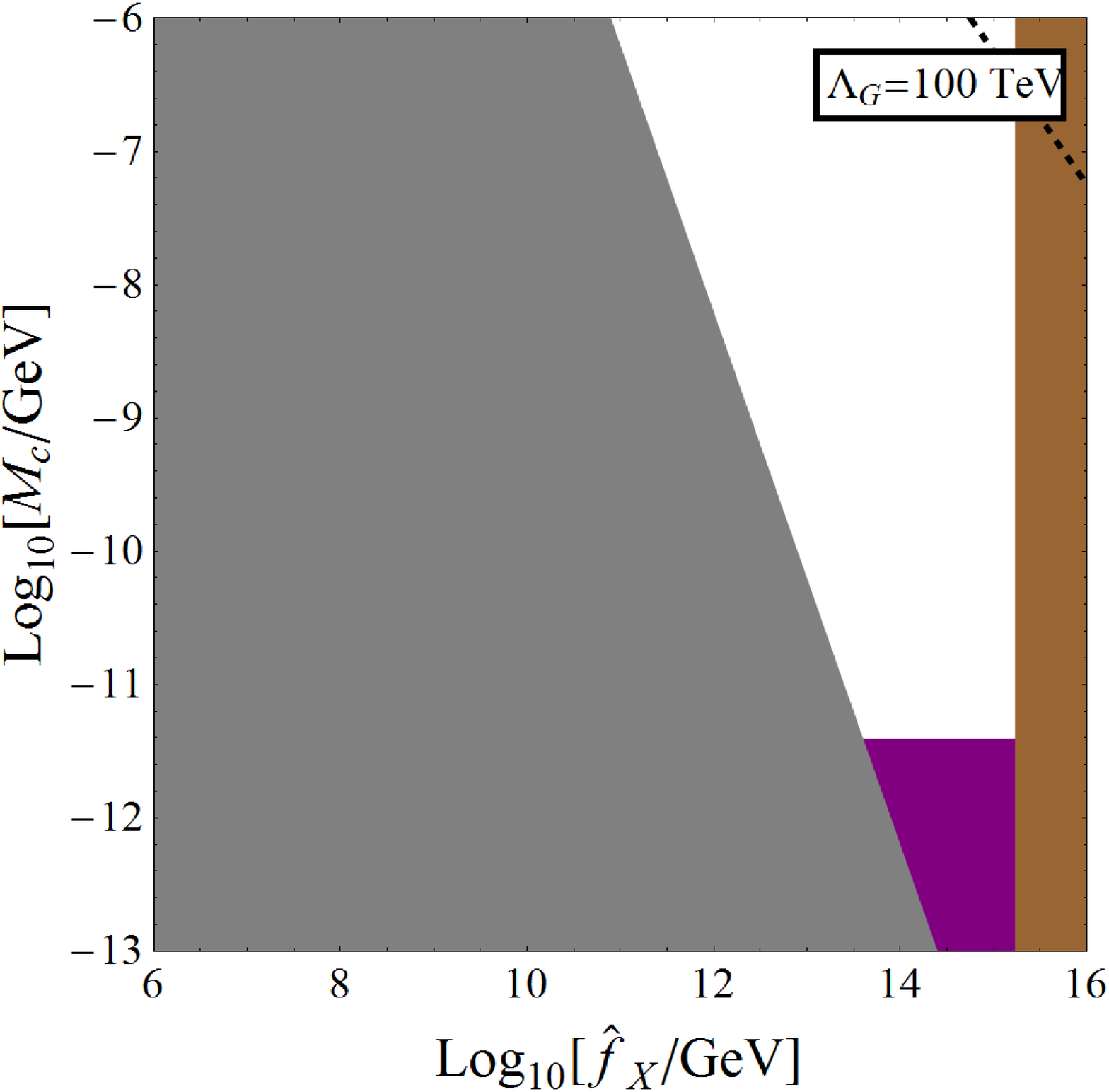}
\end{center}
\caption{Same as Fig.~\protect\ref{fig:MasterConstraintPlotPhotonic}, but for a 
``hadronic'' axion --- \ie, an axion which couples to both photons and gluons 
(and hence to pions, nucleons, and other hadrons), but not directly to SM quarks 
or leptons.  For these panels, we have taken $c_g = c_\gamma = 1$.
\label{fig:MasterConstraintPlotHadronic}}    
\end{figure} 

In Fig.~\ref{fig:MasterConstraintPlotPhotonic}, we show the combined exclusion 
regions for a purely photonic axion with $c_\gamma = 1$ for 
$\Lambda_G = 1$~GeV (left panel), $\Lambda_G = 1$~TeV (center panel), and 
$\Lambda_G = 100$~TeV (right panel).  The shaded regions displayed in each of the 
plots are excluded by the various considerations discussed in Sect.~\ref{sec:Bounds}. 
Specifically, the exclusion regions appearing in these panels are those associated with 
helioscope limits on solar axion production (red), collider considerations (magenta), 
tests of Newton's-law modifications via E\"{o}tv\"{o}s-type experiments (purple), 
measurements of the diffuse extragalactic X-ray and gamma-ray spectra (orange), 
observations of the lifetimes of globular-cluster stars (yellow), energy-loss
limits from supernova SN1987A (cyan), the model-consistency requirement 
$\Lambda_G < f_X$ discussed in Ref.~\cite{DynamicalDM2} (gray), and the $3\sigma$ 
upper bound on the dark-matter relic abundance from WMAP (brown).  
The additional requirement that the relic abundance be primarily determined by 
the misalignment mechanism (as envisioned in our DDM model) excludes the 
green-shaded region, within which a substantial population of $a_\lambda$ is 
generated via interactions with SM particles in the thermal bath.  
The remaining unshaded regions of parameter space are the regions 
within which our DDM model is consistent with all of these constraints.
The dashed black line indicates the contour $y = \pi$; smaller values of $y$ 
correspond to the region below and to the left of this line.  As discussed in
Ref.~\cite{DynamicalDM2}, we are particularly
interested in the unshaded region of parameter space which falls below and to the
left of this line, since this is the region within which not only are all of the
aforementioned constraints satisfied, but also the full tower of $a_\lambda$ 
contributes non-trivially to $\Omegatot$.      

As we see in Fig.~\ref{fig:MasterConstraintPlotPhotonic},    
for small $\Lambda_G$ the most stringent constraint on the parameter space of our 
model is the one derived from energy-loss limits from SN1987A.  The constraint from 
globular-cluster stars is also reasonably stringent, and the constraint derived 
from missing-energy processes such as $pp\rightarrow \gamma + \met$ at the LHC is 
estimated to be of roughly the same order.  However, as the $y=\pi$ contour 
superimposed over each panel in Fig.~\ref{fig:MasterConstraintPlotPhotonic} 
indicates, the full tower of $a_\lambda$ contributes significantly to $\Omegatot$
for all $\Lambda_G \gtrsim 100$~GeV.  Indeed, this is precisely the $\Lambda_G$ 
regime associated with the preferred region of parameter space for our model.
We therefore conclude that within this region of parameter space, 
a photonic bulk-axion DDM ensemble constitutes a viable dark-matter 
candidate.  

In Fig.~\ref{fig:MasterConstraintPlotHadronic}, we consider all of the same 
constraints as in Fig.~\ref{fig:MasterConstraintPlotPhotonic}, but for the
case of a hadronic axion with $c_g = c_\gamma=1$.  In this case, 
since the $a_\lambda$ couple to hadrons, the constraints from SN1987A and 
from axion production via interactions among the SM particles in the radiation 
bath both become even more stringent.  Again, as in the photonic-axion case, 
we find that the leading constraint for small $\Lambda_G$ is that from 
SN1987A, and that as $\Lambda_G$ increases, the model-consistency constraint 
becomes increasingly stringent.  However, as in the photonic-axion case, we 
see that within the preferred region of parameter space for our model, a hadronic 
bulk axion is also consistent with experimental and observational limits.  Thus 
a hadronic bulk-axion DDM ensemble is a viable dark-matter candidate as well.     

We also observe that the exclusion contours in 
Figs.~\ref{fig:MasterConstraintPlotPhotonic} 
and~\ref{fig:MasterConstraintPlotHadronic} associated 
with SN1987A energy-loss limits, globular-cluster-star evolution, collider constraints, 
and axion overproduction from SM particles in the radiation bath have the same slope.
This is because all of these constraints involve the production of light 
axions which are never directly detected, and thus involve physical processes whose 
amplitudes include a single coupling factor between the $a_\lambda$ and a 
pair of SM fields.  By contrast, the slopes of the constraint contours associated 
with other classes of physical processes can be quite different.  The
helioscope-constraint contour, for example, is related to processes in which 
axions are both produced and subsequently detected via their interactions with 
SM fields.  Likewise, the contour associated with 
limits on features in the diffuse X-ray and gamma-ray backgrounds is due to
processes involving the decays of a preexisting cosmological population of axions,
and therefore depends not only on the couplings of the $a_\lambda$ to SM fields,
but to their relative abundances as well.  The slopes of these constraint 
contours consequently differ from those which characterize the contours 
associated with SN1987A energy-loss limits, globular-cluster-star evolution, and 
so forth.     
   

\section{Discussion and Conclusions\label{sec:Conclusions}}


In Ref.~\cite{DynamicalDM1}, we proposed a new framework for dark-matter physics
which we call ``dynamical dark matter'' (DDM).  The fundamental idea underpinning
DDM is that the requirement of stability is replaced by a delicate balancing between 
lifetimes and cosmological abundances across a vast ensemble of individual 
dark-matter components.  If Ref.~\cite{DynamicalDM1}, we developed the general 
theoretical features of this new framework.  By contrast, in Ref.~\cite{DynamicalDM2},
we presented a ``proof of concept,'' namely an explicit realization of the DDM 
framework in which the DDM ensemble is realized as the infinite tower of KK 
excitations of an axion-like field propagating in the bulk of large extra spacetime 
dimensions. 

In this paper, we have completed this study by systematically investigating all of 
the experimental, astrophysical, and cosmological constraints which apply to this
DDM model.  Some of these constraints pertain to theories with large extra dimensions
in general, while others pertain specifically to our model.  Among the bounds
we have considered are constraints from limits on $a_\lambda$ production by 
astrophysical sources such as stars and supernovae; constraints related to 
the effects of late relic-axion decays on BBN, the CMB, and the diffuse 
X-ray and gamma-ray backgrounds; 
collider constraints on missing-energy processes such as $pp\rightarrow j + \met$ and
$pp\rightarrow \gamma + \met$; constraints on isocurvature perturbations generated 
as a consequence of misalignment production; constraints on the production of 
relativistic axions due to interactions in the thermal bath after inflation; and 
constraints on the direct detection of dark axions by microwave-cavity detectors and 
other, similar instruments.  We have verified that all of these constraints are 
satisfied within the preferred region of parameter space for our model --- namely, 
that in which the bulk-axion DDM ensemble accounts for the observed dark-matter 
relic abundance, while at the same time the full tower of axion modes contributes 
meaningfully to that abundance.  We therefore conclude that this bulk-axion DDM 
model is indeed phenomenologically viable, and that the overall DDM framework  
is a self-consistent alternative to traditional approaches to the dark-matter
problem.

While the focus of this paper has been on the specific bulk-axion 
DDM model presented in Ref.~\cite{DynamicalDM2}, we note that many of our 
results, and in many places our entire methodology, have a far  
wider range of applicability.   
For example, much of the formalism developed in Sect.~\ref{sec:Bounds} 
for evaluating the cosmological constraints on decaying dark matter in 
our bulk-axion DDM model is applicable to any model in which the dark 
sector comprises a large number of fields.   This is true for issues as diverse 
as BBN, diffuse photon backgrounds, or stellar cooling.  
Likewise, irrespective of issues pertaining to dark-matter physics,
many of our results and techniques may have applicability to theories 
with large numbers of axions, such as the recently discussed ``axiverse''
theories~\cite{StringAxiverse,StringAxiversePheno}.
Thus, we believe that the methods developed and employed in this paper can 
serve as a prototype for future phenomenological studies of not only the 
DDM framework, but also, more generally, any theories in which there exist 
large numbers of interacting and decaying particles.
    

\section*{Acknowledgments}


We would like to thank K.~Abazajian, Z.~Chacko, M.~Drees, J.~Feng, J.~Kumar, 
R.~Mohapatra, M.~Ramsey-Musolf, S.~Su, T.~Tait, S.-H.~H.~Tye, 
X.~Tata, and N. Weiner for discussions.
KRD is supported in part by the U.S. Department of Energy
under Grant DE-FG02-04ER-41298 and by the National Science Foundation through
its employee IR/D program.  BT is supported in part by DOE grant 
DE-FG02-04ER41291.  The opinions and conclusions expressed here are those of
the authors, and do not represent either the Department of Energy or 
the National Science Foundation.



\begin{references}


\bibitem{DynamicalDM1}
  K.~R.~Dienes and B.~Thomas,
  ``Dynamical Dark Matter: I. Theoretical Overview,''
  arXiv:1106.4546 [hep-ph], to appear in Phys.\ Rev.\  D.

\bibitem{DynamicalDM2}
  K.~R.~Dienes and B.~Thomas,
  ``Dynamical Dark Matter: II. An Explicit Model,''
  arXiv:1107.0721 [hep-ph], to appear in Phys.\ Rev.\  D.

\bibitem{DDGAxions}
  K.~R.~Dienes, E.~Dudas and T.~Gherghetta,
  Phys.\ Rev.\  D {\bf 62}, 105023 (2000)
  [arXiv:hep-ph/9912455].

\bibitem{PecceiQuinn}
  R.~D.~Peccei and H.~R.~Quinn,
  Phys.\ Rev.\ Lett.\  {\bf 38}, 1440 (1977);
  Phys.\ Rev.\  D {\bf 16}, 1791 (1977).

\bibitem{WeinbergWilczekAxion}
  S.~Weinberg,
  Phys.\ Rev.\ Lett.\  {\bf 40}, 223 (1978); \\
  F.~Wilczek,
  Phys.\ Rev.\ Lett.\  {\bf 40}, 279 (1978).

\bibitem{ADD}
  N.~Arkani-Hamed, S.~Dimopoulos and G.~R.~Dvali,
  Phys.\ Lett.\  B {\bf 429}, 263 (1998)
  [arXiv:hep-ph/9803315].

\bibitem{ADDPhenoBounds}
  N.~Arkani-Hamed, S.~Dimopoulos and G.~R.~Dvali,
  Phys.\ Rev.\  D {\bf 59}, 086004 (1999)
  [arXiv:hep-ph/9807344].

\bibitem{KSVZ}
  J.~E.~Kim,
  Phys.\ Rev.\ Lett.\  {\bf 43}, 103 (1979);\\
  M.~A.~Shifman, A.~I.~Vainshtein and V.~I.~Zakharov,
  Nucl.\ Phys.\  B {\bf 166}, 493 (1980).

\bibitem{StringAxiverse}
  A.~Arvanitaki, S.~Dimopoulos, S.~Dubovsky, N.~Kaloper and J.~March-Russell,
  Phys.\ Rev.\  D {\bf 81}, 123530 (2010)
  [arXiv:0905.4720 [hep-th]].

\bibitem{WMAP}
  E.~Komatsu {\it et al.}  [WMAP Collaboration],
  Astrophys.\ J.\ Suppl.\  {\bf 180}, 330 (2009)
  [arXiv:0803.0547 [astro-ph]].

\bibitem{GravFeynmanRulesGiudice}
  G.~F.~Giudice, R.~Rattazzi and J.~D.~Wells,
  Nucl.\ Phys.\  B {\bf 544}, 3 (1999)
  [arXiv:hep-ph/9811291].

\bibitem{GravFeynmanRulesHan}
  T.~Han, J.~D.~Lykken and R.~J.~Zhang,
  Phys.\ Rev.\  D {\bf 59}, 105006 (1999)
  [arXiv:hep-ph/9811350].

\bibitem{QGPAxionProductionRate}
  P.~Graf and F.~D.~Steffen,
  arXiv:1008.4528 [hep-ph].

\bibitem{ChangChoiThermalNucleonRate}
  S.~Chang and K.~Choi,
  Phys.\ Lett.\  B {\bf 316}, 51 (1993)
  [arXiv:hep-ph/9306216].

\bibitem{RaffeltPionAxionCrossSection}
  S.~Hannestad, A.~Mirizzi and G.~Raffelt,
  JCAP {\bf 0507}, 002 (2005)
  [arXiv:hep-ph/0504059].

\bibitem{LTRAxionsKamionkowski}
  D.~Grin, T.~L.~Smith and M.~Kamionkowski,
  Phys.\ Rev.\  D {\bf 77}, 085020 (2008)
  [arXiv:0711.1352 [astro-ph]].

\bibitem{QEDPlasmaAxionProductionRate}
  M.~Bolz, A.~Brandenburg and W.~Buchmuller,
  Nucl.\ Phys.\  B {\bf 606}, 518 (2001)
  [Erratum-ibid.\  B {\bf 790}, 336 (2008)]
  [arXiv:hep-ph/0012052].

\bibitem{RaffeltSubMeV}
  D.~Cadamuro, S.~Hannestad, G.~Raffelt and J.~Redondo,
  JCAP {\bf 1102}, 003 (2011)
  [arXiv:1011.3694 [hep-ph]].

\bibitem{KapnerModGrav}
  D.~J.~Kapner, T.~S.~Cook, E.~G.~Adelberger, J.~H.~Gundlach, B.~R.~Heckel, 
  C.~D.~Hoyle and H.~E.~Swanson,
  Phys.\ Rev.\ Lett.\  {\bf 98}, 021101 (2007)
  [arXiv:hep-ph/0611184].

\bibitem{HannestadRaffeltNeutronStar}
  S.~Hannestad and G.~G.~Raffelt,
  Phys.\ Rev.\ Lett.\  {\bf 88}, 071301 (2002)
  [arXiv:hep-ph/0110067].

\bibitem{KKGravitons1987A}
  C.~Hanhart, J.~A.~Pons, D.~R.~Phillips and S.~Reddy,
  Phys.\ Lett.\  B {\bf 509}, 1 (2001)
  [arXiv:astro-ph/0102063].

\bibitem{HannestadRaffeltSupernovae}
  S.~Hannestad and G.~Raffelt,
  Phys.\ Rev.\ Lett.\  {\bf 87}, 051301 (2001)
  [arXiv:hep-ph/0103201].

\bibitem{EDHubbleCline}
  J.~M.~Cline, C.~Grojean and G.~Servant,
  Phys.\ Rev.\ Lett.\  {\bf 83}, 4245 (1999)
  [arXiv:hep-ph/9906523].

\bibitem{EDHubbleBinetruy}
  P.~Binetruy, C.~Deffayet, U.~Ellwanger and D.~Langlois,
  Phys.\ Lett.\  B {\bf 477}, 285 (2000)
  [arXiv:hep-th/9910219].

\bibitem{EDHubbleShiromizu}
  T.~Shiromizu, K.~i.~Maeda and M.~Sasaki,
  Phys.\ Rev.\  D {\bf 62}, 024012 (2000)
  [arXiv:gr-qc/9910076].

\bibitem{CosmoConstraintsLargeED}
  L.~J.~Hall and D.~Tucker-Smith,
  Phys.\ Rev.\  D {\bf 60}, 085008 (1999)
  [arXiv:hep-ph/9904267].

\bibitem{TReheatLimits}
  S.~Hannestad,
  Phys.\ Rev.\  D {\bf 70}, 043506 (2004)
  [arXiv:astro-ph/0403291].

\bibitem{KawasakiLTR2}
  M.~Kawasaki, K.~Kohri and N.~Sugiyama,
  Phys.\ Rev.\ Lett.\  {\bf 82}, 4168 (1999)
  [arXiv:astro-ph/9811437].

\bibitem{ATLASMonojet33pb}
  G.~Aad {\it et al.}  [ATLAS Collaboration],
  Phys.\ Lett.\ B {\bf 705}, 294 (2011)
  [arXiv:1106.5327 [hep-ex]].

\bibitem{ATLASMonojet1fb}
  ATLAS Collaboration, ATLAS-CONF-2011-096.

\bibitem{CMSMonojet}
  CMS Collaboration, CMS-PAS-EXO-11-059.

\bibitem{CMSLargeEDDiphoton}
  S.~Chatrchyan {\it et al.}  [CMS Collaboration],
  JHEP {\bf 1105}, 085 (2011)
  [arXiv:1103.4279 [hep-ex]].

\bibitem{CMSLargeEDDimuon}
  A.~Ferapontov,
  arXiv:1109.1187 [hep-ex].

\bibitem{KimReview2}
  J.~E.~Kim and G.~Carosi,
  Rev.\ Mod.\ Phys.\  {\bf 82}, 557 (2010)
  [arXiv:0807.3125 [hep-ph]].

\bibitem{JaeckelReview}
  J.~Jaeckel and A.~Ringwald,
  Ann.\ Rev.\ Nucl.\ Part.\ Sci.\  {\bf 60}, 405 (2010)
  [arXiv:1002.0329 [hep-ph]].

\bibitem{CAST}
  D.~M.~Lazarus, G.~C.~Smith, R.~Cameron, A.~C.~Melissinos, G.~Ruoso, 
  Y.~K.~Semertzidis and F.~A.~Nezrick,
  Phys.\ Rev.\ Lett.\  {\bf 69}, 2333 (1992).

\bibitem{DAMAAxion} 
  R.~Bernabei, P.~Belli, R.~Cerulli, F.~Montecchia, F.~Nozzoli, A.~Incicchitti, D.~Prosperi, C.~J.~Dai {\it et al.},
  Phys.\ Lett.\ B {\bf 515}, 6 (2001);\\
  R.~Bernabei, P.~Belli, F.~Cappella, R.~Cerulli, F.~Montecchia, F.~Nozzoli, A.~Incicchitti, D.~Prosperi {\it et al.},
  Riv.\ Nuovo Cim.\  {\bf 26N1}, 1 (2003).
  [astro-ph/0307403].

\bibitem{TEXONOAxion} 
  H.~M.~Chang {\it et al.}  [TEXONO Collaboration],
  Phys.\ Rev.\ D {\bf 75}, 052004 (2007)
  [hep-ex/0609001].

\bibitem{SOLAXAxion} 
  F.~T.~Avignone, III {\it et al.}  [SOLAX Collaboration],
  Phys.\ Rev.\ Lett.\  {\bf 81}, 5068 (1998)
  [astro-ph/9708008].

\bibitem{COSMEAxion} 
  A.~Morales {\it et al.}  [COSME Collaboration],
  Astropart.\ Phys.\  {\bf 16}, 325 (2002)
  [hep-ex/0101037].

\bibitem{ADMX}
  S.~J.~Asztalos {\it et al.}  [The ADMX Collaboration],
  Phys.\ Rev.\ Lett.\  {\bf 104}, 041301 (2010)
  [arXiv:0910.5914 [astro-ph.CO]].

\bibitem{CARRACK}
  K.~Yamamoto, M.~Tada, Y.~Kishimoto, M.~Shibata, K.~Kominato, T.~Ooishi, 
  S.~Yamada and T.~Saida {\it et al.},
  hep-ph/0101200.

\bibitem{RaffeltSN1987ABound}
  G.~G.~Raffelt,
  Lect.\ Notes Phys.\  {\bf 741}, 51 (2008)
  [arXiv:hep-ph/0611350].

\bibitem{MassoALPsLongPaper}
  E.~Masso and R.~Toldra,
  Phys.\ Rev.\  D {\bf 52}, 1755 (1995)
  [arXiv:hep-ph/9503293].

\bibitem{PDG}
  C.~Amsler {\it et al.}  [Particle Data Group],
  Phys.\ Lett.\  B {\bf 667}, 1 (2008).

\bibitem{SolarLifetimeBounds}
  P.~Gondolo and G.~Raffelt,
  Phys.\ Rev.\  D {\bf 79}, 107301 (2009)
  [arXiv:0807.2926 [astro-ph]].

\bibitem{GiudiceGravitonColliders}
  G.~F.~Giudice, T.~Plehn and A.~Strumia,
  Nucl.\ Phys.\  B {\bf 706}, 455 (2005)
  [arXiv:hep-ph/0408320].

\bibitem{HuAndSilkLong}
  W.~Hu and J.~Silk,
  Phys.\ Rev.\  D {\bf 48}, 485 (1993).

\bibitem{HuAndSilkShort}
  W.~Hu and J.~Silk,
  Phys.\ Rev.\ Lett.\  {\bf 70}, 2661 (1993).

\bibitem{DeZotti}
  L.~Danese and G.~De~Zotti,
  Riv. Nuovo Cimento Soc. Ital. Fis. 7, 227 (1977).

\bibitem{HEAOdiffXRB}
  R.~L.~Kinzer, G.~V.~Jung, D.~E.~Gruber, J.~L.~Matteson and L.~E.~Peterson,
  1997, Astrophys.\ J.\ {\bf 475} 361 (1997);\\
  D.~E.~Gruber, J.~L.~Matteson, L.~E.~Peterson and G.~V.~Jung,
  arXiv:astro-ph/9903492.

\bibitem{COMPTELdiffXRB}
  S.~C.~Kappadath {\it et al.},
  BAAS 30 (2), 926 (1998);\\
  S.~C.~Kappadath, Ph.D. Thesis,
  {\tt http://wwwgro.sr.unh.edu/users/ckappada/ckappada.html}. 

\bibitem{ChandraDeepFieldData}
  W.~N.~Brandt {\it et al.},
  Astron.\ J.\  {\bf 122}, 2810 (2001)
  [arXiv:astro-ph/0108404];\\
  R.~Giacconi {\it et al.},
  Astrophys.\ J.\ Suppl.\  {\bf 139}, 369 (2002)
  [arXiv:astro-ph/0112184].

\bibitem{DMDecayChenKamionkowski1}
  X.~L.~Chen and M.~Kamionkowski,
  Phys.\ Rev.\  D {\bf 70}, 043502 (2004)
  [arXiv:astro-ph/0310473];\\
  L.~Zhang, X.~Chen, M.~Kamionkowski, Z.~g.~Si and Z.~Zheng,
  Phys.\ Rev.\  D {\bf 76}, 061301 (2007)
  [arXiv:0704.2444 [astro-ph]].

\bibitem{EGRETdiffGRB}
  A.~W.~Strong, I.~V.~Moskalenko and O.~Reimer,
  Astrophys.\ J.\  {\bf 613}, 956 (2004)
  [arXiv:astro-ph/0405441].

\bibitem{FERMIdiffGRB}
  A.~A.~Abdo {\it et al.}  [The Fermi-LAT collaboration],
  Phys.\ Rev.\ Lett.\  {\bf 104}, 101101 (2010)
  [arXiv:1002.3603 [astro-ph.HE]].

\bibitem{ChandraPowerLawFitHickox}
  R.~C.~Hickox and M.~Markevitch,
  Astrophys.\ J.\  {\bf 645}, 95 (2006)
  [arXiv:astro-ph/0512542].

\bibitem{CyburtEllisBBN}
  R.~H.~Cyburt, J.~R.~Ellis, B.~D.~Fields and K.~A.~Olive,
  Phys.\ Rev.\  D {\bf 67}, 103521 (2003)
  [arXiv:astro-ph/0211258].

\bibitem{KawasakiHadronicBBN}
  M.~Kawasaki, K.~Kohri and T.~Moroi,
  Phys.\ Lett.\  B {\bf 625}, 7 (2005)
  [arXiv:astro-ph/0402490];
  Phys.\ Rev.\  D {\bf 71}, 083502 (2005)
  [arXiv:astro-ph/0408426].

\bibitem{HertzbergAxionCosmology}
  M.~P.~Hertzberg, M.~Tegmark and F.~Wilczek,
  Phys.\ Rev.\  D {\bf 78}, 083507 (2008)
  [arXiv:0807.1726 [astro-ph]].

\bibitem{LiddleAndLyth}
  A.~R.~Liddle and D.~H.~Lyth,
  ``Cosmological inflation and large scale structure,''
  Cambridge, UK: Univ. Pr. (2000).

\bibitem{WMAPInflation} 
  H.~V.~Peiris {\it et al.} [WMAP Collaboration],
  Astrophys.\ J.\ Suppl.\ \ {\bf 148}, 213  (2003)
  [astro-ph/0302225].

\bibitem{LindeInflationReview} 
  A.~D.~Linde,
  Chur, Switzerland: Harwood (1990) 362 p. (Contemporary concepts in physics, 5)
  [hep-th/0503203].

\bibitem{lowscaleinflation}
  L.~Randall and S.~D.~Thomas,
  Nucl.\ Phys.\ B {\bf 449}, 229 (1995)
  [arXiv:hep-ph/9407248];\\
  G.~German, G.~G.~Ross and S.~Sarkar,
  Nucl.\ Phys.\  B {\bf 608}, 423 (2001)
  [arXiv:hep-ph/0103243];\\
  M.~Giovannini,
  Phys.\ Rev.\ D {\bf 67}, 123512 (2003)
  [arXiv:hep-ph/0301264];\\
  K.~Dimopoulos, D.~H.~Lyth and Y.~Rodriguez,
  JHEP {\bf 0502}, 055 (2005)
  [arXiv:hep-ph/0411119];\\
  P.~Q.~Hung, E.~Masso and G.~Zsembinszki,
  JCAP {\bf 0612}, 004 (2006)
  [arXiv:astro-ph/0609777];\\
  R.~Allahverdi, K.~Enqvist, J.~Garcia-Bellido, A.~Jokinen and A.~Mazumdar,
  JCAP {\bf 0706}, 019 (2007)
  [arXiv:hep-ph/0610134];\\
  R.~Allahverdi, B.~Dutta and K.~Sinha,
  Phys.\ Rev.\ D {\bf 81}, 083538 (2010)
  [arXiv:0912.2324 [hep-th]];\\
  G.~G.~Ross and G.~German,
  Phys.\ Lett.\  B {\bf 691}, 117 (2010)
  [arXiv:1002.0029 [hep-ph]].

\bibitem{TensorToScalarNonGeneric}
  L.~A.~Boyle, P.~J.~Steinhardt and N.~Turok,
  Phys.\ Rev.\ Lett.\  {\bf 96}, 111301 (2006)
  [arXiv:astro-ph/0507455].

\bibitem{StringAxiversePheno}
  K.~J.~Mack and P.~J.~Steinhardt,
  JCAP {\bf 1105}, 001 (2011)
  [arXiv:0911.0418 [astro-ph.CO]];\\
  A.~Arvanitaki and S.~Dubovsky,
  Phys.\ Rev.\ D {\bf 83}, 044026 (2011)
  [arXiv:1004.3558 [hep-th]];\\
  B.~S.~Acharya, K.~Bobkov and P.~Kumar,
  JHEP {\bf 1011}, 105 (2010)
  [arXiv:1004.5138 [hep-th]];\\
  B.~S.~Acharya, G.~Kane and E.~Kuflik,
  arXiv:1006.3272 [hep-ph];\\
  D.~J.~E.~Marsh and P.~G.~Ferreira,
  Phys.\ Rev.\ D {\bf 82}, 103528 (2010)
  [arXiv:1009.3501 [hep-ph]];\\
  D.~J.~E.~Marsh,
  Phys.\ Rev.\ D {\bf 83}, 123526 (2011)
  [arXiv:1102.4851 [astro-ph.CO]].


\end{references}
\end{document}